%% file: device_design.tex
\documentclass[amsmath,notitlepage,twocolumn,pra,aps]{revtex4-1}
\usepackage{hhline}
\input{device_design_header}
\usepackage{fancyhdr}
\begin{document}
\title{Optically Loaded Semiconductor Quantum Memory Register}
\author{Danny Kim}
\author{Andrey A. Kiselev, Richard S. Ross, Matthew T. Rakher, Cody Jones}
\author{Thaddeus D. Ladd}\email{tdladd@hrl.com}
\affiliation{HRL Laboratories, LLC, 3011 Malibu Canyon Road, Malibu, CA 90265, USA}
\date\today

\begin{abstract}
We propose and analyze an optically loaded quantum memory exploiting capacitive coupling between self-assembled quantum dot molecules and electrically gated quantum dot molecules.  The self-assembled dots are used for spin-photon entanglement, which is transferred to the gated dots for long-term storage or processing via a teleportation process heralded by single-photon detection.  We illustrate a device architecture enabling this interaction and we outline its operation and fabrication.  We
provide self-consistent Poisson-Schr\"odinger simulations to establish the design viability and refine the design, and to estimate the physical coupling parameters and their sensitivities to dot placement.  The device we propose generates heralded copies of an entangled state between a photonic qubit and a solid-state qubit with a rapid reset time upon failure.  The resulting fast rate of entanglement generation is of high utility for heralded quantum networking scenarios involving lossy optical channels.
\end{abstract}

\maketitle
\thispagestyle{fancy}

\section{Introduction}
    \input{device_design_introduction}

    \label{operation}
\section{Principle of Operation}
    \input{device_design_operation}
\section{Sketch of Fabrication Process}
    \input{device_design_fabrication}

\section{Simulation}
    \label{simulation}
    \input{device_design_simulation}
\section{Discussion}
\label{discussion}
    \input{device_design_discussion}

\section{Conclusion and Outlook}
\label{conclusion}

Hybrid technologies that exploit the salient properties of various systems and combine their functionalities is a research thrust that we believe will become more prevalent in future quantum device technology.  Here, we have proposed a hybridization of two semiconductor technologies for an optically interfaced quantum memory device.  It utilizes the brightest and fastest single photon source known and potentially long registers of gated, coupled electron-spin qubits.  The device principles presented here are not meant to be specific to  III-V semiconductors, but instead provide a framework that can readily be transferable to other systems.  However, at this time, this device is most readily envisioned using III-V materials due to their mature fabrication techniques in both the optical and electrical quantum dots.

We envision that the device proposed here will be one of many available technologies in a wide-area quantum network.  The speed and the brightness of single photons generated is the advantage here, where this device would excel in lossy environments.  However, as this device requires cryogenics, other technologies  may be more optimized for say, ground-to-satellite communication, or more storage-focused implementations.

Future work involves research at all levels of the device.  The constituent elements---self assembled dots and gated quantum dots---are continually under research scrutiny around the world and advances will improve coherence times, photon out-coupling, material quality, and hybridization techniques. In the immediate future, rudimentary versions of this device would be of strong interest in order to investigate coupling between the two types of dots and address the issues presented in this paper.  The overall aim is to then further advance the use and capabilities of semiconductors for coherent optoelectronics.

\acknowledgments
We acknowledge fruitful discussions with Jake Taylor, Duncan Steel, Dan Gammon, and Sven H\"ofling.

\bibliographystyle{apsrev4-1}
\bibliography{device_design}

\end{document}

%% file: device_design_introduction.tex
\label{sketch}
\begin{figure}
	\includegraphics[width=\columnwidth]{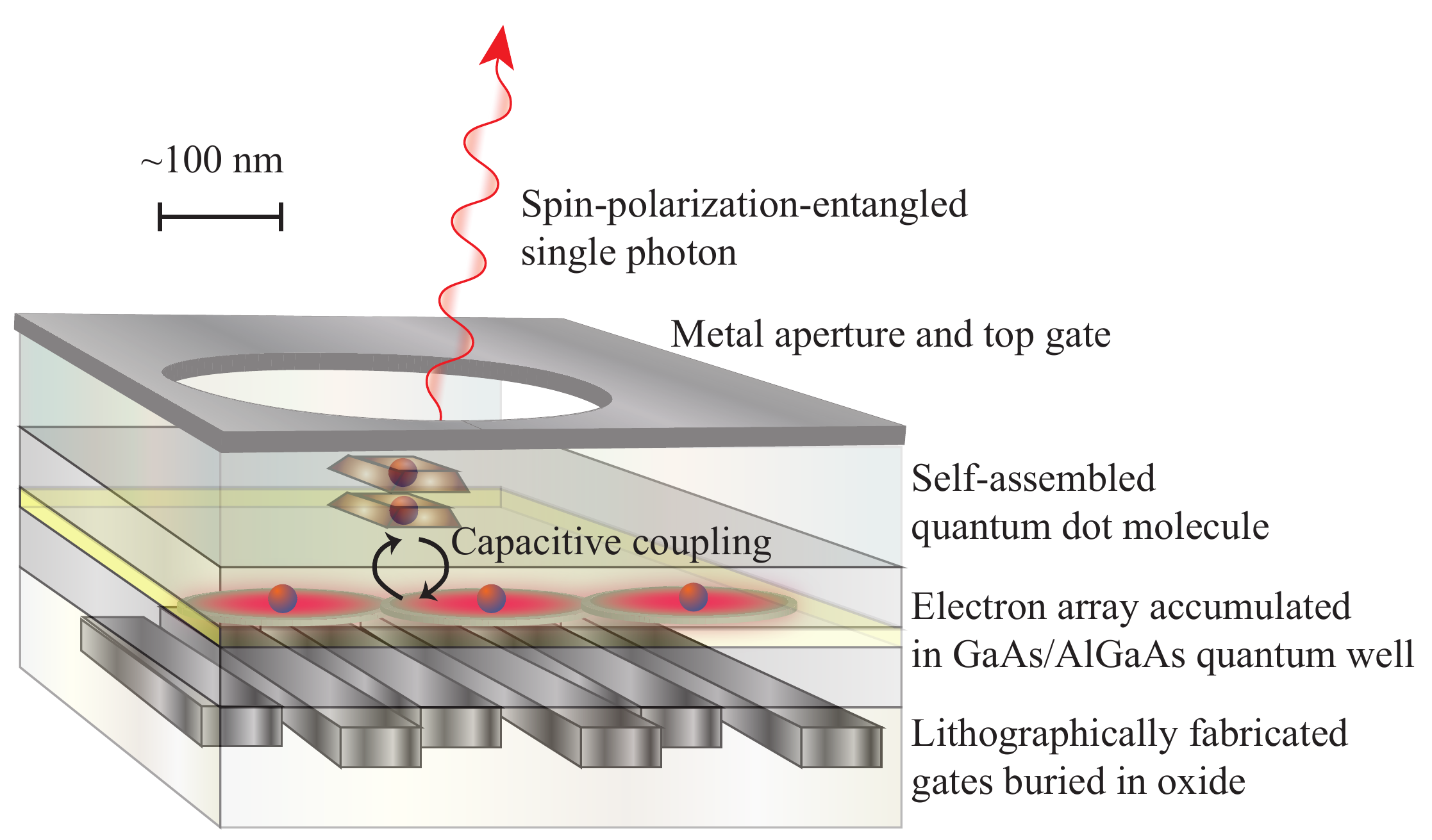}
	\caption{A sketch of the proposed device structure in which a self-assembled quantum dot molecule, whose internal spins are entangled to an emitted photon, is capacitively coupled to a gated triple quantum dot.}
	\label{devicesketch}
\end{figure}

Quantum communication refers to the ability to connect distant parties using coherent carriers of quantum information.  The fundamental quantum properties of those carriers, especially with respect to measurement, enable applications in cryptography \cite{gisinrmp}, metrology \cite{metrology}, and computing~\cite{Barz20012012}.  Quantum Key Distribution (QKD) is perhaps the principal application of interest, and has been experimentally implemented using systems including single photon sources and detectors~\cite{waks_nature}, weak coherent light \cite{GrangierQKD,PhysRevLett.98.010503,TakesueQKD}, and entangled photon sources \cite{QKDentangled}.  All of these systems, however, have a fundamental limit in either communication rate or communication distance, ultimately limited by the rate of photon loss in the communication channel.  Such photon loss cannot be naively compensated for using amplification, as this would require copying or dephasing the quantum information carried by the channel.  Solutions to compensate for loss instead use quantum teleportation and error management routines.  Single channel systems are referred to as quantum repeaters, while more complicated connections and applications are referred to as quantum networks~\cite{RodText}.

Quantum repeater and quantum networking systems depend on the ability to catch, buffer, and process quantum information.  For photonic channels, this means providing an interface between single photons and stationary quantum memories, and an ability to perform some level of quantum logic on those quantum memories.  Proposals to provide this functionality exist in many media, ranging from atomic gases~\cite{dlcz} to trapped ions~\cite{monroesci} to superconductor optomechanics~\cite{painterncomms,lehnertsci}.  A completely solid-state approach would be desirable for manufacturability, stability, and potentially superior performance.  Most notably, solid-state approaches based on rare-earth crystalline impurities are superb optical buffers, holding the coherent information from photons for as long as hours~\cite{zhong_optically_2015}.  At present, all these technologies have properties best suited for one task over another. Their limitations have inspired work into hybridizing these systems in order to exploit complementary strengths in disparate systems. Some examples include coupling single dots to atomic vapors~\cite{akopian_natphot}, single dots to single trapped ions~\cite{meyer_qdandion}, and superconducting qubits to diamond NV-centers~\cite{zhu_nature}.  Research of this type is in its nascent stage and significant issues of coupling and efficiency have created opportunity for further innovation.

In this article we present a proposal for an all semiconductor approach to a quantum optical interface and quantum memory leveraging the strong optical interface of self-assembled quantum dots (SAQDs) and the highly-controllable long-lived memory in gate-defined quantum dots (GQDs).  The optical interface is provided by the ability of a SAQD or pair of such dots to emit a single photon whose polarization is entangled to the persistent fine-structure of the dot complex \cite{degrevenat,gaonat,shaibleyprl}.  The quantum memory is provided by arrays of gate-defined quantum dots employing encoded exchange interactions \cite{medford_self-consistent_2013,HRL}.  The interface between the two is provided by a capacitive controlled-phase operation enabled by charge dipoles formed by tunneling in both a self-assembled quantum dot molecule and a gate-defined quantum dot molecule~\cite{shulmansci}. This device, roughly sketched in \reffig{devicesketch} and elaborated in the remainder of this article, hybridizes two separate but related technologies and aims to exploit the salient properties of each.

The structure of this article is as follows.  In \refsec{rol} we first motivate our scheme in the context of the substantial existing work in quantum coherent semiconductor optoelectronics.  In \refsec{operation} we introduce the elements of the basic device, imagined to be implemented based on III-V semiconductor heterostructure engineering, and provide a protocol for entangling spins in a gated quantum dot array with the polarization of a single photon.  We refer to Ref.~\onlinecite{cody2015} to illustrate how larger chains of repeaters or networks might be assembled from such devices.  In \refsec{fabrication}, we briefly discuss a possible route toward fabrication of the device, and in \refsec{simulation} we present results of self-consistent Poisson-Sch\"odinger simulations for estimating the charge stability of the device and the strength of the required interactions.  In Section~\ref{discussion} we survey likely sources of decoherence and other error mechanisms.

\section{Relation to Other Approaches in Quantum Coherent Semiconductor Optoelectronics}
\label{rol}

\begin{table*}
	\begin{tabular}{|r|l|}
		\hline\hline
		QKD   & Quantum Key Distribution\\
		QW    & Quantum Well\\
		SAQD  & Self Assembled Quantum Dot \\
		SAQDM & Self Assembled Quantum Dot Molecule \\
		GQD   & Gated Quantum Dot \\
		GQDM  & Gated Quantum Dot Molecule \\
        2DEG  & 2-Dimensional Electron Gas\\
		B,T	  & Subscripts denoting occupation of bottom or top dot in the SAQDM\\
		O     & Subscript referring to SAQDM, which is \textbf{O}ptically controlled \\
		E     & Subscript referring to GQDM, which is \textbf{E}lectrically controlled\\
		$\gamma$ & Subscript referring to the single photon emitted by the SAQDM\\
        ?     & Subscript referring to other unmeasured degrees of freedom in the quantum network\\
		CZ    & Controlled-phase\\
        DD    & Subscript indicating dipole-dipole\\
        ALD   & Atomic Layer Deposition\\
        DCS   & Dot Charge Sensor\\
        AG    & Aperture Gate\\
		\hline\hline
	\end{tabular}
	\caption{Acronyms used in this article.}
    \label{acronyms}
\end{table*}

Classical semiconductor optoelectronics is a highly mature field, featuring a large number of semiconductor-based technologies for buffering, switching, and routing optical signals.  However, providing coherent connections between single photons and semiconductor devices is substantially more challenging, and the number of demonstrated technologies is far fewer.  Still, the tremendous progress in semiconductor device engineering developed for classical optoelectronics has provided and will continue to provide substantial assistance in engineering methods to couple single, coherent photons to quantum memories.  In the existing research on quantum coherent semiconductor optoelectronics relevant to the present proposal, we distinguish two types of schemes: those that employ self-assembled quantum dots in conjunction with gated structures, and those that employ gated structures only.

Schemes employing only gated structures typically involve photon absorption via creation of a quantum-well or quantum-dot exciton.  Due to the engineered band structure, the photon polarization coherently transfers to an electron spin-state and the hole is lost~\cite{vrijen_spin-coherent_2001,rao_single_2005,croke2008quantum,kosaka_spin_2009,tarucha}.  One key challenge in these schemes is the small absorption cross section of a shallow, gated quantum dot, which limits efficiency.  In contrast, the single-photon extraction efficiency from single self-assembled quantum dots may be engineered to be quite high~\cite{arcari_prl}, motivating the combination of gated dots with self-assembled quantum dot single photon sources.

The combination of self-assembled quantum dots and gated quantum dots for coherent optoelectronics have been under consideration for some time; see in particular Ref.~\onlinecite{engel2006}. A key differentiating feature of the approach we describe here is the use of double self-assembled quantum dots and triple gated quantum dots.  Since quantum dots behave like atoms, these coupled-dot complexes behave like molecules; we refer to self-assembled quantum dot molecules as SAQDMs, and gated quantum dot molecules as GQDMs.  See Table~\ref{acronyms} for a list of acronyms used in this article. The use of engineered artificial molecules add initial complexity in fabrication, but they enhance functionality in two key ways.

One function provided by quantum-dot molecules is complete coherent control of a qubit without depending on large or inhomogeneous magnetic fields.  For both dot types, qubits are formed via the singlet and triplet spin subspaces, with a single axis of qubit control provided by the exchange interaction.  In the case of a SAQDM, full control is then rendered by the use of optically accessed exciton states which enable coherent single-spin rotations on an orthogonal axis~\cite{dkimnphys}, thus mixing the singlet and triplet states.  Similarly for the GQDM, exchange interaction with a third electron in a third dot enables a second axis of spin rotation, giving full electrical control~\cite{divincenzo_universal_2000,lairdprb,medford_self-consistent_2013,laddprb,HRL}.

The second function provided by dot molecules is the ability to coherently couple the dot types while avoiding direct tunneling between the two species, except during initial loading of electrons.  This is important because the charging energies for the two types of dots will typically be very different, making coherent, near-resonant tunneling difficult.  Further, a physical distance provided by an epitaxial semiconductor buffer is desired to prevent too strong a disorder potential in the gated quantum dot layer, as it is crucial that our buried gates define the dots, not uncontrolled disorder.  Fortunately, direct tunneling is not required for coherent transfer of quantum information; it has long been proposed~\cite{taylornphys} and recently demonstrated~\cite{shulmansci} that capacitive coupling between exchange-coupled pairs of quantum dots can accomplish multi-qubit logic.  This interaction may be thought of as an electric dipole-dipole interaction, in which the dipole is defined by charge tunneling between a pair of quantum dots.  This charge motion is coupled to spins encoding qubits by virtue of Pauli blockade in which, within a certain energy range, electrons in singlet spin states are capable of occupying the same dot and are therefore free to tunnel, while electrons in triplet states are energetically forbidden from occupying the same dot and therefore do not tunnel.

A number of pioneering experimental studies combining self-assembled quantum dots with buried-channel, gated structures provide guidance to our scheme.  There have been significant prior efforts to understand the effect of SAQDs on carriers in a nearby 2-dimensional electron gas (2DEG) and vice versa.  Early efforts were motivated by the possibility of creating a transistor where the gate bias modified the mobility rather than the carrier density of the 2DEG.  Thus, many experiments have investigated the influence of nearby SAQDs on the mobility of carriers in the 2DEG~\cite{Sakaki_APL_1995, Kim_APL_1998, Wang_APL_2000, Dettwiler_ArXiv_2014, Kurzmann_JAP_2015, Marquardt_PhysE_2008, Marquardt_APL_2011, Russ_PRB_2006, Zhukov_PRB_2003}.  These experiments measure slight reductions in mobility due to the SAQDs, however values as high as $0.5\times10^6$ cm$^2$/Vs are observed even in the presence of a SAQD layer 45~nm away~\cite{Dettwiler_ArXiv_2014}.  In fact, some groups have used the ability to controllably charge the SAQDs to explore the nature of the metal-insulator transition in the 2DEG~\cite{Kim_PRB_2000, Kim_PRB_2004, Ribeiro_PRB_1998, Ribeiro_PRL_1999}.  These experiments are sensitive to deformations of the potential energy landscape of the 2DEG through local density modifications and charge scattering from the SAQDs.  Limited studies have also examined the effect of the 2DEG on the optical properties of the SAQDs and shown that for barriers of 30~nm, ensemble emission spectra are largely unperturbed~\cite{Ribeiro_JAP_2000,Kim_APL_2005}.  In addition, transport through the 2DEG has been used to detect charge occupation in the SAQD layer~\cite{Finley_APL_1998,Yusa_APL_1997,Muller_APL_2008,Marquardt_APL_2009,Marquardt_NComm_2011,Nowozin_APL_2014}, and enable single photon detection by extension \cite{shields_detection_2000,rowe_single-photon_2006}.  More recent investigations have started to explore the coupling between SAQDs and microcavity polaritons~\cite{Puri_PRB_2014}.  Taken together, these experiments pave the way toward quantum coherent optoelectronic devices and provide an important foundation to the work described here.

%% file: device_design_operation.tex
\label{operation}
\begin{figure}
\includegraphics[width=\columnwidth]{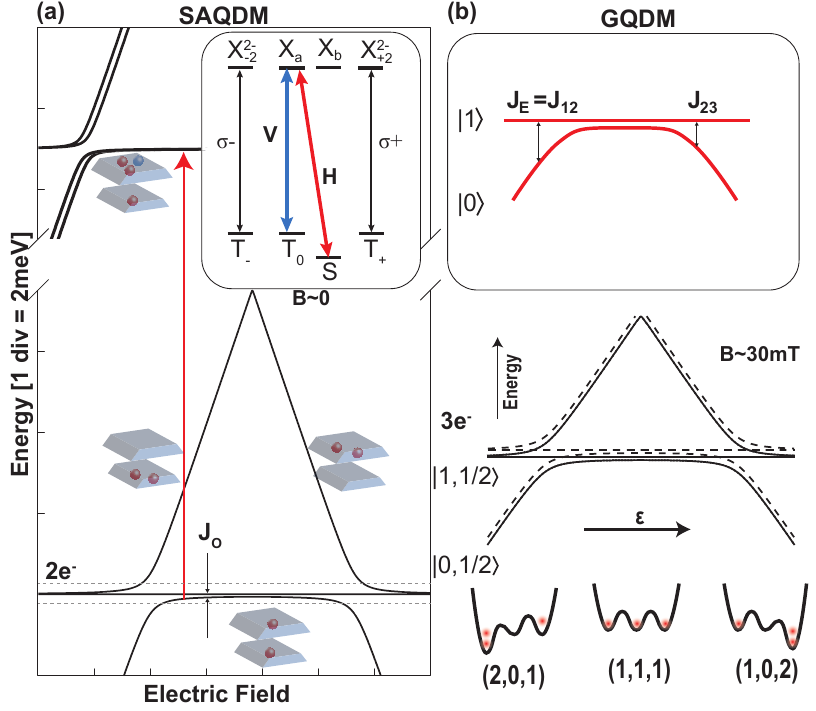}
\caption{(a) Hund-Mulliken model of the SAQDM system showing the energy levels of the ground and excited states. $\JO$ is the exchange energy separating the singlet and triplet states. Optical transitions from the ground to excited states have energies of about 1.3~eV.  Dotted lines show the lifting of the triplet degeneracy in a magnetic field. (inset) selection rules at zero B-field, written in the basis where the singlet/triplet states are excited to a a superpositon of trion states with orthogonal polarization.  (b) GQDM: (lower) energy  level diagram of a three-electron qubit in a GQDM- for the basis states see text. Dotted lines represent a separate manifold, offset in energy by the electron Zeeman splitting.   There are two exchange couplings that rotate the qubit on orthogonal axis, $\JE=J_{12}$ and $J_{23}$.  The upper inset is the simplified diagram where only relevant states are shown.}
\label{levels}
\end{figure}

\begin{figure*}
\includegraphics[width=\textwidth]{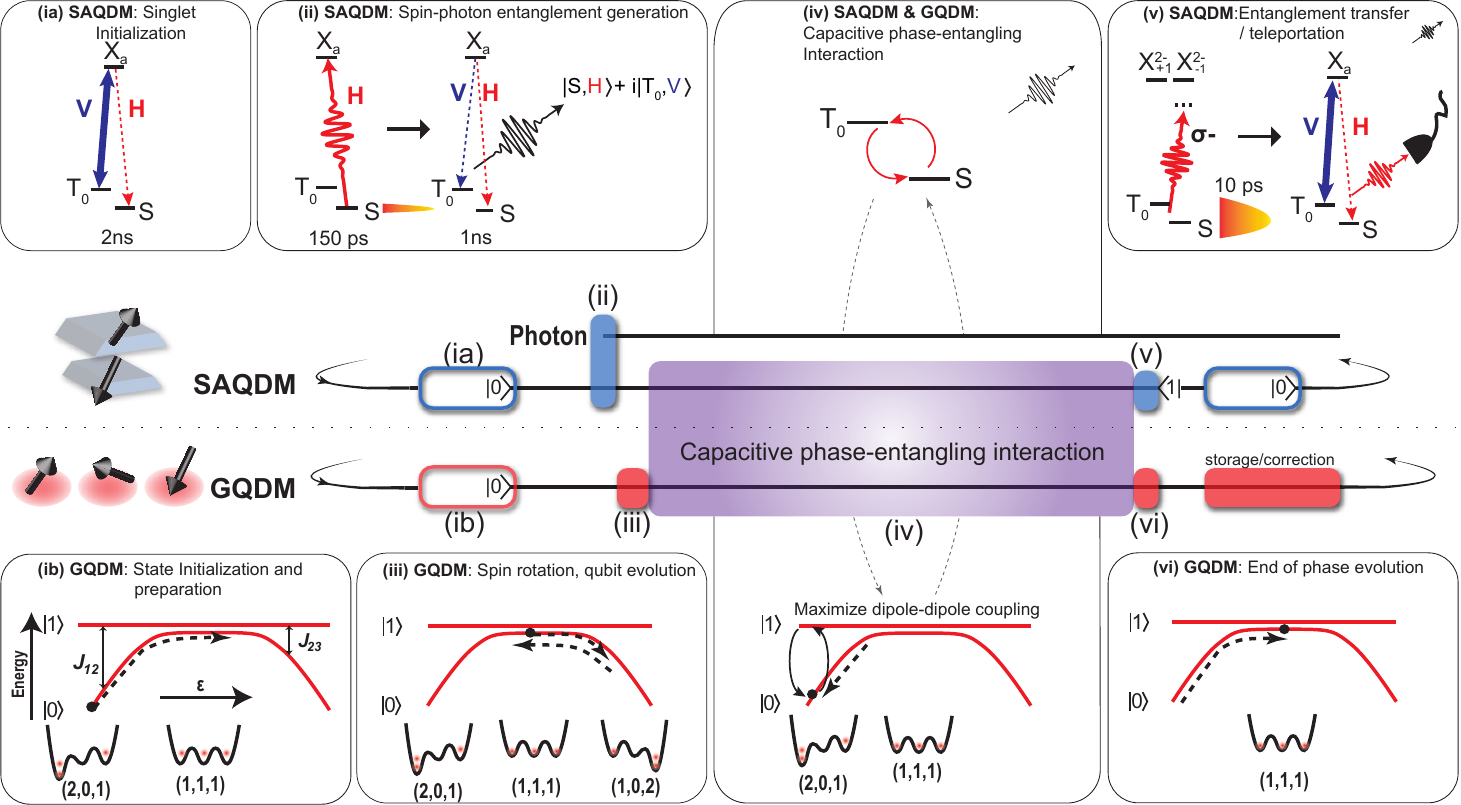}
\caption{Entanglement transfer scheme.  Time proceeds from left-to-right and the top/bottom row of boxes pertain to processes for the SAQDM/GQDM, respectively.  The middle section displays the timeline of the entanglement scheme.  Optical pumping (steps ia, and v) completes in a few nanoseconds, optical pulses (ii and v) are shorter than a nanosecond, GQDM pulses can be subnanosecond duration, and the time for the capacitive phase-entangling interaction, $\tCZ$, is also of order nanoseconds depending on fabrication alignment (see \refsec{simulation}).  The full cycle is therefore on the order of 10 nanoseconds to complete, and is likely to be limited by the dead-time of the single-photon detector that heralds success in step (v).}
\label{scheme}
\end{figure*}

Our general strategy is to generate spin-photon entanglement in self-assembled quantum dots and then coherently couple those quantum dots with electrically gated quantum dots.  The self-assembled dots strongly bind optically active excitons and therefore provide a strong optical interface,  while the gated dots are more straightforward to fabricate in large arrays and have well demonstrated mechanisms for high-speed multi-qubit logic.  The physical pairing between the two can be envisioned as a heterostructure stack, schematized in \reffig{devicesketch}, in which the self-assembled dots are in close proximity to a high-mobility quantum well, which is modified by buried metal gates to accumulate electrons in the gated quantum dots.  A buffer on top of the quantum well protects this layer from the strained layer of self-assembled quantum dots.  The SAQDM is near the top of the stack to facilitate the emission of photons through metal apertures.  Our scheme creates spin-photon entanglement in the SAQDM (following a similar procedure as in Refs. \cite{degrevenat,gaonat,shaibleyprl}), then uses the coherent capacitive coupling to implement a controlled-phase operation between SAQDM and GQDM qubits, and finally applies quantum erasure to the SAQDM to yield a bit entangled state between the photon and the GQDM, heralded by the detection of a single photon.  Schemes without heralding are possible, since the coherent coupling could be used to achieve a ``SWAP" operation, but such schemes require longer and more complex control.  The scheme we present here, a form of partial one-bit teleportation~\cite{zhou00}, requires a relatively small number of timed optical and electrical pulses and succeeds with a maximum probability bounded by 1/2 for successful heralding.  In practice, the success probability will be substantially lower due to limitations in collection and detector efficiency, etc.  A critical advantage of quantum dots, however, is the speed by which they may be reset for a repeat-until-success scheme such as this one.  Natural radiative lifetimes are about 1~ns and can be shortened by Purcell enhancement using an integrated microcavity structure.  Recent demonstrations in SAQDs have exploited their speed to show four-orders of magnitude improvement in heralded entanglement generation rate~\cite{imamoglu_entanglement} over similar schemes implemented in trapped ion or NV diamond systems, as well as picosecond-speed quantum control via single photons in strongly coupled photonic crystal cavities~\cite{waks_strong}, clearly highlighting the potential speed of our proposed system.

We work in a small magnetic field of order 30 mT.  Neither the direction nor the value of the field need be precise; the main role of this magnetic field is to suppress hyperfine-induced electron spin flips in the GQDM layer, which require an energy of order $|\ts{g}{E}\ts\mu{B}B|$ where $\ts{g}{E}$ is the $g$-factor for the GQDMs (about that of GaAs, about $-0.4$), $\ts\mu{B}$ is the Bohr magneton, and $B$ is the amplitude of the applied field.  Except at certain detunings which provide anticrossings between singlet and polarized triplet states, it is a very rare event for the GaAs hyperfine bath to provide this energy of order $\mu$eV.  In principle, the small magnetic field would suppress hyperfine-induced spin-flips in the SAQDMs as well, since both the electrons and holes have larger $g$-factors than in GaAs.  However, the introduction of optical fields provides an injection of energy into the system enabling more rapid hyperfine mixing even at very large magnetic fields, as observed in a number of experiments \cite{xunat,latta09,laddprl}.  In fact, we rely on such mixing for spin initialization, which we will describe after introducing the states and the Hamiltonian.

To describe how our proposed device works, we consider a Hubbard-model approach for the quantum dot molecule ground states; here the standard model is the Hamiltonian
\begin{multline}
\label{HOE}
 \ts{H}{OE} =
\frac{1}{2}\sum_{jk\sigma} t_{jk} (c_{j\sigma}^\dag c_{k\sigma}^\nodag+c_{k\sigma}^\dag c_{j\sigma})
\\
   +  \frac{1}{2}\sum_{jk}\sum_{\sigma\sigma'} U_{jk}
        c_{j\sigma}^\dag c_{k\sigma'}^\dag c_{k\sigma'}^\nodag c_{j\sigma}^\nodag,
\end{multline}
where $c_{j\sigma}^\dag$ creates an electron into the ground-state wavefunction $\phi_j(\vec{r})$ of dot $j$ with spin $\sigma$ from the empty quantum dot vacuum $\vac$.  The transition elements are
\be
t_{jk} = \int d^3\vec{r} \ \phi^*_j(\vec{r})\biggl[
    -\frac{\hbar^2}{2m^*}\nabla^2+V(\vec{r})\biggr]
    \phi_k^\nostar(\vec{r}),
\ee
and the Coulomb matrix elements are
\be
U_{jk} = \int d^3\vec{r} d^3\vec{r}' \frac{e^2}{4\pi\epsilon}
    \frac{|\phi_j(\vec{r})|^2|\phi_k(\vec{r}')|^2}{|\vec{r}-\vec{r}'|}.
\ee
Note that direct exchange is neglected in this tight-binding approximation; in practice the physics of SAQDMs and GQDMs is well-described by kinetic exchange only.  We will refer to this basic form of the Hamiltonian in the next few sections. In addition to these couplings, there will be optical couplings to exciton states.

\subsection{Self-assembled quantum dot molecules}

The expected efficacy of SAQDMs for our scheme results from multiple recent advances in the understanding of their level structure \cite{Stinaff,scheibner} and methods for control \cite{tureciprb,economouprb,cohenarxiv}.
The functionality realized by going from a single quantum dot to two tunnel-coupled quantum dots is significant with regards to spin manipulation and control.  The key reason is the rich level structure realized by spin exchange through the tunneling interaction.  Through electric/magnetic field, level anticrossings, and polarized cavities, these levels can be further tuned for specific purposes.  Capabilities of SAQDMs unavailable to single dots include non-destructive cycling transitions \cite{dkimprl}, and single shot QND measurement~\cite{vamivakasnat}.

A key resource for this device design is the SAQDM with one electron in each dot forming a singlet-triplet qubit in the ground state~\cite{dkimnphys,greilichnatphoton} requiring no magnetic field.  A qubit is available in the decoherence free subspace provided by the $m_z=0$ spin projection of the singlet and triplet states; this subspace is invariant to global magnetic field fluctuations.

The main parameter of interest in this device for the SAQDM is the exchange energy $\ts{J}{O}$, which determines the clock rate between the singlet/triplet states and the dipole magnitude.  Here, the ``O" subscript reminds us that $\ts{J}{O}$ refers to exchange for \textbf{O}ptically controlled quantum dots.  This exchange energy is set by the tunneling rate between quantum dots, which is roughly determined by growth parameters (barrier height and thickness between the two dots, and carrier type).  Fine tuning of this splitting post-growth is done by applying an electric-field bias to the device that changes the relative detuning between the dots. Previous reports of $\ts{J}{O}$ range from 18--30~GHz~\cite{greilichnatphoton,weissprl} or larger~\cite{elzermanprl} and  have shown a post-growth tuning span of 10~GHz~\cite{dkimnphys}.

To provide a simple formula for the exchange energy, let us first review the basic physics of kinetic exchange using \refeq{HOE}, considering only the pair of dots in the SAQDM, using $T$ and $B$ subscripts for top and bottom.  Referring to \refeq{HOE}, we define the optical tunnel coupling $\ts{t}{O}=\ts{t}{TB}/\sqrt{2}$.  We abbreviate the detuning of these states $\ts\epsilon{O}=\ts{U}{BB}-\ts{U}{TB}+(\ts{t}{BB}-\ts{t}{TT})/2$ and presume the device is configured so that $\ts\epsilon{O} \ll \ts{U}{BB}-\ts{U}{TB},\ts{U}{TT}-\ts{U}{TB}$, meaning that the state in which two electrons live in the bottom dot is reasonably close to resonant to the energy in which one electron lives in each dot, and we may neglect mixing of the doubly-occupied top dot $c_{\text{T}\uparrow}^\dag c_{\text{T}\downarrow}^\dag\vac$ state.  (An additional offset due to Coulomb interactions with the third dot of the GQDM, $\ts{U}{B3}-\ts{U}{T3}$ might be added to $\ts\epsilon{O}$ for completeness; interactions with dots 1 and 2 will be handled perturbatively below.)  The singlet and triplet states for the SAQDM are then
\begin{align}
\ts{\ket{S}}{O}&=
    \biggl[\cos\biggl(\frac{\ts\theta{O}}{2}\biggr)\frac{c_{T\uparrow}^\dag c_{B\downarrow}^\dag-c_{T\downarrow}^\dag c_{B\uparrow}^\dag}{\sqrt{2}}
\notag\\
&\hspace{1in}
+\sin\biggl(\frac{\ts\theta{O}}{2}\biggr)c_{B\uparrow}^\dag c_{B\downarrow}^\dag\biggr]\vac,
\\
\ts{\ket{T_0}}{O}&=\biggl[
    \frac{c_{T\uparrow}^\dag c_{B\downarrow}^\dag
         +c_{T\downarrow}^\dag c_{B\uparrow}^\dag}{\sqrt{2}}\biggr]\vac,
\\
\ts{\ket{T_{\pm}}}{O}&=c_{T,\pm m/2}^\dag c_{B,\pm m/2}^\dag\vac,
\end{align}
where the mixing angle satisfies $\tan\ts\theta{O}=2\ts{t}{O}/\ts\epsilon{O}.$  The energy difference between these two states is
\begin{multline}
\JO^{(0)}(\ts\epsilon{O}) = \Tbra\ts{H}{O}\Tket-\ts{\bra{S}\ts{H}{O}\ket{S}}{O}
\\
= \frac{\sqrt{\ts\epsilon{O}^2+4\ts{t}{O}^2}}{2}-\frac{\ts\epsilon{O}}{2},
\label{JOdef}
\end{multline}
where $\ts{H}{O}$ is explicitly omitting the GQDM and focussing on the double-dot.  We define a qubit based on the states $\ts{\ket{S}}{O}$ and $\ts{\ket{T}}{O}$; in particular the logical Pauli-$Z$ operator is
\be
\ts{Z}{O} = \ts{\ketbra{S}{S}}{O}-\ts{\ketbra{T_0}{T_0}}{O}.
\ee
The exchange energy $\ts{J}{O}$ is ``always on" for this qubit; it is set to large degree by growth parameters and to a smaller degree by a vertical bias field, which is presumed to be held constant during operation. Hence this qubit constantly accumulates phase.

There are four possible excited states which are optically connected to the ground states of the SAQDM.  These are used as auxiliary states to the qubit states for initialization, control, and readout.  The excited states add a bright exciton to one of the two dots; we will assume the top dot here.  Hence there are three particles in this top dot, referred to as a trion, with an electron in the bottom dot.  The two electrons in the top dot form a singlet, and the paired hole is of predominantly heavy-hole character, with spin projection $m_z=\pm 3/2$ in the growth direction.  In our notation, the hole is created by $h^\dag_{Tm}$.  Hence we consider the SAQDM trion states to be
\begin{align}
\ket{X^{2-}_{\pm 2}}
 &=c^\dag_{B,\pm 1/2}c^\dag_{T\uparrow}c^\dag_{T\downarrow}h^\dag_{T,\pm 3/2}\vac,
\\
\ket{X^{2-}_{\pm 1}}
 &=c^\dag_{B,\mp 1/2}c^\dag_{T\uparrow}c^\dag_{T\downarrow}h^\dag_{T,\pm 3/2}\vac.
\end{align}
At zero or near-zero magnetic field and neglecting electron-hole exchange, the four states $\ket{X^{2-}_{m}}$ are degenerate.  Given the bandwidths of optical pulsed control and rapid spontaneous emission, small effects lifting the degeneracy such as a weak magnetic field or electron-hole exchange interactions may be neglected.

The optical dipole Hamiltonian may be written
\be
\ts{H}{dip} = \sum_{k} \Omega_k a^\dag_{k,\pm} c_{T,\pm 1/2}^\nodag
h^\nodag_{T,\pm 3/2}
+\hc
\ee
which may be regarded as a consequence of the conservation of angular momentum.  Here, the operators $a^\dag_{k,\pm}$ each create a photon in emission mode $k$ with $\sigma^\pm$ circular polarization.  An optical pulse from the $\Sket$ state which is horizontally polarized will therefore create a coherent superposition of the effectively degenerate $\ket{X^{2-}_{\pm1}}$ states; we refer to this state as $\ket{X_a}$.  The matrix element for the decay of $\ket{X_a}$ is then
\begin{multline}
\ts{H}{dip}\ket{X_a}=
\\
\sum_k \Omega_k\biggr[ \frac{a_{k,+}^\dag+a_{k,-}^\dag}{2}\Sket + \frac{a_{k,+}^\dag-a_{k,-}^\dag}{2}\Tket\biggr].
\end{multline}
We define horizontal and vertically polarized photon states as
\be
\ket{H}_\gamma = \frac{a_{k,+}^\dag+a_{k,-}^\dag}{\sqrt{2}}\vac,\quad
\ket{V}_\gamma = \frac{a_{k,+}^\dag-a_{k,-}^\dag}{\sqrt{2}}\vac.
\ee
A photon emitted from the superposition $\ket{X_a}$ will create $\ket{H}_\gamma$ for the $\Sket$ final state or $\ket{V}_\gamma$ for the $\Tket$ state.
The emitted photon is therefore polarization-entangled to the spin.  The splitting of $\Sket$ and $\Tket$ by $\JO$ means the photon's energy will also be split by $\JO$, and hence the photon's energy is also entangled to the spin.

\subsection{Gated Quantum Dot Molecules}

While SAQDMs have excellent properties for interfacing semiconductor spins with optical photons, building arrays of more than a few coupled SAQDMs presents a substantial challenge.  Stacks of quantum dots cannot easily scale past a few dots, and methods for the controlled lateral coupling of self-assembled dots are poorly established.  While a number of theoretical proposals for transverse dot coupling mediated by cavity photons, excitons, or exciton-polaritons \cite{imaprl99,shamprl02,laddyamamotoprb,economouprb,Puri_PRB_2014} have been published, these mechanisms are highly limited by the loss mechanisms of their associated mediating fields.  Proposals for arrays of coupled self-assembled quantum dots~\cite{van2010distributed,jones2012layered} depend critically on fabrication advances, especially microcavity engineering.  In contrast, gated quantum dot molecules (GQDMs) in both GaAs and silicon systems have successfully demonstrated pairs~\cite{pettasci,folleti,bluhm_dd,maunenat}, triples~\cite{gaudreaunphys,lairdprb,medford_self-consistent_2013,HRL}, and quadruples~\cite{shulmansci}, showing little fundamental impediment to the fabrication of larger arrays.  Moreover, the ability to control these dots using only voltages substantially eases the burden of distributing synchronous laser pulses across an array of SAQDMs~\cite{van2010distributed,jones2012layered}, or engineering magnetic field gradients or microwave fields to control arrays.

Exchange-only qubits based on GQDMs rely on decoherence-free subspaces or subsystems.  The smallest single qubit results from the two subspaces of total angular momentum 1/2 for three coupled spins.  We refer to this as a Decoherence Free Subsystem (DFS) qubit in the GQDM.  As shown in \cite{divincenzo_universal_2000}, this DFS qubit may be controlled entirely using exchange between dots 1 and 2, with rate $J_{12}=\ts{J}{E}$, and exchange between dots 2 and 3 with rate $J_{23}$.  The ``E" subscript in $\ts{J}{E}$ reminds us that this is exchange for \textbf{E}lectrically controlled quantum dots. This exchange rate is in turn controlled by detuning voltages which alter the relative energies of the potentials across all dots, enabling $\ts{J}{E}$ values as high as GHz near zero detuning and nearly zero at high detuning.

The spin states for the GQDM are comparable to the SAQDM, except we have a third spin degree of freedom (see Fig. \ref{levels}b).  The third spin is labelled by its projection value $m$ which we leave unspecified.  Here we neglect the spin states with total angular momentum 3/2, as these cannot be created using our initialization scheme, even after exchange interactions or capacitive couplings to the SAQDM.  These states can, however, be populated via hyperfine interactions~\cite{laddprb}, an error which requires mitigation.  Let us at first consider the Hamiltonian neglecting hyperfine interactions to better explain the scheme, and we will return to the mitigation of hyperfine errors in \refsec{discussion}.

The model for kinetic exchange is the same as in the SAQDM case discussed above.  We number these dots $1-3$ and define
\begin{align}
\ts{\ket{S,m}}{E}&=
    \biggl[\cos\biggl(\frac{\ts\theta{E}}{2}\biggr)\frac{c_{1\uparrow}^\dag c_{2\downarrow}^\dag-c_{1\downarrow}^\dag c_{2\uparrow}^\dag}{\sqrt{2}}
\notag\\&\hspace{1in}
+\sin\biggl(\frac{\ts\theta{E}}{2}\biggr)
    c_{1\uparrow}^\dag
    c_{1\downarrow}^\dag\biggr]c_{3m}^\dag\vac,
\notag\\
\ts{\ket{T_0,m}}{E}&=\biggl[
    \frac{c_{1\uparrow}^\dag c_{2\downarrow}^\dag+c_{1\downarrow}^\dag c_{2\uparrow}^\dag}{\sqrt{2}}\biggr]c_{3m}^\dag\vac,
\notag\\
\ts{\ket{T_{2m},-m}}{E}&=c_{1m}^\dag c_{2m}^\dag c_{3,-m}^\dag\vac,
\end{align}
where $m$ is an extra degree of freedom corresponding to the spin projection of the third, always occupied dot. One important difference for the GQDM is that the detuning $\ts\epsilon{E}$ is presumed to be highly controllable, and in particular it can be set to a high value where $\ts{J}{E}$ (and similarly $\ts\theta{E}$) can be brought to zero.  Hence the singlets and triplets can be brought to degeneracy, and this qubit does not continue to accumulate phase.

Another important difference for the GQDM is the presence of the third electron and the possibility of detuning the device to bring $\epsilon_{23}\approx U_{33}-U_{23}+U_{13}-U_{12}+(t_{33}-t_{22})/2$ close to resonance.  To clarify, we have independent control over the detuning for dots 1 and 2, which we notate $\ts\epsilon{E}=\ts\epsilon{12},$ and for the detuning between dots 2 and 3, $\ts\epsilon{23}$.  The exchange energy between dots 1 and 2, notated $\ts{J}{E}=\ts{J}{12}$, effectively reduces the energy of $\ts{\ket{S,m}}{E}$.  Exchange between dots 2 and 3, notated $\ts{J}{23}$, reduces the energy of the singlet on dots 2 and 3 hybridized with $c_{1m}^\dag c_{2\uparrow}^\dag c_{3\downarrow}^\dag\vac$.  This singlet state overlaps the two states we treat as our logical qubit,
\begin{align}
\ts{\ket{0m}}{E} &= \ts{\ket{S,m}}{E}\\
\ts{\ket{1m}}{E} &= \frac{\ts{\ket{T_0,m}}{E}-\sqrt{2}\ts{\ket{T_{2m},-m}}{E}}{\sqrt{3}}.
\end{align}
The third state with the same spin-projection, $\ts{\ket{Qm}}{E}$, is fully spin-symmetric and inert to either exchange operation~\cite{laddprb}.  Our logical Pauli-$Z$ is then
\be
\ts{Z}{E} = \sum_m \ts{\ketbra{0m}{0m}}{E}-\ts{\ketbra{1m}{1m}}{E}.
\ee
Reducing $\epsilon_{12}$ causes a rotation for this qubit about $\ts{Z}{E}$, while reducing $\epsilon_{23}$ causes a rotation about $\ts{Z}{E}\cos\phi+\ts{X}{E}\sin\phi$, for $\phi=2\pi/3$.

The GQDM may be initialized into the state $\ket{0m}$ by changing gate voltages to a configuration where $(1,0,1)$ is the stable state, and then ramping voltage to load a second electron into the first dot to form (2,0,1).  Near the charge transition regime, relaxation phenomena will load the two electrons in the first dot into a singlet.  The third electron's spin state sets $m$ which need not be initialized in our scheme.  This procedure loads a total angular momentum of 1/2 which is conserved by all exchange and capacitive interactions.  Measurement is accomplished via Pauli Spin Blockade, as in Refs.~\onlinecite{lairdprb,medford_self-consistent_2013,HRL}.

\subsection{The coupling Hamiltonian}

The capacitive or dipole-dipole coupling between the SAQDM and the GQDM may be understood using the model of \refeq{HOE}.  We neglect direct tunneling between the two kinds of dots here;  it is the direct Coulomb terms which are important.  These shift the energies of the singlet and triplet states depending on whether they are singlet-states, for which tunneling is allowed, or triplet states, for which tunneling is suppressed.
Some of these shifts depend on one charge state only, and contribute to $\JO$ or $\JE$.  However, there is an energy shift that depends on the qubit state of both molecules, and therefore on the product of logical-$Z$ operators for the singlet-triplet or DFS qubits.  Hence
the system Hamiltonian can be written
\be
\ts{H}{OE} \approx -\frac{1}{2}\biggl[\JO\ts{Z}{O}+\JE\ts{Z}{E} + \ts{J}{OE}\ts{Z}{O}\ts{Z}{E}\biggr].
\ee
Here, $\JO$ differs from $\JO^{(0)}$ as in \refeq{JOdef} by the first-order Coulomb coupling to the GQDM, i.e.
\begin{multline}
\JO = \JO^{(0)}(\ts\epsilon{O}) +
\\\sin^2\frac{\ts\theta{O}}{2}\biggl[
    \frac{\ts{U}{T1}+\ts{U}{T2}-\ts{U}{B1}-\ts{U}{B2}}{2}
+\frac{\ts\Delta{DD}}{4}\sin^2\frac{\ts\theta{E}}{2}\biggr]
\end{multline}
\label{JOmod}
and similarly for \JE.  The coupling term is
\be
\JOE = \sin^2\frac{\ts\theta{E}}{2}\sin^2\frac{\ts\theta{O}}{2}\frac{\ts\Delta{DD}}{4},
\ee
where
\be
\ts\Delta{DD}=U_{T1}-U_{T2}-U_{B1}+U_{B2}.
\ee
The intuition for this coupling term is that the detuning of one molecule is shifted by the qubit state of the other; hence the $Z$-state of one qubit applies a $Z$ operator to the other.  The effect is therefore most sensitive at the peak derivative of exchange with respect to detuning, and indeed  $\JOE$ scales as $\sin^2\theta_x/2 = |\partial J_x/\partial\epsilon_x|$ for both $x=$O and E.

With this interaction term, a controlled-phase (CZ) gate (along with single-qubit phase evolutions) can be implemented by allowing the system to evolve for a time $\ts{t}{CZ} = \pi/2\ts{J}{OE}$.  The coupling is maximized at $\ts\theta{O}=\ts\theta{E}=\pi$, which corresponds to the singlet state being entirely a two-electron state in one dot, for maximum dipole coupling.  In practice, the GQDM would be gated as far as reasonable into the regime where the singlet has the (2,0,1) charge configuration (i.e. two electrons in dot 1, one electron in dot 3).  Eventually the size of this regime is limited by orbital states, but prior to encountering those the primary limitation in how far into this regime we may go will likely be reliable pulsing while tracking the phase due to the increasing $\ts{J}{E}\ts{Z}{E}$ term.  The SAQDM is presumed to have fixed gate values during this interaction, and this device will likely stay close to the (1,1) regime, as the optical dipole for the excitonic states will be substantially reduced far into the (2,0) regime.  Very approximately, then,
\be
\JOE \approx \frac{\ts\Delta{DD} t_O^2}{4\epsilon_O^2}.
\label{approxJOE}
\ee
Numerical estimates of $\ts\Delta{DD}$ will be made in \refsec{simulation}.

Note that this model makes a number of simplifying assumptions, such as the notion that the tunnel coupling remains constant under changes in detuning.  Checking these assumptions requires a more detailed treatment, which we also address in \refsec{simulation}.

\subsection{The entanglement procedure}

Following \reffig{scheme}, the basic entanglement process occurs as follows.

The SAQDM is initialized by applying a long nanosecond-order laser pulse on the $\ts{\ket{T_0}}{O}$ to $\ket{X^{2-}_{\pm 1}}$ transitions.  These excite the system into the $\ket{X^{2-}_{\pm 1}}$ trion states until it is shelved into the $\ket{S}_O$ state.  This pumping process continues as long as necessary to initialize this state with high fidelity. In previous reports~\cite{dkimnphys} this was done with high fidelity, owing in part to the weak hyperfine mixing of the triplet states, which allows complete pumping out of the triplet manifold.  Small DC magnetic fields and microwave fields could also enhance this mixing if more rapid initialization is required, however our scheme generally maintains the $m_z=0$ subspace for the SAQDM so hyperfine initialization may only be important occasionally.  A $\pi$ pulse then excites into the $\ket{X_a}$ excited state followed by radiative decay that emits a single photon.  This recombination of an electron-hole pair emits a single photon entangled to the spin state of the two electrons:
\be
\ts{\ket{\psi(0)}}{O,$\gamma$} = \frac{1}{\sqrt{2}}\left(\Sket\ket{H}_\gamma+i\Tket\ket{V}_\gamma\right).
\ee
Note that this is an energy eigenstate of our system, as the photon's energy compensates for the $\Sket/\Tket$ splitting.  Phase evolution in our scheme is therefore not affected by delays between the creation of this photon and the next event in our entanglement scheme.

Likewise, we initialize the GQDM; the exact timing between the GQDM initialization and the spin-photon entanglement is not critical.  Initialization occurs by placing the double-dot with highest capacitive coupling to the SAQDM into a two-electron singlet state.  This is done by tunneling from a thermal electron bath into the ground state of a single dot whose energy has been lowered (i.e. into the (2,0,1) charge ground state).  The spin-singlet is guaranteed after thermal relaxation via fermion asymmetry.  This creates the state $\ts{\ket{0,m}}{E}$.

At a well clocked time which we label here as $t=0$, a fast exchange pulse lowers the energy of the third dot in the GQDM, reducing the energy of the (1,0,2) singlet.  This activates kinetic exchange of amplitude $J_{23}$ between dots 2 and 3, and if applied for a time $\tau$ such that $J_{23}\tau = \pi-\tan^{-1}\sqrt{8}$, the pulse generates the state
\be
\ts{\ket{\psi(0)}}{E} = \ts{R}{E}\ts{\ket{0m}}{E}
= \frac{1}{\sqrt{2}}\left(\ts{\ket{0m}}{E}-e^{-i\xi}\ts{\ket{1m}}{E}\right),
\ee
where $\xi=\tan^{-1}\sqrt{2}$, for either spin-state of the third dot, $m=\pm 1/2$.  The rotation we have performed, which we notate $\ts{R}{E}$, rotates the DFS qubit into the equator of its Bloch sphere.  For simplicity of the present analysis, we neglect capacitive couplings during the duration of the pulse; as we argue in \refsec{simulation} it will be small near the (1,0,2) regime where this rotation takes place.

To load our GQDM memory, we transfer the photon-entangled state from the SAQDM to the GQDM using a form of quantum teleportation.  Immediately following $\ts{R}{E}$ at $t=0$, the GQDM is tuned to the regime that maximizes $\JOE$, where the singlet is almost entirely in the (2,0,1) charge configuration.  This turns on the SAQDM-GQDM controlled-phase gate for time $\tCZ=\pi/2\JOE$.  As a result, we perform a controlled-$Z$ gate between the SAQDM and GQDM qubits, yielding the state
\begin{multline}
\ts{\ket{\psi(\tCZ^-)}}{OE$\gamma$}=
\\
\frac{1}{2}\biggl[
     \Sket\ts{\ket{0m}}{E}\ket{H}_\gamma
    +e^{-i\delta\JO\tCZ}\Tket\ts{\ket{0m}}{E}\ket{V}_\gamma
\\
    +ie^{-i(\JE\tCZ+\xi)}\biggl(\Sket\ts{\ket{1m}}{E}\ket{H}_\gamma
\\
\hspace{0.5in}
    -e^{-i\delta\JO\tCZ}\Tket\ts{\ket{1m}}{E}\ket{V}_\gamma\biggr)\biggr].
\end{multline}
Here, $\delta\JO$ is the difference between the SAQDM exchange splitting when the photon was created, roughly $\JO^{(0)}$, and the SAQDM exchange splitting during the CZ gate.  This phase difference is comparable to \JOE\ and therefore must be calibrated, along with the substantially larger $\JE\tCZ$.

The evolution ends when we disentangle the SAQDM qubit.  This begins with a basis rotation by applying a detuned, circularly polarized, ultrafast optical pulse to the SAQDM, causing an AC Stark shift for the bright-state superposition $(\Sket+\Tket)/\sqrt{2}$ by dressing it with the exciton state $\ket{X^{2-}_{-1}}$.   For an appropriately tuned power, this amounts to a $\pi/2$ rotation around the $X$-axis for the SAQDM qubit, resulting in
\begin{multline}
\ts{\ket{\psi(\ts{t}{CZ}^+)}}{OE$\gamma$}=
\frac{1}{2\sqrt{2}}\biggl\{\\
    \Sket\biggl[\biggl(
        \ts{\ket{0m}}{E}+ie^{-i(\JE\tCZ+\xi)}\ts{\ket{1m}}{E}\biggr)\ket{H}_\gamma
\\
    -ie^{-i\delta\JO\tCZ}\biggl(
        \ts{\ket{0m}}{E}-ie^{-i(\JE\tCZ+\xi)}\ts{\ket{1m}}{E}\biggr)\ket{V}_\gamma\biggr]
\\
    -i\Tket\biggl[\biggl(
        \ts{\ket{0m}}{E}+ie^{-i(\JE\tCZ+\xi)}\ts{\ket{1m}}{E}\biggr)\ket{H}_\gamma
\\
    +e^{-i\delta\JO\tCZ}\biggl(
        \ts{\ket{0m}}{E}-ie^{-i(\JE\tCZ+\xi)}\ts{\ket{1m}}{E}\biggr)\ket{V}_\gamma\biggr]\biggr\}.
\end{multline}
At the same time, we rapidly gate the GQDM into the (1,1,1) regime where $\JE$ and $\delta\JO$ vanish, rendering the DFS qubit degenerate.  Following these actions, the pumping laser between $\ts{\ket{T_0}}{O}$ and the exciton states is again turned on and a detector is gated open to seek a spontaneously emitted photon from $\ket{X^{2-}_{\pm1}}$ to $\ts{\ket{S}}{O}$, whose energy differs from the pumping energy by $\ts{J}{O}$.  If this spontaneously emitted photon is detected at some time after $\ts{t}{CZ}$, which will occur with maximum probability 1/2, we project the $\Tket$ component.  This detection completes a quantum erasure of the SAQDM qubit, disentangling itself from the entangled state between the photon and the GQDM memory qubit.  The degeneracy of the DFS qubit protects our entanglement from the uncertain delay of this detection.
The resulting state after renormalization is
\be
\ket{\psi(t>\tCZ)}_{\text{E}\gamma} = \ts{R}{E}'\frac{\ts{\ket{0m}}{E}\ket{H}_\gamma+\ts{\ket{1m}}{E}\ket{V}_\gamma}{\sqrt{2}},
\label{state}
\ee
where $\ts{R}{E}'=e^{i\eta_2\ts{Z}{E}/2} R_E e^{i\eta_1\ts{Z}{E}/2}$ for $\eta_1=\pi/2-\xi+\delta\JO\tCZ$ and $\eta_2=\pi/2+\JE\tCZ.$  The two-qubit state in \refeq{state} is maximally entangled and differs from a Bell state by the three local exchange rotations on the DFS qubit, which comprise $\ts{R}{E}'$.  Note also that any dephasing in the $\ts{\ket{S}}{O}$/$\ts{\ket{T_0}}{O}$ basis during the single-photon measurement procedure will not perturb the memory/photon entanglement, so the measurement time only needs to be short relative to the $\ts{\ket{S}}{O}$/$\ts{\ket{T_0}}{O}$ population relaxation timescale and the GQDM dephasing time; these are both substantially longer than the $T_2^*$ dephasing time in the SAQDM.

After creation of the state in \refeq{state}, some corrections will need to be tracked.  For one, the single-qubit exchange evolution of the GQDM during the interval $\ts{t}{CZ}$ is a phase which requires correction.  As the timescale is set by fast, applied pulses, the timing can be accurately clocked, although the effects of finite ramping time, especially of the GQDM, must be calibrated. The complete control of the triple-dot GQDM assures a pulse sequence exists to invert, if desired, any calibrated $Z$-rotation as well the $\ts{R}{E}$ operation.  Further, however, there is also exchange evolution of the SAQDM at rate $\JO$.  This energy difference also corresponds to the non-degenerate photon energy to which the dots are entangled.  Exactly how this phase evolution manifests depends on what happens to the entangled photon.

One possibility is that this photon is immediately detected in some polarization basis.  This would need to happen to tomographically analyze the spin-photon entanglement of the system, as in Refs. \onlinecite{degrevenat}, \onlinecite{gaonat}, and \onlinecite{shaibleyprl}.  In these cases, the detection time of that photon needs to be recorded and included in the phase correction for the GQDM qubit.

The intended use case for the device we propose, however, is for this photon to link the GQDM (to which it is entangled) to other degrees of freedom in a quantum network.  One way this may be performed is using two-photon interference with a second photon which is itself entangled to other degrees of freedom, and may itself be energy-entangled.  Such interference results from interfering two photons on a non-polarizing beam-splitter, whose output ports enter polarizing beam-splitters with single-photon detectors on their output ports.   Let us label the state of the second photon from the network as $\ts{\ket\alpha}{?}\ket{H}_{\gamma'}+\ts{\ket\beta}{?}\ket{V}_{\gamma'}$.  Here, $\alpha$ and $\beta$ are s other degree of freedom labelled ``?", perhaps another GQDM elsewhere in the network or perhaps another entangled photon.  (These cases are analyzed for link entanglement speed in Ref.~\onlinecite{cody2015}.)  We adopt the spatio-temporal mode notation of Ref.~\onlinecite{legero03}, in which $\zeta_{Pj}(t)$ represents the complex mode function for a photon wavepacket of polarization $P$ entering into mode $j$ of the beam-splitter.  Consider, for example, the case that a horizontal photon is detected at time $t_1$ from one port of the non-polarizing beam-splitter, and a second, vertically polarized photon is detected at time $t_2$ at the second port of the beam-splitter.  Assuming the detectors are perfect and instantaneous, the resulting conditional entangled state between our GQDM and the $\ket\alpha/\ket\beta$ degree of freedom would be~\cite{legero03}
\begin{multline}
\ts{\ket{\Psi}}{E,?}=
\frac{1}{4}\ts{R}{E}'\times
\\
\biggl(\zeta_{H2}(t_1)\zeta_{V1}(t_2)\ts{\ket{0m}}{E}\ts{\ket\beta}{?}
      +\zeta_{H1}(t_1)\zeta_{V2}(t_2)\ts{\ket{1m}}{E}\ts{\ket\alpha}{?}\biggr).
\end{multline}
This state is not normalized because of the reduced probability of detecting this polarization configuration at these particular times.  Now let us suppose our wavepackets are non-degenerate, so
\be
\zeta_{Pj}(t)=e^{-i(\omega+\delta_{Pj})t}|\zeta_{Pj}(t)|.
\ee
Our state may then be written
\begin{multline}
\ts{\ket{\Psi}}{E,?}=
\frac{1}{4}
\ts{R}{E}'e^{-i\phi_+(t_1,t_2)}e^{-i\phi_-(t_1,t_2)\ts{Z}{E}/2}\times
\\
\biggl(|\zeta_{H2}(t_1)\zeta_{V1}(t_2)\ts{\ket{0m}}{E}\ts{\ket\beta}{?}
      +|\zeta_{H1}(t_1)\zeta_{V2}(t_2)\ts{\ket{1m}}{E}\ts{\ket\alpha}{?}\biggr)
\label{entangled}
\end{multline}
in terms of an overall phase
\begin{multline}
\phi_+(t_1,t_2)=(\omega+\delta_{H2}/2+\delta_{H1}/2+\delta_{V1}/2+\delta_{V2}/2)t_1
\\
+(\omega+\delta_{V1}/2+\delta_{V2}/2)t_2,
\end{multline}
which we may ignore, and a relative phase
\be
\phi_-(t_1,t_2)=
(\delta_{H1}-\delta_{H2})t_1/2+(\delta_{V2}-\delta_{V1})t_2/2,
\ee
which, in principle, we can correct using electrically controlled gates on the GQDM.  The remaining $|\zeta_{pj}(t)|$ terms in \refeq{entangled} may describe finite wavepacket bandwidths and associated reductions in fidelity, as in Ref.~\onlinecite{legero03}.

We emphasize two important points about the ability of two-photon entanglement to handle non-degenerate photons.  First, the two-photon interference as described here may be performed local to the memory-containing node in the network and corrections may be performed or tracked locally based on the triple-photon clicks (one for the ``teleportation" photon heralding success, and two at $t_1$ and $t_2$ in two of the 4 detectors of the two-photon interference apparatus) \emph{without} the necessity of communication to other nodes of the network.  After the entanglement procedure, the GQDM may be gated deep into the (1,1,1) regime where its qubit states are degenerate and no clock evolution need be considered.  In this regard, the use of fast detectors and DFS qubits eliminates the need for nonlocal clock synchronization in a quantum network built of these devices.  A second important point, however, is that if highly non-degenerate qubits are employed, then real, finite-bandwidth detectors will severely degrade the fidelity of the resulting entanglement.  The overlap of $\ts{\ket{\Psi}}{E,?}$ with a maximally-entangled Bell-state $\ket{\Psi^+}$, neglecting imperfect inversion of $\ts{R}{E}'$ via exchange pulses or exchange evolution during the detection process, would be
\begin{multline}
\left\langle|\ts{\bra{\Psi^+}{\ts{R}{E}'}^\dag{\Psi}}{E,?}|^2\right\rangle=\\
\frac{1}{2}\biggl[1+
G(t_1,t_2)e^{-(\delta_{H1}-\delta_{H2})^2\sigma_1^2/2-(\delta_{V1}-\delta_{V2})^2\sigma_2^2/2}\biggr],
\end{multline}
where $\langle\cdot\rangle$ refers to averaging over the detector bandwidth and ensembled-averaged jitter, and $\sigma_j$ is the resulting root-mean-square error in our correction for detector time $t_j$.  The function $G(t_1,t_2)$ is given in this formalism by
\be
G(t_1,t_2)=
2\frac{|\zeta_{H2}(t_1)\zeta_{V1}(t_2)\zeta_{H1}(t_1)\zeta_{V2}(t_2)|}%
{|\zeta_{H2}(t_1)\zeta_{V1}(t_2)|^2+|\zeta_{H1}(t_1)\zeta_{V2}(t_2)|^2},
\ee
which captures the influence of the wavepacket arrival differences.  For example, consider the simple model that wavepackets are defined by a pure, lifetime-limited exponential decay with rate $\kappa_j$ and arrival delay $\tau_j$ which is independent of polarization, so $\zeta_{Pj}(t)=\exp[-\kappa_j(t-\tau_{j})]$ for $t>\tau_{j}$ and zero otherwise.  Then $G(t_1,t_2)=\sech[(\kappa_2-\kappa_1)(t_2-t_1)]$ for $t_1$ and $t_2$ both greater than $\tau_1$ and $\tau_2$, and zero otherwise.  Arrival jitter (i.e. randomness in $\tau_j$) cancels for $G(t_1,t_2)$ in this model.  Obviously, the entanglement fidelity is improved with real detectors if the photon frequency offsets $\delta_{H1}-\delta_{H2}$ and $\delta_{V1}-\delta_{V2}$, as well as the pulse bandwidth difference $\kappa_1-\kappa_2$, are brought to zero.  A combination of spectral filtering, frequency conversion \cite{PhysRevLett.109.147405}, and electric tuning of $\JO$ may help enable high fidelity entanglement with slow detectors. 

%% file: device_design_fabrication.tex
\label{fabrication}

\begin{figure}
	\centering
	\includegraphics[width=\columnwidth]{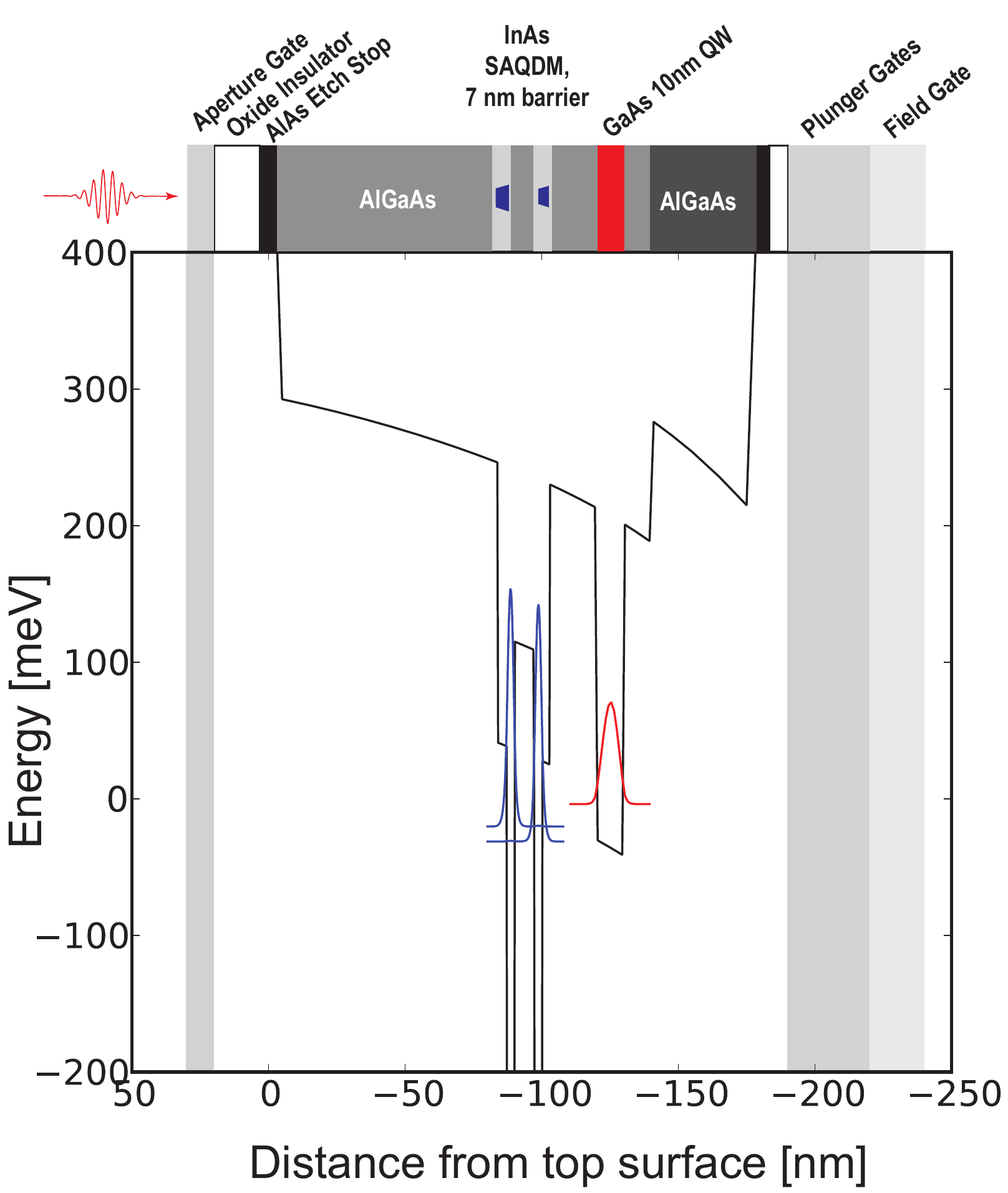}
	\caption{Self-consistent calcuation of conduction band profile of the device with layers depicted above. The bottom of the SAQDM layer is 20~nm from the top of the QW. Blue and red curves represent the normalized electron density envelope for the SAQDM and GQDM electrons, respectively.}
	\label{fig:devicecartoon}
\end{figure}

This section is meant to give a rough outline of the device fabrication to illustrate that the design is feasible, not as a precise or detailed recipe.   The device proposed here leverages existing and proven fabrication techniques for III-V semiconductors.

The process is divided into two steps:  the first is the molecular-beam epitaxial growth of the heterostructure stack, and the second is the metal gate deposition on both sides.  The final structure is depicted in \reffig{fig:devicecartoon}.  The heterostructure stack is grown on a GaAs wafer with AlAs etch stops on either ends, with other III-V materials available in common epitaxy systems.

The structure after growth from top to bottom might be:
\begin{itemize} \itemsep1pt \parskip0pt \parsep0pt
	\item AlAs etch stop 10nm
	\item Al$_{0.3}$Ga$_{0.7}$As cap  100nm
	\item InAs quantum dots 3nm
	\item GaAs/Al$_{x}$Ga$_{1-x}$As/GaAs tunnel barrier 7nm
	\item InAs quantum dots 3nm
	\item Al$_{0.3}$Ga$_{0.7}$As spacer 20nm
	\item GaAs QW     10nm
	\item Al$_{0.3}$Ga$_{0.7}$As spacer 10nm
	\item Al$_{0.4}$Ga$_{0.6}$As layer  40nm
	\item AlAs etch stop 10nm
	\item GaAs substrate
\end{itemize}

Once the structure is grown, the metal gates on the top and bottom surfaces need to be lithographically defined and deposited.  The layout of the gates in the GQDM layer is inspired by the orthogonal-control triple quantum dot design explained in detail in Refs.~\onlinecite{Borselli_ArXiv_2014} and \onlinecite{HRL}, with the chosen scaling to account for the smaller electron effective mass in GaAs vs. Si.  See \reffig{simulationcolormap} for the particular geometry used in our simulations.  As in Refs.~\onlinecite{Borselli_ArXiv_2014} and \onlinecite{HRL}, the GQDM operates in accumulation mode. When properly tuned, the QW layer accumulates a 2DEG that is partitioned to form three gated quantum dots and a dot charge sensor (DCS) circuit with controllable source and drain barriers.  The dots are defined by three QD plunger gates, inter-dot exchange gates, and couplings to the supply baths. The electrostatic potential of the area outside the individual gates is controlled by an overlapping global field gate.  Gate deposition is first accomplished by transferring to a sacrificial ``handle'' wafer, with the top side down in order to etch the GaAs substrate away to the AlAs etch stop.  A large-gap dielectric oxide (such as Al$_2$O$_3$)  is then deposited and patterned over the dot areas.  The gates for the plunger gates are patterned and metallized.  The plunger electrodes are then encapsulated in a thick planarizing oxide.  The sample is transferred to another handle wafer with the new oxide face-down.  This way the first handle wafer can be etched away, where the gates with optical access apertures of about 1~$\mu m$ are then deposited.  Note that no explicit doping is included; this is intended to be an entirely gated structure to reduce the disorder due to random dopants, as in Ref.~\onlinecite{Borselli_ArXiv_2014}.

A significant challenge for this device is maintaining high mobility in the QW 2DEG layer, with the SAQD layer above it.   Having the SAQD layer closer to the QW would increase capacitive coupling but at the cost of mobility.  In previous reports mobilities as high as $0.5\times 10^6\text{~cm}^2/\text{Vs}$ were reported for a separation of 45~nm~\cite{Dettwiler_ArXiv_2014}.  Our device here proposes a shorter separation of 20 to 30~nm.

Another challenge is the lateral placement of the SAQDM and GDQM with respect to each other.  The first layer of the SAQDM is randomly nucleated (the top dot has high probability of nucleating directly above due to strain correlation) with standard growth techniques.  The random positioning can be addressed in several ways.  A brute-force method would be to grow a high areal density of SAQDM's, and spectrally select the ones that only interact with the GQDM.  Other options use lithography: gates are deposited deterministically by first discovering a ``good'' SAQDM through photoluminescence, similar to how photonic crystal cavities are made around quantum dots~\cite{thon_apl}.  Using these techniques gives a lateral positioning tolerance of less than 30~nm.  Finally, the SAQDM could be deterministically grown in predetermined array through surface patterning techniques~\cite{schnano}.

%% file: device_design_simulation.tex
\label{simulation}

\begin{figure*}
	\centering
	\includegraphics[width=5in]{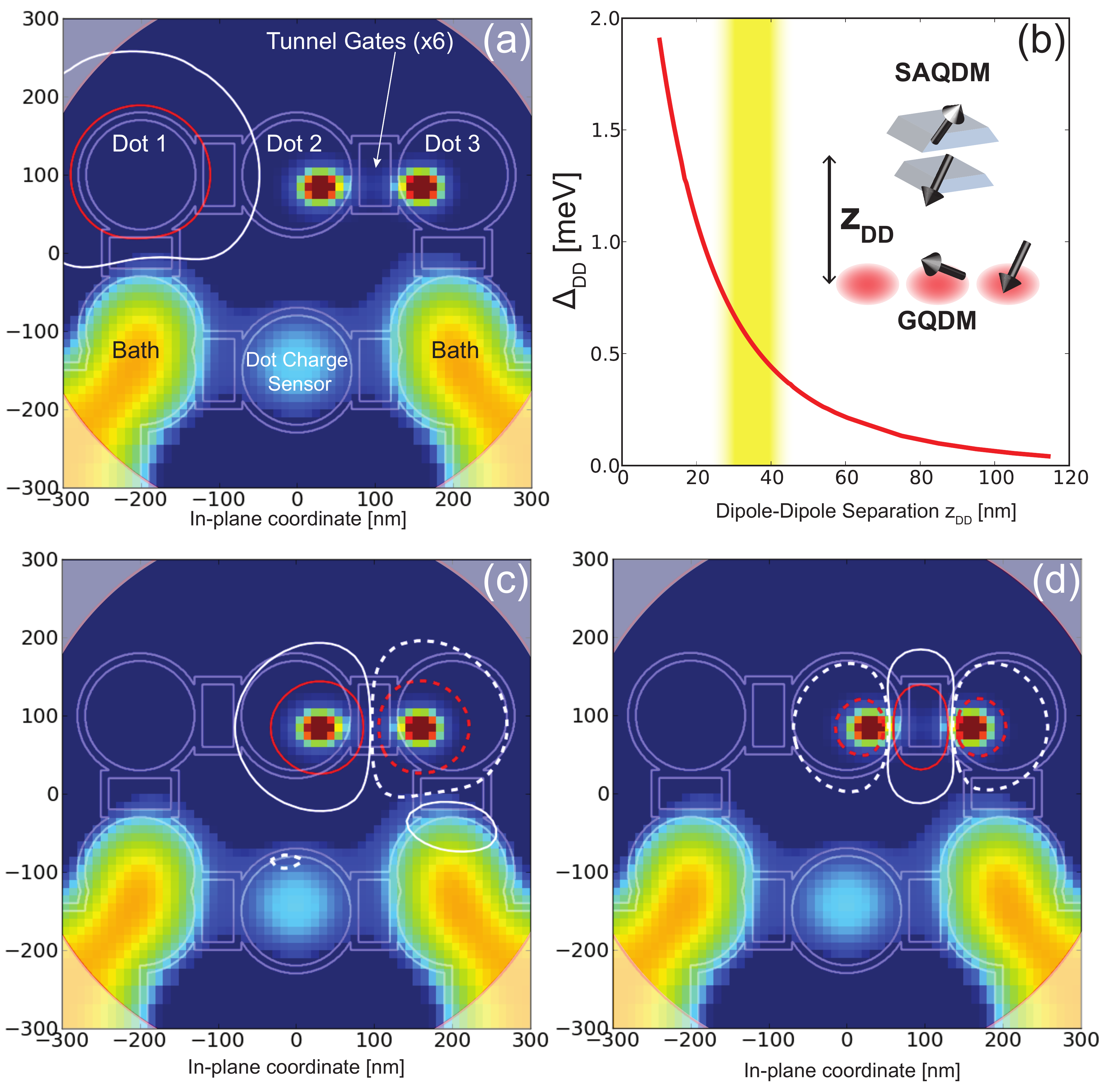}
	\caption{Simulation results.  Subfigures (a),(c),and (d) show a colormap of the simulated electron density superimposed on thin white lines indicating the edges of the metal gates.  For better visibility, the electron density of the GQDM (0,1,1) state is artificially enhanced 10-fold versus the remainder of the 2DEG.  The translucent white corners indicate the aperture.  (a) Lever arm of dot 1 plunger gate on the detuning of the SAQDM as a function of the SAQDM in-plane position at dipole separation $\ts{z}{DD}$=30~nm. Contours of $\alpha_{P1}=10$ and $1$~meV/V are shown by red and white lines, respectively.  (b) Dipole-dipole coupling $\ts\Delta{DD}$ as a function of separation between the center of the SAQDM and the center of QW ($\ts{z}{DD}$) shown for the case of perfect in-plane alignment of the SAQDM with the middle GQD. The highlighted bar indicates the range of vertical dot spacing where tunnel-loading of all dots is estimated by simulation to occur on a timescale between about 1~ms and 1 second.  Outside this range, device stability or the charge loading pathways would need re-evaluation.  (c) Dipole-dipole coupling $\ts\Delta{DD}$ as function of the SAQDM in-plane position at a vertical spacing of dipoles of $\ts{z}{DD}=30$~nm. (d) Modulation of the GQDM interdot barrier height $\ts{X}{DD}$ by the SAQDM dipole 30~nm away in the growth direction. In (c) and (d) contours where $|\ts\Delta{DD}|$ and $|\ts{X}{DD}|$ are equal to $100$ and $10$ $\mu$eV are shown by red and white, respectively, with solid/dashed lines coding petals with opposite sign of $\ts\Delta{DD}$ and $\ts{X}{DD}$.  }
	\label{simulationcolormap}
\end{figure*}

Our goal of hybridizing an optically active SAQDM to a GQDM critically depends on the magnitude of the capacitive coupling between their dipoles.  As the orientations of both dipoles are fixed (and orthogonal to each other), the remaining parameter is the relative spatial placement of the dots.  In practice, the vertical separation between dots has high precision (due to atomic layer growth) whereas the in-plane positioning is limited by fabrication techniques. Then, the question remains as to what the tolerances are in positioning this SAQDM above the gated dot in order to have substantial dipole coupling.  Choosing the vertical separation between dots is a balance between maximizing dipole coupling, maximizing QW mobility, and having sufficient tunneling to the SAQDM for initial loading.  This balance is constrained by the requirement of maintaining a device charge regime which is stable against reasonable amounts of disorder.  Some level of disorder in potentials will always be present in such a heterostructure stack, and a key goal for our simulation is to assure that stable charge regimes will be large enough in voltage space with reasonable values of exchange and capacitive coupling to survive reasonable amounts of disorder and charge noise.

To address these questions we perform 3D self-consistent Schr\"{o}dinger-Poisson simulations of the whole device (see Figure \ref{fig:devicecartoon}), employing a Thomas-Fermi approximation for the partitioned 2DEG and quantum-mechanical description of QD states in the inner area.  Detailed calculations of the level structure of the SAQDM are beyond the scope of this paper.  We focus our calculations on cases where the separation between the bottom of the self-assembled dot layer and the QW upper heterointerface is 20--30~nm (or 30--40nm dipole separation);  this is limited by the loading rate of the SAQDM from the QW 2DEG (with a 30\% Al barrier content).

Figure \ref{fig:devicecartoon} shows the device stack as entered into the solver.  Top and bottom Atomic Layer Deposition (ALD) dielectrics isolate the heterostack from the top aperture gate, and a set of gates at the back are used to form and manipulate GQDs, as briefly outlined in \refsec{fabrication}. Our simulations indicate that, for the specific dimensions chosen, the DCS sensitivity to the electron moving between the inner GQD and one of the outer GQD's (e.g. when discriminating between (0,1,1) and (0,0,2) states), is equivalent to $\sim$0.22~mV in the DCS bias. Modulation of a current through the DCS then is a function of the characteristics of a particular Coulomb peak, which is dependent on the electron effective temperature, barrier transparencies, and the presence and structure of low-lying excited states.  Simulations indicate that a specific desired charge configuration of the SAQDM can be maintained easily, owing to large addition energies of 30 meV or more and singlet-triplet splitting of a few meV.

There is a substantial amount of screening in the device both due to presence of gate metal and the 2DEG; thus, gate lever arms are strong functions of SAQDM spatial position. For example, the aperture gate (AG) lever arm varies from ~0.1~eV/V for a molecule position directly above the bath-accumulated 2DEG up to $\sim$0.4~eV/V outside the bath. It is important to note that the same AG also presents a fairly strong control of the potential in the non-accumulated regions of the QW, requiring additional back-gate bias compensation to restore the potential profile of the GQDs. Alternatively, GQDs can be formed after the SAQDM initial charging. Another important control is the capability to retune the SAQDM between (1,1) and (0,2) configurations. While the growth parameters are chosen so that at a particular bias the SAQDM is occupied with one electron in each dot in the vicinity of (1,1) to (0,2) transition, the same AG can be used to fine tune the levels (and $\JO$).  With $\sim$30~meV/V control of SAQDM detuning by the AG in the center of aperture, $\sim$50~mV AG bias should be sufficient to tune the desired span safely within a single addition energy interval.

The required spatial placement of the SAQDM in close proximity to the GQDM would render its own levels sensitive to other gates. We estimate this effect numerically. For example, a plunger gate would substantially affect the detuning of a SAQDM placed immediately above it with a lever arm of $\gtrsim 10$~meV/V; fortunately the range of influence of the gate falls off rapidly outside of its perimeter as it is strongly screened (see Fig. \ref{simulationcolormap}a).  In particular, it drops by a factor of 100 for a SAQDM placed above the center of a neighboring plunger gate 200~nm away.  We find it should be feasible to delegate control of the GQDM mostly to those remote gates.

The largest possible electrostatic coupling $\ts\Delta{DD}$ would result from the SAQDM directly above one of the electrons in the GQDM, and the dipole coupling decays rapidly as a function of the SAQDM-to-QW separation (see Fig. \ref{simulationcolormap}b).  Also, $\ts\Delta{DD}$ decays rapidly when the in-plane offset exceeds the vertical separation both due to its dipolar nature and screening.  It is worth mentioning that due to the nature of the SAQD system and for the gate geometries considered, GQDM electrons are positioned in the far field of the SAQDM dipole while the SAQDM is in near field of the GQDM.  In a stable charge configuration regime, our simulations find that the magnitude of $\ts\Delta{DD}$ between the SAQDM positioned in a 30~nm plane above two electrons in the GQDM is of order 1~meV as shown in Fig.~\ref{simulationcolormap}c. For an SAQDM close to its singlet anticrossing ($\ts{t}{O}\sim\ts\epsilon{O}$), this order of magnitude indicates a controlled-$Z$ timescale of order $\sim\hbar/\ts\Delta{DD}$, which lower-bounds $\ts{t}{CZ}$ by picoseconds.  In practice, $\ts\Delta{DD}$ will be weaker due to imperfect placement of the SAQDM with respect to the GQDM. A roughly 100~nm diameter contour line is shown in Fig.~\ref{simulationcolormap}c indicating where this drops to 100~$\mu eV$ coupling, and a roughly 150 nm line is shown indicating a drop to 10~$\mu$eV.  Since $\ts{t}{CZ}$ is tunable and will be most conveniently of order nanoseconds to match pulse-timing hardware while still being substantially less than nuclear-induced $T_2$ times, this indicates a substantial tolerance of over 100~nm in the SAQDM dot placement.  The value of $\ts{t}{CZ}$ will also be lengthened further from the SAQDM anticrossing, as indicated by \refeq{approxJOE}.

The dipole of the SAQDM does not only affect detuning; the SAQDM may also significantly modulate the interdot tunnel-barrier of the GQDM if positioned in the vicinity of an exchange gate.   The result of barrier-height modulation may be understood in the context of the WKB approximation, where for a 1D parabolic barrier $t_E\propto\exp(-\pi\ts\chi{E}/\ts{C}{E})$ for barrier height $\ts{\chi}{E}$ and barrier curvature $\ts{C}{E}$, which is of the order of a few meV.  The modulation of the GQDM interdot barrier height, $\ts{X}{DD}$, (i.e. the variation of $\ts\chi{E}$ due to a single electron displacement in the SAQDM) as a function of the SAQDM in-plane position is presented in Fig. \ref{simulationcolormap}(d). This modulation is most important if the SAQDM is placed symmetrically between the dots, in which case the inflicted GQDM detuning ($\propto \ts\Delta{DD}$) is minimal and barrier transparency dominates $\JOE$.

%% file: device_design_discussion.tex
A number of imperfections will reduce the fidelity of entanglement in the scheme we have presented.  We have already discussed the fact that finite photon detection bandwidth (in comparison to $\ts{J}{O}$) will ultimately dephase our created entanglement.  We have also indicated that the probability of success for each round of our entanglement scheme depends on the heralded detection of a single photon, the probability of which reduces with each reduction in collection efficiency in the ultimate system.  Failure requires reset, and although this may be rapid (limited by a few nanoseconds), these failures are likely to provide a severe rate-limiting bottleneck in most realizations.  Although it is beyond the scope of our design and simulations, SAQDMs may be integrated into distributed-Bragg-reflector, photonic-crystal, waveguide, or plasmonic cavities to improve photon extraction efficiency and enhanced the radiative decay rate through the Purcell effect~\cite{vorancomms}.  Engineering such a microcavity could push the heralded success probability reasonably close to the limit of 1/2.  Detection of the SAQDM charge state via the DCS in the GQD layer is not considered in this scheme since the SAQDM detuning is envisioned to remain constant and far from the (2,0) charge regime in which single-shot singlet detection could occur.

Other sources of decoherence have yet to be addressed.  A major such source is hyperfine effects from nuclear spins in the III-V semiconductors of the heterostructure.  These are known to cause dephasing times comparable to the expected entanglement interval $\ts{t}{CZ}$, an error source which absolutely requires mitigation.  Fortunately, several mitigations exist.  For both self-assembled and gated quantum dots, spin-echo techniques show that decoherence times after refocusing are several microseconds at worst, and can be extended to 100s of $\mu$s using multi-pulse dynamical decoupling~\cite{bluhm_dd}.  For the SAQDM, DD can be performed using optical pulses which periodically rotate the $\ts{\ket{S}}{O}/\ts{\ket{T_0}}{O}$ qubit with respect to the nuclear magnetization.  Improved dephasing is also available using nuclear spin locking techniques~\cite{xunat,latta09,laddprl}.  For the GQDM, exchange-pulse DD is possible using at least 6 pulses to permute the three dots~\cite{westfongnjp}.  Another form of dynamical decoupling is also possible for the GQDM, due to the encoding of this qubit into a subspace of fixed total angular momentum with angular momentum projection treated as a gauge freedom.  This encoding renders the DFS qubit immune to global magnetic fields.  However, the field gradients due to substrate nuclei impact the DFS qubit with opposite signs depending on this gauge freedom.  Therefore, periodically rotating this gauge freedom using microwave fields resonant with the Zeeman splitting of electrons in the GQDMs in a small magnetic field will decouple nuclear spin effects without adverse effect on the desired exchange dynamics~\cite{laddprb}.  This may be done without impacting the SAQDM layer due to the different electron $g$-factors in the different semiconductor alloys. Further, nuclei may be directly decoupled at finite magnetic field via RF fields for randomizing nuclear polarization.  More speculatively, self-assembled III-V quantum dots have been successfully grown on silicon substrates \cite{benyoucef13}.  It may be possible then to construct the GQDM layer out of an isotopically purified Si/SiGe heterostructure, eliminating the nuclear problem for the long-term memory layer of our device entirely.

While many solutions exist for mitigating the effects of hyperfine noise, the ubiquitous problem of charge noise, typically with a $1/f$ spectral density, may be more problematic.  The capacitive coupling we have presented deliberately exploits the sensitivity of exchange couplings to fluctuating electric dipoles, including those resulting from defects in the heterostructure.  The buried oxide layer and gates are likely to be the primary source of such noise in our proposed design.  This type of noise limits the fidelity of demonstrated capacitive entangling gates in GQDM experiments \cite{shulmansci}.  The two counters to this noise source are improvement in materials and the use of noise-compensating pulse sequences, some of which are derived in Ref.~\onlinecite{wang_sarma_pra}. Noise compensation in the presence of $1/f$ noise may particularly be improved by changing the entanglement strategy to use microwaves, modulated optical control, and resonant effects in the GQDM as in Ref.~\onlinecite{medfordprl}.

%% file: device_design.bbl
\begin{thebibliography}{96}%
\makeatletter
\providecommand \@ifxundefined [1]{%
 \@ifx{#1\undefined}
}%
\providecommand \@ifnum [1]{%
 \ifnum #1\expandafter \@firstoftwo
 \else \expandafter \@secondoftwo
 \fi
}%
\providecommand \@ifx [1]{%
 \ifx #1\expandafter \@firstoftwo
 \else \expandafter \@secondoftwo
 \fi
}%
\providecommand \natexlab [1]{#1}%
\providecommand \enquote  [1]{``#1''}%
\providecommand \bibnamefont  [1]{#1}%
\providecommand \bibfnamefont [1]{#1}%
\providecommand \citenamefont [1]{#1}%
\providecommand \href@noop [0]{\@secondoftwo}%
\providecommand \href [0]{\begingroup \@sanitize@url \@href}%
\providecommand \@href[1]{\@@startlink{#1}\@@href}%
\providecommand \@@href[1]{\endgroup#1\@@endlink}%
\providecommand \@sanitize@url [0]{\catcode `\\12\catcode `\$12\catcode
  `\&12\catcode `\#12\catcode `\^12\catcode `\_12\catcode `\%12\relax}%
\providecommand \@@startlink[1]{}%
\providecommand \@@endlink[0]{}%
\providecommand \url  [0]{\begingroup\@sanitize@url \@url }%
\providecommand \@url [1]{\endgroup\@href {#1}{\urlprefix }}%
\providecommand \urlprefix  [0]{URL }%
\providecommand \Eprint [0]{\href }%
\providecommand \doibase [0]{http://dx.doi.org/}%
\providecommand \selectlanguage [0]{\@gobble}%
\providecommand \bibinfo  [0]{\@secondoftwo}%
\providecommand \bibfield  [0]{\@secondoftwo}%
\providecommand \translation [1]{[#1]}%
\providecommand \BibitemOpen [0]{}%
\providecommand \bibitemStop [0]{}%
\providecommand \bibitemNoStop [0]{.\EOS\space}%
\providecommand \EOS [0]{\spacefactor3000\relax}%
\providecommand \BibitemShut  [1]{\csname bibitem#1\endcsname}%
\let\auto@bib@innerbib\@empty
\bibitem [{\citenamefont {Gisin}\ \emph {et~al.}(2002)\citenamefont {Gisin},
  \citenamefont {Ribordy}, \citenamefont {Tittel},\ and\ \citenamefont
  {Zbinden}}]{gisinrmp}%
  \BibitemOpen
  \bibfield  {author} {\bibinfo {author} {\bibfnamefont {N.}~\bibnamefont
  {Gisin}}, \bibinfo {author} {\bibfnamefont {G.}~\bibnamefont {Ribordy}},
  \bibinfo {author} {\bibfnamefont {W.}~\bibnamefont {Tittel}}, \ and\ \bibinfo
  {author} {\bibfnamefont {H.}~\bibnamefont {Zbinden}},\ }\href {\doibase
  10.1103/RevModPhys.74.145} {\bibfield  {journal} {\bibinfo  {journal} {Rev.
  Mod. Phys.}\ }\textbf {\bibinfo {volume} {74}},\ \bibinfo {pages} {145}
  (\bibinfo {year} {2002})}\BibitemShut {NoStop}%
\bibitem [{\citenamefont {Giovannetti}\ \emph {et~al.}(2011)\citenamefont
  {Giovannetti}, \citenamefont {Lloyd},\ and\ \citenamefont
  {Maccone}}]{metrology}%
  \BibitemOpen
  \bibfield  {author} {\bibinfo {author} {\bibfnamefont {V.}~\bibnamefont
  {Giovannetti}}, \bibinfo {author} {\bibfnamefont {S.}~\bibnamefont {Lloyd}},
  \ and\ \bibinfo {author} {\bibfnamefont {L.}~\bibnamefont {Maccone}},\
  }\href@noop {} {\bibfield  {journal} {\bibinfo  {journal} {Nature Photon.}\
  }\textbf {\bibinfo {volume} {5}},\ \bibinfo {pages} {222} (\bibinfo {year}
  {2011})}\BibitemShut {NoStop}%
\bibitem [{\citenamefont {Barz}\ \emph {et~al.}(2012)\citenamefont {Barz},
  \citenamefont {Kashefi}, \citenamefont {Broadbent}, \citenamefont
  {Fitzsimons}, \citenamefont {Zeilinger},\ and\ \citenamefont
  {Walther}}]{Barz20012012}%
  \BibitemOpen
  \bibfield  {author} {\bibinfo {author} {\bibfnamefont {S.}~\bibnamefont
  {Barz}}, \bibinfo {author} {\bibfnamefont {E.}~\bibnamefont {Kashefi}},
  \bibinfo {author} {\bibfnamefont {A.}~\bibnamefont {Broadbent}}, \bibinfo
  {author} {\bibfnamefont {J.~F.}\ \bibnamefont {Fitzsimons}}, \bibinfo
  {author} {\bibfnamefont {A.}~\bibnamefont {Zeilinger}}, \ and\ \bibinfo
  {author} {\bibfnamefont {P.}~\bibnamefont {Walther}},\ }\href {\doibase
  10.1126/science.1214707} {\bibfield  {journal} {\bibinfo  {journal}
  {Science}\ }\textbf {\bibinfo {volume} {335}},\ \bibinfo {pages} {303}
  (\bibinfo {year} {2012})}\BibitemShut {NoStop}%
\bibitem [{\citenamefont {Waks}\ \emph {et~al.}(2002)\citenamefont {Waks},
  \citenamefont {Inoue}, \citenamefont {Santori}, \citenamefont {Fattal},
  \citenamefont {Vuckovic}, \citenamefont {Solomon},\ and\ \citenamefont
  {Yamamoto}}]{waks_nature}%
  \BibitemOpen
  \bibfield  {author} {\bibinfo {author} {\bibfnamefont {E.}~\bibnamefont
  {Waks}}, \bibinfo {author} {\bibfnamefont {K.}~\bibnamefont {Inoue}},
  \bibinfo {author} {\bibfnamefont {C.}~\bibnamefont {Santori}}, \bibinfo
  {author} {\bibfnamefont {D.}~\bibnamefont {Fattal}}, \bibinfo {author}
  {\bibfnamefont {J.}~\bibnamefont {Vuckovic}}, \bibinfo {author}
  {\bibfnamefont {G.~S.}\ \bibnamefont {Solomon}}, \ and\ \bibinfo {author}
  {\bibfnamefont {Y.}~\bibnamefont {Yamamoto}},\ }\href
  {http://dx.doi.org/10.1038/420762a} {\bibfield  {journal} {\bibinfo
  {journal} {Nature}\ }\textbf {\bibinfo {volume} {420}},\ \bibinfo {pages}
  {762} (\bibinfo {year} {2002})}\BibitemShut {NoStop}%
\bibitem [{\citenamefont {Grosshans}\ \emph {et~al.}(2002)\citenamefont
  {Grosshans}, \citenamefont {{Van Assche}}, \citenamefont {Wenger},
  \citenamefont {Brouri}, \citenamefont {Cerf},\ and\ \citenamefont
  {Grangier}}]{GrangierQKD}%
  \BibitemOpen
  \bibfield  {author} {\bibinfo {author} {\bibfnamefont {F.}~\bibnamefont
  {Grosshans}}, \bibinfo {author} {\bibfnamefont {G.}~\bibnamefont {{Van
  Assche}}}, \bibinfo {author} {\bibfnamefont {J.}~\bibnamefont {Wenger}},
  \bibinfo {author} {\bibfnamefont {R.}~\bibnamefont {Brouri}}, \bibinfo
  {author} {\bibfnamefont {N.~J.}\ \bibnamefont {Cerf}}, \ and\ \bibinfo
  {author} {\bibfnamefont {P.}~\bibnamefont {Grangier}},\ }\href@noop {}
  {\bibfield  {journal} {\bibinfo  {journal} {Nature}\ }\textbf {\bibinfo
  {volume} {421}},\ \bibinfo {pages} {238} (\bibinfo {year}
  {2002})}\BibitemShut {NoStop}%
\bibitem [{\citenamefont {Rosenberg}\ \emph {et~al.}(2007)\citenamefont
  {Rosenberg}, \citenamefont {Harrington}, \citenamefont {Rice}, \citenamefont
  {Hiskett}, \citenamefont {Peterson}, \citenamefont {Hughes}, \citenamefont
  {Lita}, \citenamefont {Nam},\ and\ \citenamefont
  {Nordholt}}]{PhysRevLett.98.010503}%
  \BibitemOpen
  \bibfield  {author} {\bibinfo {author} {\bibfnamefont {D.}~\bibnamefont
  {Rosenberg}}, \bibinfo {author} {\bibfnamefont {J.~W.}\ \bibnamefont
  {Harrington}}, \bibinfo {author} {\bibfnamefont {P.~R.}\ \bibnamefont
  {Rice}}, \bibinfo {author} {\bibfnamefont {P.~A.}\ \bibnamefont {Hiskett}},
  \bibinfo {author} {\bibfnamefont {C.~G.}\ \bibnamefont {Peterson}}, \bibinfo
  {author} {\bibfnamefont {R.~J.}\ \bibnamefont {Hughes}}, \bibinfo {author}
  {\bibfnamefont {A.~E.}\ \bibnamefont {Lita}}, \bibinfo {author}
  {\bibfnamefont {S.~W.}\ \bibnamefont {Nam}}, \ and\ \bibinfo {author}
  {\bibfnamefont {J.~E.}\ \bibnamefont {Nordholt}},\ }\href {\doibase
  10.1103/PhysRevLett.98.010503} {\bibfield  {journal} {\bibinfo  {journal}
  {Phys. Rev. Lett.}\ }\textbf {\bibinfo {volume} {98}},\ \bibinfo {pages}
  {010503} (\bibinfo {year} {2007})}\BibitemShut {NoStop}%
\bibitem [{\citenamefont {Takesue}\ \emph {et~al.}(2007)\citenamefont
  {Takesue}, \citenamefont {Nam}, \citenamefont {Zhang}, \citenamefont
  {Hadfield}, \citenamefont {Honjo}, \citenamefont {Tamaki},\ and\
  \citenamefont {Yamamoto}}]{TakesueQKD}%
  \BibitemOpen
  \bibfield  {author} {\bibinfo {author} {\bibfnamefont {H.}~\bibnamefont
  {Takesue}}, \bibinfo {author} {\bibfnamefont {S.~W.}\ \bibnamefont {Nam}},
  \bibinfo {author} {\bibfnamefont {Q.}~\bibnamefont {Zhang}}, \bibinfo
  {author} {\bibfnamefont {R.~H.}\ \bibnamefont {Hadfield}}, \bibinfo {author}
  {\bibfnamefont {T.}~\bibnamefont {Honjo}}, \bibinfo {author} {\bibfnamefont
  {K.}~\bibnamefont {Tamaki}}, \ and\ \bibinfo {author} {\bibfnamefont
  {Y.}~\bibnamefont {Yamamoto}},\ }\href@noop {} {\bibfield  {journal}
  {\bibinfo  {journal} {Nature Photon.}\ }\textbf {\bibinfo {volume} {1}},\
  \bibinfo {pages} {343} (\bibinfo {year} {2007})}\BibitemShut {NoStop}%
\bibitem [{\citenamefont {Ursin}\ \emph {et~al.}(2007)\citenamefont {Ursin},
  \citenamefont {Tiefenbacher}, \citenamefont {Schmitt-Manderbach},
  \citenamefont {Weier}, \citenamefont {Scheidl}, \citenamefont {Lindenthal},
  \citenamefont {Blauensteiner}, \citenamefont {Jennewein}, \citenamefont
  {Perdigues}, \citenamefont {Trojek}, \citenamefont {Omer}, \citenamefont
  {Furst}, \citenamefont {Meyenburg}, \citenamefont {Rarity}, \citenamefont
  {Sodnik}, \citenamefont {Barbieri}, \citenamefont {Weinfurter},\ and\
  \citenamefont {Zeilinger}}]{QKDentangled}%
  \BibitemOpen
  \bibfield  {author} {\bibinfo {author} {\bibfnamefont {R.}~\bibnamefont
  {Ursin}}, \bibinfo {author} {\bibfnamefont {F.}~\bibnamefont {Tiefenbacher}},
  \bibinfo {author} {\bibfnamefont {T.}~\bibnamefont {Schmitt-Manderbach}},
  \bibinfo {author} {\bibfnamefont {H.}~\bibnamefont {Weier}}, \bibinfo
  {author} {\bibfnamefont {T.}~\bibnamefont {Scheidl}}, \bibinfo {author}
  {\bibfnamefont {M.}~\bibnamefont {Lindenthal}}, \bibinfo {author}
  {\bibfnamefont {B.}~\bibnamefont {Blauensteiner}}, \bibinfo {author}
  {\bibfnamefont {T.}~\bibnamefont {Jennewein}}, \bibinfo {author}
  {\bibfnamefont {J.}~\bibnamefont {Perdigues}}, \bibinfo {author}
  {\bibfnamefont {P.}~\bibnamefont {Trojek}}, \bibinfo {author} {\bibfnamefont
  {B.}~\bibnamefont {Omer}}, \bibinfo {author} {\bibfnamefont {M.}~\bibnamefont
  {Furst}}, \bibinfo {author} {\bibfnamefont {M.}~\bibnamefont {Meyenburg}},
  \bibinfo {author} {\bibfnamefont {J.}~\bibnamefont {Rarity}}, \bibinfo
  {author} {\bibfnamefont {Z.}~\bibnamefont {Sodnik}}, \bibinfo {author}
  {\bibfnamefont {C.}~\bibnamefont {Barbieri}}, \bibinfo {author}
  {\bibfnamefont {H.}~\bibnamefont {Weinfurter}}, \ and\ \bibinfo {author}
  {\bibfnamefont {A.}~\bibnamefont {Zeilinger}},\ }\href@noop {} {\bibfield
  {journal} {\bibinfo  {journal} {Nature Phys.}\ }\textbf {\bibinfo {volume}
  {3}},\ \bibinfo {pages} {481} (\bibinfo {year} {2007})}\BibitemShut {NoStop}%
\bibitem [{\citenamefont {{Van Meter}}(2014)}]{RodText}%
  \BibitemOpen
  \bibfield  {author} {\bibinfo {author} {\bibfnamefont {R.}~\bibnamefont {{Van
  Meter}}},\ }\href@noop {} {\emph {\bibinfo {title} {Quantum Networking}}}\
  (\bibinfo  {publisher} {Wiley},\ \bibinfo {year} {2014})\BibitemShut
  {NoStop}%
\bibitem [{\citenamefont {Duan}\ \emph {et~al.}(2001)\citenamefont {Duan},
  \citenamefont {Lukin}, \citenamefont {Cirac},\ and\ \citenamefont
  {Zoller}}]{dlcz}%
  \BibitemOpen
  \bibfield  {author} {\bibinfo {author} {\bibfnamefont {L.~M.}\ \bibnamefont
  {Duan}}, \bibinfo {author} {\bibfnamefont {M.~D.}\ \bibnamefont {Lukin}},
  \bibinfo {author} {\bibfnamefont {J.~I.}\ \bibnamefont {Cirac}}, \ and\
  \bibinfo {author} {\bibfnamefont {P.}~\bibnamefont {Zoller}},\ }\href@noop {}
  {\bibfield  {journal} {\bibinfo  {journal} {Nature}\ }\textbf {\bibinfo
  {volume} {414}},\ \bibinfo {pages} {413} (\bibinfo {year}
  {2001})}\BibitemShut {NoStop}%
\bibitem [{\citenamefont {Monroe}\ and\ \citenamefont {Kim}(2013)}]{monroesci}%
  \BibitemOpen
  \bibfield  {author} {\bibinfo {author} {\bibfnamefont {C.}~\bibnamefont
  {Monroe}}\ and\ \bibinfo {author} {\bibfnamefont {J.}~\bibnamefont {Kim}},\
  }\href {\doibase 10.1126/science.1231298} {\bibfield  {journal} {\bibinfo
  {journal} {Science}\ }\textbf {\bibinfo {volume} {339}},\ \bibinfo {pages}
  {1164} (\bibinfo {year} {2013})}\BibitemShut {NoStop}%
\bibitem [{\citenamefont {Hill}\ \emph {et~al.}(2012)\citenamefont {Hill},
  \citenamefont {Safavi-Naeini}, \citenamefont {Chan},\ and\ \citenamefont
  {Painter}}]{painterncomms}%
  \BibitemOpen
  \bibfield  {author} {\bibinfo {author} {\bibfnamefont {J.~T.}\ \bibnamefont
  {Hill}}, \bibinfo {author} {\bibfnamefont {A.~H.}\ \bibnamefont
  {Safavi-Naeini}}, \bibinfo {author} {\bibfnamefont {J.}~\bibnamefont {Chan}},
  \ and\ \bibinfo {author} {\bibfnamefont {O.}~\bibnamefont {Painter}},\
  }\href@noop {} {\bibfield  {journal} {\bibinfo  {journal} {Nat. Commun.}\
  }\textbf {\bibinfo {volume} {3}},\ \bibinfo {pages} {1196} (\bibinfo {year}
  {2012})}\BibitemShut {NoStop}%
\bibitem [{\citenamefont {Palomaki}\ \emph {et~al.}(2013)\citenamefont
  {Palomaki}, \citenamefont {Teufel}, \citenamefont {Simmonds},\ and\
  \citenamefont {Lehnert}}]{lehnertsci}%
  \BibitemOpen
  \bibfield  {author} {\bibinfo {author} {\bibfnamefont {T.~A.}\ \bibnamefont
  {Palomaki}}, \bibinfo {author} {\bibfnamefont {J.~D.}\ \bibnamefont
  {Teufel}}, \bibinfo {author} {\bibfnamefont {R.~W.}\ \bibnamefont
  {Simmonds}}, \ and\ \bibinfo {author} {\bibfnamefont {K.~W.}\ \bibnamefont
  {Lehnert}},\ }\href {\doibase 10.1126/science.1244563} {\bibfield  {journal}
  {\bibinfo  {journal} {Science}\ }\textbf {\bibinfo {volume} {342}},\ \bibinfo
  {pages} {710} (\bibinfo {year} {2013})}\BibitemShut {NoStop}%
\bibitem [{\citenamefont {Zhong}\ \emph {et~al.}(2015)\citenamefont {Zhong},
  \citenamefont {Hedges}, \citenamefont {Ahlefeldt}, \citenamefont
  {Bartholomew}, \citenamefont {Beavan}, \citenamefont {Wittig}, \citenamefont
  {Longdell},\ and\ \citenamefont {Sellars}}]{zhong_optically_2015}%
  \BibitemOpen
  \bibfield  {author} {\bibinfo {author} {\bibfnamefont {M.}~\bibnamefont
  {Zhong}}, \bibinfo {author} {\bibfnamefont {M.~P.}\ \bibnamefont {Hedges}},
  \bibinfo {author} {\bibfnamefont {R.~L.}\ \bibnamefont {Ahlefeldt}}, \bibinfo
  {author} {\bibfnamefont {J.~G.}\ \bibnamefont {Bartholomew}}, \bibinfo
  {author} {\bibfnamefont {S.~E.}\ \bibnamefont {Beavan}}, \bibinfo {author}
  {\bibfnamefont {S.~M.}\ \bibnamefont {Wittig}}, \bibinfo {author}
  {\bibfnamefont {J.~J.}\ \bibnamefont {Longdell}}, \ and\ \bibinfo {author}
  {\bibfnamefont {M.~J.}\ \bibnamefont {Sellars}},\ }\href@noop {} {\bibfield
  {journal} {\bibinfo  {journal} {Nature}\ }\textbf {\bibinfo {volume} {517}},\
  \bibinfo {pages} {177} (\bibinfo {year} {2015})}\BibitemShut {NoStop}%
\bibitem [{\citenamefont {Akopian}\ \emph {et~al.}(2011)\citenamefont
  {Akopian}, \citenamefont {Wang}, \citenamefont {Rastelli}, \citenamefont
  {Schmidt},\ and\ \citenamefont {Zwiller}}]{akopian_natphot}%
  \BibitemOpen
  \bibfield  {author} {\bibinfo {author} {\bibfnamefont {N.}~\bibnamefont
  {Akopian}}, \bibinfo {author} {\bibfnamefont {L.}~\bibnamefont {Wang}},
  \bibinfo {author} {\bibfnamefont {A.}~\bibnamefont {Rastelli}}, \bibinfo
  {author} {\bibfnamefont {O.~G.}\ \bibnamefont {Schmidt}}, \ and\ \bibinfo
  {author} {\bibfnamefont {V.}~\bibnamefont {Zwiller}},\ }\href@noop {}
  {\bibfield  {journal} {\bibinfo  {journal} {Nature Photon.}\ }\textbf
  {\bibinfo {volume} {5}},\ \bibinfo {pages} {230} (\bibinfo {year}
  {2011})}\BibitemShut {NoStop}%
\bibitem [{\citenamefont {Meyer}\ \emph {et~al.}(2015)\citenamefont {Meyer},
  \citenamefont {Stockill}, \citenamefont {Steiner}, \citenamefont {Le~Gall},
  \citenamefont {Matthiesen}, \citenamefont {Clarke}, \citenamefont {Ludwig},
  \citenamefont {Reichel}, \citenamefont {Atature},\ and\ \citenamefont
  {Kohl}}]{meyer_qdandion}%
  \BibitemOpen
  \bibfield  {author} {\bibinfo {author} {\bibfnamefont {M.}~\bibnamefont
  {Meyer}, \bibfnamefont {H.}}, \bibinfo {author} {\bibfnamefont
  {R.}~\bibnamefont {Stockill}}, \bibinfo {author} {\bibfnamefont
  {M.}~\bibnamefont {Steiner}}, \bibinfo {author} {\bibfnamefont
  {C.}~\bibnamefont {Le~Gall}}, \bibinfo {author} {\bibfnamefont
  {C.}~\bibnamefont {Matthiesen}}, \bibinfo {author} {\bibfnamefont
  {E.}~\bibnamefont {Clarke}}, \bibinfo {author} {\bibfnamefont
  {A.}~\bibnamefont {Ludwig}}, \bibinfo {author} {\bibfnamefont
  {J.}~\bibnamefont {Reichel}}, \bibinfo {author} {\bibfnamefont
  {M.}~\bibnamefont {Atature}}, \ and\ \bibinfo {author} {\bibfnamefont
  {M.}~\bibnamefont {Kohl}},\ }\href {\doibase 10.1103/PhysRevLett.114.123001}
  {\bibfield  {journal} {\bibinfo  {journal} {Phys. Rev. Lett.}\ }\textbf
  {\bibinfo {volume} {114}},\ \bibinfo {pages} {123001} (\bibinfo {year}
  {2015})}\BibitemShut {NoStop}%
\bibitem [{\citenamefont {Zhu}\ \emph {et~al.}(2011)\citenamefont {Zhu},
  \citenamefont {Saito}, \citenamefont {Kemp}, \citenamefont {Kakuyanagi},
  \citenamefont {Karimoto}, \citenamefont {Nakano}, \citenamefont {Munro},
  \citenamefont {Tokura}, \citenamefont {Everitt}, \citenamefont {Nemoto},
  \citenamefont {Kasu}, \citenamefont {Mizuochi},\ and\ \citenamefont
  {Semba}}]{zhu_nature}%
  \BibitemOpen
  \bibfield  {author} {\bibinfo {author} {\bibfnamefont {X.}~\bibnamefont
  {Zhu}}, \bibinfo {author} {\bibfnamefont {S.}~\bibnamefont {Saito}}, \bibinfo
  {author} {\bibfnamefont {A.}~\bibnamefont {Kemp}}, \bibinfo {author}
  {\bibfnamefont {K.}~\bibnamefont {Kakuyanagi}}, \bibinfo {author}
  {\bibfnamefont {S.-i.}\ \bibnamefont {Karimoto}}, \bibinfo {author}
  {\bibfnamefont {H.}~\bibnamefont {Nakano}}, \bibinfo {author} {\bibfnamefont
  {W.~J.}\ \bibnamefont {Munro}}, \bibinfo {author} {\bibfnamefont
  {Y.}~\bibnamefont {Tokura}}, \bibinfo {author} {\bibfnamefont {M.~S.}\
  \bibnamefont {Everitt}}, \bibinfo {author} {\bibfnamefont {K.}~\bibnamefont
  {Nemoto}}, \bibinfo {author} {\bibfnamefont {M.}~\bibnamefont {Kasu}},
  \bibinfo {author} {\bibfnamefont {N.}~\bibnamefont {Mizuochi}}, \ and\
  \bibinfo {author} {\bibfnamefont {K.}~\bibnamefont {Semba}},\ }\href
  {http://dx.doi.org/10.1038/nature10462} {\bibfield  {journal} {\bibinfo
  {journal} {Nature}\ }\textbf {\bibinfo {volume} {478}},\ \bibinfo {pages}
  {221} (\bibinfo {year} {2011})}\BibitemShut {NoStop}%
\bibitem [{\citenamefont {De~Greve}\ \emph {et~al.}(2012)\citenamefont
  {De~Greve}, \citenamefont {Yu}, \citenamefont {McMahon}, \citenamefont
  {Pelc}, \citenamefont {Natarajan}, \citenamefont {Kim}, \citenamefont {Abe},
  \citenamefont {Maier}, \citenamefont {Schneider}, \citenamefont {Kamp},
  \citenamefont {Hofling}, \citenamefont {Hadfield}, \citenamefont {Forchel},
  \citenamefont {Fejer},\ and\ \citenamefont {Yamamoto}}]{degrevenat}%
  \BibitemOpen
  \bibfield  {author} {\bibinfo {author} {\bibfnamefont {K.}~\bibnamefont
  {De~Greve}}, \bibinfo {author} {\bibfnamefont {L.}~\bibnamefont {Yu}},
  \bibinfo {author} {\bibfnamefont {P.~L.}\ \bibnamefont {McMahon}}, \bibinfo
  {author} {\bibfnamefont {J.~S.}\ \bibnamefont {Pelc}}, \bibinfo {author}
  {\bibfnamefont {C.~M.}\ \bibnamefont {Natarajan}}, \bibinfo {author}
  {\bibfnamefont {N.~Y.}\ \bibnamefont {Kim}}, \bibinfo {author} {\bibfnamefont
  {E.}~\bibnamefont {Abe}}, \bibinfo {author} {\bibfnamefont {S.}~\bibnamefont
  {Maier}}, \bibinfo {author} {\bibfnamefont {C.}~\bibnamefont {Schneider}},
  \bibinfo {author} {\bibfnamefont {M.}~\bibnamefont {Kamp}}, \bibinfo {author}
  {\bibfnamefont {S.}~\bibnamefont {Hofling}}, \bibinfo {author} {\bibfnamefont
  {R.~H.}\ \bibnamefont {Hadfield}}, \bibinfo {author} {\bibfnamefont
  {A.}~\bibnamefont {Forchel}}, \bibinfo {author} {\bibfnamefont {M.~M.}\
  \bibnamefont {Fejer}}, \ and\ \bibinfo {author} {\bibfnamefont
  {Y.}~\bibnamefont {Yamamoto}},\ }\href@noop {} {\bibfield  {journal}
  {\bibinfo  {journal} {Nature}\ }\textbf {\bibinfo {volume} {491}},\ \bibinfo
  {pages} {421} (\bibinfo {year} {2012})}\BibitemShut {NoStop}%
\bibitem [{\citenamefont {Gao}\ \emph {et~al.}(2012)\citenamefont {Gao},
  \citenamefont {Fallahi}, \citenamefont {Togan}, \citenamefont
  {Miguel-Sanchez},\ and\ \citenamefont {Imamoglu}}]{gaonat}%
  \BibitemOpen
  \bibfield  {author} {\bibinfo {author} {\bibfnamefont {W.~B.}\ \bibnamefont
  {Gao}}, \bibinfo {author} {\bibfnamefont {P.}~\bibnamefont {Fallahi}},
  \bibinfo {author} {\bibfnamefont {E.}~\bibnamefont {Togan}}, \bibinfo
  {author} {\bibfnamefont {J.}~\bibnamefont {Miguel-Sanchez}}, \ and\ \bibinfo
  {author} {\bibfnamefont {A.}~\bibnamefont {Imamoglu}},\ }\href@noop {}
  {\bibfield  {journal} {\bibinfo  {journal} {Nature}\ }\textbf {\bibinfo
  {volume} {491}},\ \bibinfo {pages} {426} (\bibinfo {year}
  {2012})}\BibitemShut {NoStop}%
\bibitem [{\citenamefont {Schaibley}\ \emph {et~al.}(2013)\citenamefont
  {Schaibley}, \citenamefont {Burgers}, \citenamefont {McCracken},
  \citenamefont {Duan}, \citenamefont {Berman}, \citenamefont {Steel},
  \citenamefont {Bracker}, \citenamefont {Gammon},\ and\ \citenamefont
  {Sham}}]{shaibleyprl}%
  \BibitemOpen
  \bibfield  {author} {\bibinfo {author} {\bibfnamefont {J.~R.}\ \bibnamefont
  {Schaibley}}, \bibinfo {author} {\bibfnamefont {A.~P.}\ \bibnamefont
  {Burgers}}, \bibinfo {author} {\bibfnamefont {G.~A.}\ \bibnamefont
  {McCracken}}, \bibinfo {author} {\bibfnamefont {L.-M.}\ \bibnamefont {Duan}},
  \bibinfo {author} {\bibfnamefont {P.~R.}\ \bibnamefont {Berman}}, \bibinfo
  {author} {\bibfnamefont {D.~G.}\ \bibnamefont {Steel}}, \bibinfo {author}
  {\bibfnamefont {A.~S.}\ \bibnamefont {Bracker}}, \bibinfo {author}
  {\bibfnamefont {D.}~\bibnamefont {Gammon}}, \ and\ \bibinfo {author}
  {\bibfnamefont {L.~J.}\ \bibnamefont {Sham}},\ }\href {\doibase
  10.1103/PhysRevLett.110.167401} {\bibfield  {journal} {\bibinfo  {journal}
  {Phys. Rev. Lett.}\ }\textbf {\bibinfo {volume} {110}},\ \bibinfo {pages}
  {167401} (\bibinfo {year} {2013})}\BibitemShut {NoStop}%
\bibitem [{\citenamefont {Medford}\ \emph
  {et~al.}(2013{\natexlab{a}})\citenamefont {Medford}, \citenamefont {Beil},
  \citenamefont {Taylor}, \citenamefont {Bartlett}, \citenamefont {Doherty},
  \citenamefont {Rashba}, \citenamefont {DiVincenzo}, \citenamefont {Lu},
  \citenamefont {Gossard},\ and\ \citenamefont
  {Marcus}}]{medford_self-consistent_2013}%
  \BibitemOpen
  \bibfield  {author} {\bibinfo {author} {\bibfnamefont {J.}~\bibnamefont
  {Medford}}, \bibinfo {author} {\bibfnamefont {J.}~\bibnamefont {Beil}},
  \bibinfo {author} {\bibfnamefont {J.~M.}\ \bibnamefont {Taylor}}, \bibinfo
  {author} {\bibfnamefont {S.~D.}\ \bibnamefont {Bartlett}}, \bibinfo {author}
  {\bibfnamefont {A.~C.}\ \bibnamefont {Doherty}}, \bibinfo {author}
  {\bibfnamefont {E.~I.}\ \bibnamefont {Rashba}}, \bibinfo {author}
  {\bibfnamefont {D.~P.}\ \bibnamefont {DiVincenzo}}, \bibinfo {author}
  {\bibfnamefont {H.}~\bibnamefont {Lu}}, \bibinfo {author} {\bibfnamefont
  {A.~C.}\ \bibnamefont {Gossard}}, \ and\ \bibinfo {author} {\bibfnamefont
  {C.~M.}\ \bibnamefont {Marcus}},\ }\href {\doibase 10.1038/nnano.2013.168}
  {\bibfield  {journal} {\bibinfo  {journal} {Nature Nanotech.}\ }\textbf
  {\bibinfo {volume} {8}},\ \bibinfo {pages} {654} (\bibinfo {year}
  {2013}{\natexlab{a}})}\BibitemShut {NoStop}%
\bibitem [{\citenamefont {Eng}\ \emph {et~al.}(2015)\citenamefont {Eng},
  \citenamefont {Ladd}, \citenamefont {Smith}, \citenamefont {Borselli},
  \citenamefont {Kiselev}, \citenamefont {Fong}, \citenamefont {Holabird},
  \citenamefont {Hazard}, \citenamefont {Huang}, \citenamefont {Deelman},
  \citenamefont {Milosavljevic}, \citenamefont {Schmitz}, \citenamefont {Ross},
  \citenamefont {Gyure},\ and\ \citenamefont {Hunter}}]{HRL}%
  \BibitemOpen
  \bibfield  {author} {\bibinfo {author} {\bibfnamefont {K.}~\bibnamefont
  {Eng}}, \bibinfo {author} {\bibfnamefont {T.~D.}\ \bibnamefont {Ladd}},
  \bibinfo {author} {\bibfnamefont {A.}~\bibnamefont {Smith}}, \bibinfo
  {author} {\bibfnamefont {M.~G.}\ \bibnamefont {Borselli}}, \bibinfo {author}
  {\bibfnamefont {A.~A.}\ \bibnamefont {Kiselev}}, \bibinfo {author}
  {\bibfnamefont {B.~H.}\ \bibnamefont {Fong}}, \bibinfo {author}
  {\bibfnamefont {K.~S.}\ \bibnamefont {Holabird}}, \bibinfo {author}
  {\bibfnamefont {T.~M.}\ \bibnamefont {Hazard}}, \bibinfo {author}
  {\bibfnamefont {B.}~\bibnamefont {Huang}}, \bibinfo {author} {\bibfnamefont
  {P.~W.}\ \bibnamefont {Deelman}}, \bibinfo {author} {\bibfnamefont
  {I.}~\bibnamefont {Milosavljevic}}, \bibinfo {author} {\bibfnamefont {A.~E.}\
  \bibnamefont {Schmitz}}, \bibinfo {author} {\bibfnamefont {R.~S.}\
  \bibnamefont {Ross}}, \bibinfo {author} {\bibfnamefont {M.~F.}\ \bibnamefont
  {Gyure}}, \ and\ \bibinfo {author} {\bibfnamefont {A.~T.}\ \bibnamefont
  {Hunter}},\ }\href@noop {} {\bibfield  {journal} {\bibinfo  {journal} {Sci.
  Adv.}\ }\textbf {\bibinfo {volume} {1}} (\bibinfo {year} {2015})}\BibitemShut
  {NoStop}%
\bibitem [{\citenamefont {Shulman}\ \emph {et~al.}(2012)\citenamefont
  {Shulman}, \citenamefont {Dial}, \citenamefont {Harvey}, \citenamefont
  {Bluhm}, \citenamefont {Umansky},\ and\ \citenamefont {Yacoby}}]{shulmansci}%
  \BibitemOpen
  \bibfield  {author} {\bibinfo {author} {\bibfnamefont {M.~D.}\ \bibnamefont
  {Shulman}}, \bibinfo {author} {\bibfnamefont {O.~E.}\ \bibnamefont {Dial}},
  \bibinfo {author} {\bibfnamefont {S.~P.}\ \bibnamefont {Harvey}}, \bibinfo
  {author} {\bibfnamefont {H.}~\bibnamefont {Bluhm}}, \bibinfo {author}
  {\bibfnamefont {V.}~\bibnamefont {Umansky}}, \ and\ \bibinfo {author}
  {\bibfnamefont {A.}~\bibnamefont {Yacoby}},\ }\href {\doibase
  10.1126/science.1217692} {\bibfield  {journal} {\bibinfo  {journal}
  {Science}\ }\textbf {\bibinfo {volume} {336}},\ \bibinfo {pages} {202}
  (\bibinfo {year} {2012})}\BibitemShut {NoStop}%
\bibitem [{\citenamefont {Jones}\ \emph {et~al.}(2015)\citenamefont {Jones},
  \citenamefont {Kim}, \citenamefont {Rakher}, \citenamefont {Kwiat},\ and\
  \citenamefont {Ladd}}]{cody2015}%
  \BibitemOpen
  \bibfield  {author} {\bibinfo {author} {\bibfnamefont {N.~C.}\ \bibnamefont
  {Jones}}, \bibinfo {author} {\bibfnamefont {D.}~\bibnamefont {Kim}}, \bibinfo
  {author} {\bibfnamefont {M.}~\bibnamefont {Rakher}}, \bibinfo {author}
  {\bibfnamefont {P.}~\bibnamefont {Kwiat}}, \ and\ \bibinfo {author}
  {\bibfnamefont {T.~D.}\ \bibnamefont {Ladd}},\ }\href@noop {} {\  (\bibinfo
  {year} {2015})},\ \bibinfo {note} {arXiv:1505.01536}\BibitemShut {NoStop}%
\bibitem [{\citenamefont {Vrijen}\ and\ \citenamefont
  {Yablonovitch}(2001)}]{vrijen_spin-coherent_2001}%
  \BibitemOpen
  \bibfield  {author} {\bibinfo {author} {\bibfnamefont {R.}~\bibnamefont
  {Vrijen}}\ and\ \bibinfo {author} {\bibfnamefont {E.}~\bibnamefont
  {Yablonovitch}},\ }\href {\doibase 10.1016/S1386-9477(00)00296-4} {\bibfield
  {journal} {\bibinfo  {journal} {Physica E}\ }\textbf {\bibinfo {volume}
  {10}},\ \bibinfo {pages} {569} (\bibinfo {year} {2001})}\BibitemShut
  {NoStop}%
\bibitem [{\citenamefont {Rao}\ \emph {et~al.}(2005)\citenamefont {Rao},
  \citenamefont {Szkopek}, \citenamefont {Robinson}, \citenamefont
  {Yablonovitch},\ and\ \citenamefont {Jiang}}]{rao_single_2005}%
  \BibitemOpen
  \bibfield  {author} {\bibinfo {author} {\bibfnamefont {D.~S.}\ \bibnamefont
  {Rao}}, \bibinfo {author} {\bibfnamefont {T.}~\bibnamefont {Szkopek}},
  \bibinfo {author} {\bibfnamefont {H.~D.}\ \bibnamefont {Robinson}}, \bibinfo
  {author} {\bibfnamefont {E.}~\bibnamefont {Yablonovitch}}, \ and\ \bibinfo
  {author} {\bibfnamefont {H.-W.}\ \bibnamefont {Jiang}},\ }\href {\doibase
  10.1063/1.2134888} {\bibfield  {journal} {\bibinfo  {journal} {J. Appl.
  Phys.}\ }\textbf {\bibinfo {volume} {98}},\ \bibinfo {pages} {114507}
  (\bibinfo {year} {2005})}\BibitemShut {NoStop}%
\bibitem [{\citenamefont {Croke}\ and\ \citenamefont
  {Gyure}(2008)}]{croke2008quantum}%
  \BibitemOpen
  \bibfield  {author} {\bibinfo {author} {\bibfnamefont {E.}~\bibnamefont
  {Croke}}\ and\ \bibinfo {author} {\bibfnamefont {M.}~\bibnamefont {Gyure}},\
  }\href@noop {} {\enquote {\bibinfo {title} {Quantum well design for a
  coherent, single-photon detector with spin resonant transistor},}\ }
  (\bibinfo {year} {2008}),\ \bibinfo {note} {{US Patent
  7,462,859}}\BibitemShut {NoStop}%
\bibitem [{\citenamefont {Kosaka}\ \emph {et~al.}(2009)\citenamefont {Kosaka},
  \citenamefont {Inagaki}, \citenamefont {Rikitake}, \citenamefont {Imamura},
  \citenamefont {Mitsumori},\ and\ \citenamefont
  {Edamatsu}}]{kosaka_spin_2009}%
  \BibitemOpen
  \bibfield  {author} {\bibinfo {author} {\bibfnamefont {H.}~\bibnamefont
  {Kosaka}}, \bibinfo {author} {\bibfnamefont {T.}~\bibnamefont {Inagaki}},
  \bibinfo {author} {\bibfnamefont {Y.}~\bibnamefont {Rikitake}}, \bibinfo
  {author} {\bibfnamefont {H.}~\bibnamefont {Imamura}}, \bibinfo {author}
  {\bibfnamefont {Y.}~\bibnamefont {Mitsumori}}, \ and\ \bibinfo {author}
  {\bibfnamefont {K.}~\bibnamefont {Edamatsu}},\ }\href {\doibase
  10.1038/nature07729} {\bibfield  {journal} {\bibinfo  {journal} {Nature}\
  }\textbf {\bibinfo {volume} {457}},\ \bibinfo {pages} {702} (\bibinfo {year}
  {2009})}\BibitemShut {NoStop}%
\bibitem [{\citenamefont {{Fujita}}\ \emph {et~al.}(2015)\citenamefont
  {{Fujita}}, \citenamefont {{Morimoto}}, \citenamefont {{Kiyama}},
  \citenamefont {{Allison}}, \citenamefont {{Larsson}}, \citenamefont
  {{Ludwig}}, \citenamefont {{Valentin}}, \citenamefont {{Wieck}},
  \citenamefont {{Oiwa}},\ and\ \citenamefont {{Tarucha}}}]{tarucha}%
  \BibitemOpen
  \bibfield  {author} {\bibinfo {author} {\bibfnamefont {T.}~\bibnamefont
  {{Fujita}}}, \bibinfo {author} {\bibfnamefont {K.}~\bibnamefont
  {{Morimoto}}}, \bibinfo {author} {\bibfnamefont {H.}~\bibnamefont
  {{Kiyama}}}, \bibinfo {author} {\bibfnamefont {G.}~\bibnamefont {{Allison}}},
  \bibinfo {author} {\bibfnamefont {M.}~\bibnamefont {{Larsson}}}, \bibinfo
  {author} {\bibfnamefont {A.}~\bibnamefont {{Ludwig}}}, \bibinfo {author}
  {\bibfnamefont {S.~R.}\ \bibnamefont {{Valentin}}}, \bibinfo {author}
  {\bibfnamefont {A.~D.}\ \bibnamefont {{Wieck}}}, \bibinfo {author}
  {\bibfnamefont {A.}~\bibnamefont {{Oiwa}}}, \ and\ \bibinfo {author}
  {\bibfnamefont {S.}~\bibnamefont {{Tarucha}}},\ }\href@noop {} {\bibfield
  {journal} {\bibinfo  {journal} {arXiv:1504.03696}\ } (\bibinfo {year}
  {2015})}\BibitemShut {NoStop}%
\bibitem [{\citenamefont {Arcari}\ \emph {et~al.}(2014)\citenamefont {Arcari},
  \citenamefont {Sollner}, \citenamefont {Javadi}, \citenamefont
  {Lindskov~Hansen}, \citenamefont {Mahmoodian}, \citenamefont {Liu},
  \citenamefont {Thyrrestrup}, \citenamefont {Lee}, \citenamefont {Song},
  \citenamefont {Stobbe},\ and\ \citenamefont {Lodahl}}]{arcari_prl}%
  \BibitemOpen
  \bibfield  {author} {\bibinfo {author} {\bibfnamefont {M.}~\bibnamefont
  {Arcari}}, \bibinfo {author} {\bibfnamefont {I.}~\bibnamefont {Sollner}},
  \bibinfo {author} {\bibfnamefont {A.}~\bibnamefont {Javadi}}, \bibinfo
  {author} {\bibfnamefont {S.}~\bibnamefont {Lindskov~Hansen}}, \bibinfo
  {author} {\bibfnamefont {S.}~\bibnamefont {Mahmoodian}}, \bibinfo {author}
  {\bibfnamefont {J.}~\bibnamefont {Liu}}, \bibinfo {author} {\bibfnamefont
  {H.}~\bibnamefont {Thyrrestrup}}, \bibinfo {author} {\bibfnamefont
  {E.}~\bibnamefont {Lee}}, \bibinfo {author} {\bibfnamefont {J.}~\bibnamefont
  {Song}}, \bibinfo {author} {\bibfnamefont {S.}~\bibnamefont {Stobbe}}, \ and\
  \bibinfo {author} {\bibfnamefont {P.}~\bibnamefont {Lodahl}},\ }\href
  {\doibase 10.1103/PhysRevLett.113.093603} {\bibfield  {journal} {\bibinfo
  {journal} {Phys. Rev. Lett.}\ }\textbf {\bibinfo {volume} {113}},\ \bibinfo
  {pages} {093603} (\bibinfo {year} {2014})}\BibitemShut {NoStop}%
\bibitem [{\citenamefont {{Engel}}\ \emph {et~al.}(2006)\citenamefont
  {{Engel}}, \citenamefont {{Taylor}}, \citenamefont {{Lukin}},\ and\
  \citenamefont {{Imamoglu}}}]{engel2006}%
  \BibitemOpen
  \bibfield  {author} {\bibinfo {author} {\bibfnamefont {H.-A.}\ \bibnamefont
  {{Engel}}}, \bibinfo {author} {\bibfnamefont {J.~M.}\ \bibnamefont
  {{Taylor}}}, \bibinfo {author} {\bibfnamefont {M.~D.}\ \bibnamefont
  {{Lukin}}}, \ and\ \bibinfo {author} {\bibfnamefont {A.}~\bibnamefont
  {{Imamoglu}}},\ }\href@noop {} {\bibfield  {journal} {\bibinfo  {journal}
  {arXiv:cond-mat/0612700}\ } (\bibinfo {year} {2006})}\BibitemShut {NoStop}%
\bibitem [{\citenamefont {Kim}\ \emph {et~al.}(2011)\citenamefont {Kim},
  \citenamefont {Carter}, \citenamefont {Greilich}, \citenamefont {Bracker},\
  and\ \citenamefont {Gammon}}]{dkimnphys}%
  \BibitemOpen
  \bibfield  {author} {\bibinfo {author} {\bibfnamefont {D.}~\bibnamefont
  {Kim}}, \bibinfo {author} {\bibfnamefont {S.~G.}\ \bibnamefont {Carter}},
  \bibinfo {author} {\bibfnamefont {A.}~\bibnamefont {Greilich}}, \bibinfo
  {author} {\bibfnamefont {A.~S.}\ \bibnamefont {Bracker}}, \ and\ \bibinfo
  {author} {\bibfnamefont {D.}~\bibnamefont {Gammon}},\ }\href
  {http://dx.doi.org/10.1038/nphys1863} {\bibfield  {journal} {\bibinfo
  {journal} {Nature Phys.}\ }\textbf {\bibinfo {volume} {7}},\ \bibinfo {pages}
  {223} (\bibinfo {year} {2011})}\BibitemShut {NoStop}%
\bibitem [{\citenamefont {DiVincenzo}\ \emph {et~al.}(2000)\citenamefont
  {DiVincenzo}, \citenamefont {Bacon}, \citenamefont {Kempe}, \citenamefont
  {Burkard},\ and\ \citenamefont {Whaley}}]{divincenzo_universal_2000}%
  \BibitemOpen
  \bibfield  {author} {\bibinfo {author} {\bibfnamefont {D.~P.}\ \bibnamefont
  {DiVincenzo}}, \bibinfo {author} {\bibfnamefont {D.}~\bibnamefont {Bacon}},
  \bibinfo {author} {\bibfnamefont {J.}~\bibnamefont {Kempe}}, \bibinfo
  {author} {\bibfnamefont {G.}~\bibnamefont {Burkard}}, \ and\ \bibinfo
  {author} {\bibfnamefont {K.~B.}\ \bibnamefont {Whaley}},\ }\href {\doibase
  10.1038/35042541} {\bibfield  {journal} {\bibinfo  {journal} {Nature}\
  }\textbf {\bibinfo {volume} {408}},\ \bibinfo {pages} {339} (\bibinfo {year}
  {2000})}\BibitemShut {NoStop}%
\bibitem [{\citenamefont {Laird}\ \emph {et~al.}(2010)\citenamefont {Laird},
  \citenamefont {Taylor}, \citenamefont {DiVincenzo}, \citenamefont {Marcus},
  \citenamefont {Hanson},\ and\ \citenamefont {Gossard}}]{lairdprb}%
  \BibitemOpen
  \bibfield  {author} {\bibinfo {author} {\bibfnamefont {E.~A.}\ \bibnamefont
  {Laird}}, \bibinfo {author} {\bibfnamefont {J.~M.}\ \bibnamefont {Taylor}},
  \bibinfo {author} {\bibfnamefont {D.~P.}\ \bibnamefont {DiVincenzo}},
  \bibinfo {author} {\bibfnamefont {C.~M.}\ \bibnamefont {Marcus}}, \bibinfo
  {author} {\bibfnamefont {M.~P.}\ \bibnamefont {Hanson}}, \ and\ \bibinfo
  {author} {\bibfnamefont {A.~C.}\ \bibnamefont {Gossard}},\ }\href {\doibase
  10.1103/PhysRevB.82.075403} {\bibfield  {journal} {\bibinfo  {journal} {Phys.
  Rev. B}\ }\textbf {\bibinfo {volume} {82}},\ \bibinfo {pages} {075403}
  (\bibinfo {year} {2010})}\BibitemShut {NoStop}%
\bibitem [{\citenamefont {Ladd}(2012)}]{laddprb}%
  \BibitemOpen
  \bibfield  {author} {\bibinfo {author} {\bibfnamefont {T.~D.}\ \bibnamefont
  {Ladd}},\ }\href@noop {} {\bibfield  {journal} {\bibinfo  {journal} {Phys.
  Rev. B}\ }\textbf {\bibinfo {volume} {86}},\ \bibinfo {pages} {125408}
  (\bibinfo {year} {2012})}\BibitemShut {NoStop}%
\bibitem [{\citenamefont {Taylor}\ \emph {et~al.}(2005)\citenamefont {Taylor},
  \citenamefont {Engel}, \citenamefont {Dur}, \citenamefont {Yacoby},
  \citenamefont {Marcus}, \citenamefont {Zoller},\ and\ \citenamefont
  {Lukin}}]{taylornphys}%
  \BibitemOpen
  \bibfield  {author} {\bibinfo {author} {\bibfnamefont {J.~M.}\ \bibnamefont
  {Taylor}}, \bibinfo {author} {\bibfnamefont {H.~A.}\ \bibnamefont {Engel}},
  \bibinfo {author} {\bibfnamefont {W.}~\bibnamefont {Dur}}, \bibinfo {author}
  {\bibfnamefont {A.}~\bibnamefont {Yacoby}}, \bibinfo {author} {\bibfnamefont
  {C.~M.}\ \bibnamefont {Marcus}}, \bibinfo {author} {\bibfnamefont
  {P.}~\bibnamefont {Zoller}}, \ and\ \bibinfo {author} {\bibfnamefont {M.~D.}\
  \bibnamefont {Lukin}},\ }\href@noop {} {\bibfield  {journal} {\bibinfo
  {journal} {Nature Phys.}\ }\textbf {\bibinfo {volume} {1}},\ \bibinfo {pages}
  {177} (\bibinfo {year} {2005})}\BibitemShut {NoStop}%
\bibitem [{\citenamefont {{Sakaki}}\ \emph {et~al.}(1995)\citenamefont
  {{Sakaki}}, \citenamefont {{Yusa}}, \citenamefont {{Someya}}, \citenamefont
  {{Ohno}}, \citenamefont {{Noda}}, \citenamefont {{Akiyama}}, \citenamefont
  {{Kadoya}},\ and\ \citenamefont {{Noge}}}]{Sakaki_APL_1995}%
  \BibitemOpen
  \bibfield  {author} {\bibinfo {author} {\bibfnamefont {H.}~\bibnamefont
  {{Sakaki}}}, \bibinfo {author} {\bibfnamefont {G.}~\bibnamefont {{Yusa}}},
  \bibinfo {author} {\bibfnamefont {T.}~\bibnamefont {{Someya}}}, \bibinfo
  {author} {\bibfnamefont {Y.}~\bibnamefont {{Ohno}}}, \bibinfo {author}
  {\bibfnamefont {T.}~\bibnamefont {{Noda}}}, \bibinfo {author} {\bibfnamefont
  {H.}~\bibnamefont {{Akiyama}}}, \bibinfo {author} {\bibfnamefont
  {Y.}~\bibnamefont {{Kadoya}}}, \ and\ \bibinfo {author} {\bibfnamefont
  {H.}~\bibnamefont {{Noge}}},\ }\href {\doibase 10.1063/1.115274} {\bibfield
  {journal} {\bibinfo  {journal} {Appl. Phys. Lett.}\ }\textbf {\bibinfo
  {volume} {67}},\ \bibinfo {pages} {3444} (\bibinfo {year}
  {1995})}\BibitemShut {NoStop}%
\bibitem [{\citenamefont {{Kim}}\ \emph {et~al.}(1998)\citenamefont {{Kim}},
  \citenamefont {{Ritchie}}, \citenamefont {{Pepper}}, \citenamefont {{Lian}},
  \citenamefont {{Yuan}},\ and\ \citenamefont {{Brown}}}]{Kim_APL_1998}%
  \BibitemOpen
  \bibfield  {author} {\bibinfo {author} {\bibfnamefont {G.~H.}\ \bibnamefont
  {{Kim}}}, \bibinfo {author} {\bibfnamefont {D.~A.}\ \bibnamefont
  {{Ritchie}}}, \bibinfo {author} {\bibfnamefont {M.}~\bibnamefont {{Pepper}}},
  \bibinfo {author} {\bibfnamefont {G.~D.}\ \bibnamefont {{Lian}}}, \bibinfo
  {author} {\bibfnamefont {J.}~\bibnamefont {{Yuan}}}, \ and\ \bibinfo {author}
  {\bibfnamefont {L.~M.}\ \bibnamefont {{Brown}}},\ }\href {\doibase
  10.1063/1.122484} {\bibfield  {journal} {\bibinfo  {journal} {Appl. Phys.
  Lett.}\ }\textbf {\bibinfo {volume} {73}},\ \bibinfo {eid} {2468} (\bibinfo
  {year} {1998})}\BibitemShut {NoStop}%
\bibitem [{\citenamefont {{Wang}}\ \emph {et~al.}(2000)\citenamefont {{Wang}},
  \citenamefont {{Carlsson}}, \citenamefont {{Omling}}, \citenamefont
  {{Samuelson}}, \citenamefont {{Seifert}},\ and\ \citenamefont
  {{Xu}}}]{Wang_APL_2000}%
  \BibitemOpen
  \bibfield  {author} {\bibinfo {author} {\bibfnamefont {Q.}~\bibnamefont
  {{Wang}}}, \bibinfo {author} {\bibfnamefont {N.}~\bibnamefont {{Carlsson}}},
  \bibinfo {author} {\bibfnamefont {P.}~\bibnamefont {{Omling}}}, \bibinfo
  {author} {\bibfnamefont {L.}~\bibnamefont {{Samuelson}}}, \bibinfo {author}
  {\bibfnamefont {W.}~\bibnamefont {{Seifert}}}, \ and\ \bibinfo {author}
  {\bibfnamefont {H.~Q.}\ \bibnamefont {{Xu}}},\ }\href {\doibase
  10.1063/1.126142} {\bibfield  {journal} {\bibinfo  {journal} {Appl. Phys.
  Lett.}\ }\textbf {\bibinfo {volume} {76}},\ \bibinfo {eid} {1704} (\bibinfo
  {year} {2000})}\BibitemShut {NoStop}%
\bibitem [{\citenamefont {{Dettwiler}}\ \emph {et~al.}(2014)\citenamefont
  {{Dettwiler}}, \citenamefont {{Fallahi}}, \citenamefont {{Scholz}},
  \citenamefont {{Reiger}}, \citenamefont {{Schuh}}, \citenamefont
  {{Badolato}}, \citenamefont {{Wegscheider}},\ and\ \citenamefont
  {{Zumb{\"u}hl}}}]{Dettwiler_ArXiv_2014}%
  \BibitemOpen
  \bibfield  {author} {\bibinfo {author} {\bibfnamefont {F.}~\bibnamefont
  {{Dettwiler}}}, \bibinfo {author} {\bibfnamefont {P.}~\bibnamefont
  {{Fallahi}}}, \bibinfo {author} {\bibfnamefont {D.}~\bibnamefont {{Scholz}}},
  \bibinfo {author} {\bibfnamefont {E.}~\bibnamefont {{Reiger}}}, \bibinfo
  {author} {\bibfnamefont {D.}~\bibnamefont {{Schuh}}}, \bibinfo {author}
  {\bibfnamefont {A.}~\bibnamefont {{Badolato}}}, \bibinfo {author}
  {\bibfnamefont {W.}~\bibnamefont {{Wegscheider}}}, \ and\ \bibinfo {author}
  {\bibfnamefont {D.~M.}\ \bibnamefont {{Zumb{\"u}hl}}},\ }\href@noop {}
  {\bibfield  {journal} {\bibinfo  {journal} {arXiv:1403.7775}\ } (\bibinfo
  {year} {2014})}\BibitemShut {NoStop}%
\bibitem [{\citenamefont {{Kurzmann}}\ \emph {et~al.}(2015)\citenamefont
  {{Kurzmann}}, \citenamefont {{Beckel}}, \citenamefont {{Ludwig}},
  \citenamefont {{Wieck}}, \citenamefont {{Lorke}},\ and\ \citenamefont
  {{Geller}}}]{Kurzmann_JAP_2015}%
  \BibitemOpen
  \bibfield  {author} {\bibinfo {author} {\bibfnamefont {A.}~\bibnamefont
  {{Kurzmann}}}, \bibinfo {author} {\bibfnamefont {A.}~\bibnamefont
  {{Beckel}}}, \bibinfo {author} {\bibfnamefont {A.}~\bibnamefont {{Ludwig}}},
  \bibinfo {author} {\bibfnamefont {A.~D.}\ \bibnamefont {{Wieck}}}, \bibinfo
  {author} {\bibfnamefont {A.}~\bibnamefont {{Lorke}}}, \ and\ \bibinfo
  {author} {\bibfnamefont {M.}~\bibnamefont {{Geller}}},\ }\href {\doibase
  10.1063/1.4907217} {\bibfield  {journal} {\bibinfo  {journal} {J. of Appl.
  Phys.}\ }\textbf {\bibinfo {volume} {117}},\ \bibinfo {eid} {054305}
  (\bibinfo {year} {2015})}\BibitemShut {NoStop}%
\bibitem [{\citenamefont {{Marquardt}}\ \emph {et~al.}(2008)\citenamefont
  {{Marquardt}}, \citenamefont {{Russ}}, \citenamefont {{Lorke}}, \citenamefont
  {{Meier}}, \citenamefont {{Reuter}},\ and\ \citenamefont
  {{Wieck}}}]{Marquardt_PhysE_2008}%
  \BibitemOpen
  \bibfield  {author} {\bibinfo {author} {\bibfnamefont {B.}~\bibnamefont
  {{Marquardt}}}, \bibinfo {author} {\bibfnamefont {M.}~\bibnamefont {{Russ}}},
  \bibinfo {author} {\bibfnamefont {A.}~\bibnamefont {{Lorke}}}, \bibinfo
  {author} {\bibfnamefont {C.}~\bibnamefont {{Meier}}}, \bibinfo {author}
  {\bibfnamefont {D.}~\bibnamefont {{Reuter}}}, \ and\ \bibinfo {author}
  {\bibfnamefont {A.~D.}\ \bibnamefont {{Wieck}}},\ }\href {\doibase
  10.1016/j.physe.2007.09.198} {\bibfield  {journal} {\bibinfo  {journal}
  {Physica E}\ }\textbf {\bibinfo {volume} {40}},\ \bibinfo {pages} {2075}
  (\bibinfo {year} {2008})}\BibitemShut {NoStop}%
\bibitem [{\citenamefont {{Marquardt}}\ \emph {et~al.}(2011)\citenamefont
  {{Marquardt}}, \citenamefont {{Beckel}}, \citenamefont {{Lorke}},
  \citenamefont {{Wieck}}, \citenamefont {{Reuter}},\ and\ \citenamefont
  {{Geller}}}]{Marquardt_APL_2011}%
  \BibitemOpen
  \bibfield  {author} {\bibinfo {author} {\bibfnamefont {B.}~\bibnamefont
  {{Marquardt}}}, \bibinfo {author} {\bibfnamefont {A.}~\bibnamefont
  {{Beckel}}}, \bibinfo {author} {\bibfnamefont {A.}~\bibnamefont {{Lorke}}},
  \bibinfo {author} {\bibfnamefont {A.~D.}\ \bibnamefont {{Wieck}}}, \bibinfo
  {author} {\bibfnamefont {D.}~\bibnamefont {{Reuter}}}, \ and\ \bibinfo
  {author} {\bibfnamefont {M.}~\bibnamefont {{Geller}}},\ }\href {\doibase
  10.1063/1.3665070} {\bibfield  {journal} {\bibinfo  {journal} {Appl. Phys.
  Lett.}\ }\textbf {\bibinfo {volume} {99}},\ \bibinfo {eid} {223510} (\bibinfo
  {year} {2011})}\BibitemShut {NoStop}%
\bibitem [{\citenamefont {{Ru{\ss}}}\ \emph {et~al.}(2006)\citenamefont
  {{Ru{\ss}}}, \citenamefont {{Meier}}, \citenamefont {{Lorke}}, \citenamefont
  {{Reuter}},\ and\ \citenamefont {{Wieck}}}]{Russ_PRB_2006}%
  \BibitemOpen
  \bibfield  {author} {\bibinfo {author} {\bibfnamefont {M.}~\bibnamefont
  {{Ru{\ss}}}}, \bibinfo {author} {\bibfnamefont {C.}~\bibnamefont {{Meier}}},
  \bibinfo {author} {\bibfnamefont {A.}~\bibnamefont {{Lorke}}}, \bibinfo
  {author} {\bibfnamefont {D.}~\bibnamefont {{Reuter}}}, \ and\ \bibinfo
  {author} {\bibfnamefont {A.~D.}\ \bibnamefont {{Wieck}}},\ }\href {\doibase
  10.1103/PhysRevB.73.115334} {\bibfield  {journal} {\bibinfo  {journal} {Phys.
  Rev. B}\ }\textbf {\bibinfo {volume} {73}},\ \bibinfo {eid} {115334}
  (\bibinfo {year} {2006})}\BibitemShut {NoStop}%
\bibitem [{\citenamefont {{Zhukov}}\ \emph {et~al.}(2003)\citenamefont
  {{Zhukov}}, \citenamefont {{Weichsel}}, \citenamefont {{Beyer}},
  \citenamefont {{Schn{\"u}ll}}, \citenamefont {{Heyn}},\ and\ \citenamefont
  {{Hansen}}}]{Zhukov_PRB_2003}%
  \BibitemOpen
  \bibfield  {author} {\bibinfo {author} {\bibfnamefont {A.~A.}\ \bibnamefont
  {{Zhukov}}}, \bibinfo {author} {\bibfnamefont {C.}~\bibnamefont
  {{Weichsel}}}, \bibinfo {author} {\bibfnamefont {S.}~\bibnamefont {{Beyer}}},
  \bibinfo {author} {\bibfnamefont {S.}~\bibnamefont {{Schn{\"u}ll}}}, \bibinfo
  {author} {\bibfnamefont {C.}~\bibnamefont {{Heyn}}}, \ and\ \bibinfo {author}
  {\bibfnamefont {W.}~\bibnamefont {{Hansen}}},\ }\href {\doibase
  10.1103/PhysRevB.67.125310} {\bibfield  {journal} {\bibinfo  {journal} {Phys.
  Rev. B}\ }\textbf {\bibinfo {volume} {67}},\ \bibinfo {eid} {125310}
  (\bibinfo {year} {2003})}\BibitemShut {NoStop}%
\bibitem [{\citenamefont {{Kim}}\ \emph {et~al.}(2000)\citenamefont {{Kim}},
  \citenamefont {{Nicholls}}, \citenamefont {{Khondaker}}, \citenamefont
  {{Farrer}},\ and\ \citenamefont {{Ritchie}}}]{Kim_PRB_2000}%
  \BibitemOpen
  \bibfield  {author} {\bibinfo {author} {\bibfnamefont {G.~H.}\ \bibnamefont
  {{Kim}}}, \bibinfo {author} {\bibfnamefont {J.~T.}\ \bibnamefont
  {{Nicholls}}}, \bibinfo {author} {\bibfnamefont {S.~I.}\ \bibnamefont
  {{Khondaker}}}, \bibinfo {author} {\bibfnamefont {I.}~\bibnamefont
  {{Farrer}}}, \ and\ \bibinfo {author} {\bibfnamefont {D.~A.}\ \bibnamefont
  {{Ritchie}}},\ }\href {\doibase 10.1103/PhysRevB.61.10910} {\bibfield
  {journal} {\bibinfo  {journal} {Phys. Rev. B}\ }\textbf {\bibinfo {volume}
  {61}},\ \bibinfo {pages} {10910} (\bibinfo {year} {2000})}\BibitemShut
  {NoStop}%
\bibitem [{\citenamefont {{Kim}}\ \emph {et~al.}(2004)\citenamefont {{Kim}},
  \citenamefont {{Liang}}, \citenamefont {{Huang}}, \citenamefont {{Nicholls}},
  \citenamefont {{Ritchie}}, \citenamefont {{Kim}}, \citenamefont {{Oh}},
  \citenamefont {{Juang}},\ and\ \citenamefont {{Chang}}}]{Kim_PRB_2004}%
  \BibitemOpen
  \bibfield  {author} {\bibinfo {author} {\bibfnamefont {G.-H.}\ \bibnamefont
  {{Kim}}}, \bibinfo {author} {\bibfnamefont {C.-T.}\ \bibnamefont {{Liang}}},
  \bibinfo {author} {\bibfnamefont {C.~F.}\ \bibnamefont {{Huang}}}, \bibinfo
  {author} {\bibfnamefont {J.~T.}\ \bibnamefont {{Nicholls}}}, \bibinfo
  {author} {\bibfnamefont {D.~A.}\ \bibnamefont {{Ritchie}}}, \bibinfo {author}
  {\bibfnamefont {P.~S.}\ \bibnamefont {{Kim}}}, \bibinfo {author}
  {\bibfnamefont {C.~H.}\ \bibnamefont {{Oh}}}, \bibinfo {author}
  {\bibfnamefont {J.~R.}\ \bibnamefont {{Juang}}}, \ and\ \bibinfo {author}
  {\bibfnamefont {Y.~H.}\ \bibnamefont {{Chang}}},\ }\href {\doibase
  10.1103/PhysRevB.69.073311} {\bibfield  {journal} {\bibinfo  {journal} {Phys.
  Rev. B}\ }\textbf {\bibinfo {volume} {69}},\ \bibinfo {eid} {073311}
  (\bibinfo {year} {2004})}\BibitemShut {NoStop}%
\bibitem [{\citenamefont {{Ribeiro}}\ \emph {et~al.}(1998)\citenamefont
  {{Ribeiro}}, \citenamefont {{M{\"u}ller}}, \citenamefont {{Heinzel}},
  \citenamefont {{Auderset}}, \citenamefont {{Ensslin}}, \citenamefont
  {{Medeiros-Ribeiro}},\ and\ \citenamefont {{Petroff}}}]{Ribeiro_PRB_1998}%
  \BibitemOpen
  \bibfield  {author} {\bibinfo {author} {\bibfnamefont {E.}~\bibnamefont
  {{Ribeiro}}}, \bibinfo {author} {\bibfnamefont {E.}~\bibnamefont
  {{M{\"u}ller}}}, \bibinfo {author} {\bibfnamefont {T.}~\bibnamefont
  {{Heinzel}}}, \bibinfo {author} {\bibfnamefont {H.}~\bibnamefont
  {{Auderset}}}, \bibinfo {author} {\bibfnamefont {K.}~\bibnamefont
  {{Ensslin}}}, \bibinfo {author} {\bibfnamefont {G.}~\bibnamefont
  {{Medeiros-Ribeiro}}}, \ and\ \bibinfo {author} {\bibfnamefont {P.~M.}\
  \bibnamefont {{Petroff}}},\ }\href {\doibase 10.1103/PhysRevB.58.1506}
  {\bibfield  {journal} {\bibinfo  {journal} {Phys. Rev. B}\ }\textbf {\bibinfo
  {volume} {58}},\ \bibinfo {pages} {1506} (\bibinfo {year}
  {1998})}\BibitemShut {NoStop}%
\bibitem [{\citenamefont {{Ribeiro}}\ \emph {et~al.}(1999)\citenamefont
  {{Ribeiro}}, \citenamefont {{J{\"a}ggi}}, \citenamefont {{Heinzel}},
  \citenamefont {{Ensslin}}, \citenamefont {{Medeiros-Ribeiro}},\ and\
  \citenamefont {{Petroff}}}]{Ribeiro_PRL_1999}%
  \BibitemOpen
  \bibfield  {author} {\bibinfo {author} {\bibfnamefont {E.}~\bibnamefont
  {{Ribeiro}}}, \bibinfo {author} {\bibfnamefont {R.~D.}\ \bibnamefont
  {{J{\"a}ggi}}}, \bibinfo {author} {\bibfnamefont {T.}~\bibnamefont
  {{Heinzel}}}, \bibinfo {author} {\bibfnamefont {K.}~\bibnamefont
  {{Ensslin}}}, \bibinfo {author} {\bibfnamefont {G.}~\bibnamefont
  {{Medeiros-Ribeiro}}}, \ and\ \bibinfo {author} {\bibfnamefont {P.~M.}\
  \bibnamefont {{Petroff}}},\ }\href {\doibase 10.1103/PhysRevLett.82.996}
  {\bibfield  {journal} {\bibinfo  {journal} {Phys. Rev. Lett.}\ }\textbf
  {\bibinfo {volume} {82}},\ \bibinfo {pages} {996} (\bibinfo {year}
  {1999})}\BibitemShut {NoStop}%
\bibitem [{\citenamefont {{Ribeiro}}\ \emph {et~al.}(2000)\citenamefont
  {{Ribeiro}}, \citenamefont {{Cerdeira}}, \citenamefont {{Brasil}},
  \citenamefont {{Heinzel}}, \citenamefont {{Ensslin}}, \citenamefont
  {{Medeiros-Ribeiro}},\ and\ \citenamefont {{Petroff}}}]{Ribeiro_JAP_2000}%
  \BibitemOpen
  \bibfield  {author} {\bibinfo {author} {\bibfnamefont {E.}~\bibnamefont
  {{Ribeiro}}}, \bibinfo {author} {\bibfnamefont {F.}~\bibnamefont
  {{Cerdeira}}}, \bibinfo {author} {\bibfnamefont {M.~J.~S.~P.}\ \bibnamefont
  {{Brasil}}}, \bibinfo {author} {\bibfnamefont {T.}~\bibnamefont {{Heinzel}}},
  \bibinfo {author} {\bibfnamefont {K.}~\bibnamefont {{Ensslin}}}, \bibinfo
  {author} {\bibfnamefont {G.}~\bibnamefont {{Medeiros-Ribeiro}}}, \ and\
  \bibinfo {author} {\bibfnamefont {P.~M.}\ \bibnamefont {{Petroff}}},\ }\href
  {\doibase 10.1063/1.373485} {\bibfield  {journal} {\bibinfo  {journal} {J.
  Appl. Phys.}\ }\textbf {\bibinfo {volume} {87}},\ \bibinfo {pages} {7994}
  (\bibinfo {year} {2000})}\BibitemShut {NoStop}%
\bibitem [{\citenamefont {{Kim}}\ \emph {et~al.}(2005)\citenamefont {{Kim}},
  \citenamefont {{Kim}}, \citenamefont {{Lee}}, \citenamefont {{Lee}},\ and\
  \citenamefont {{Kim}}}]{Kim_APL_2005}%
  \BibitemOpen
  \bibfield  {author} {\bibinfo {author} {\bibfnamefont {T.~W.}\ \bibnamefont
  {{Kim}}}, \bibinfo {author} {\bibfnamefont {J.~H.}\ \bibnamefont {{Kim}}},
  \bibinfo {author} {\bibfnamefont {H.~S.}\ \bibnamefont {{Lee}}}, \bibinfo
  {author} {\bibfnamefont {J.~Y.}\ \bibnamefont {{Lee}}}, \ and\ \bibinfo
  {author} {\bibfnamefont {M.~D.}\ \bibnamefont {{Kim}}},\ }\href {\doibase
  10.1063/1.1849853} {\bibfield  {journal} {\bibinfo  {journal} {Appl. Phys.
  Lett.}\ }\textbf {\bibinfo {volume} {86}},\ \bibinfo {eid} {021916} (\bibinfo
  {year} {2005})}\BibitemShut {NoStop}%
\bibitem [{\citenamefont {{Finley}}\ \emph {et~al.}(1998)\citenamefont
  {{Finley}}, \citenamefont {{Skalitz}}, \citenamefont {{Arzberger}},
  \citenamefont {{Zrenner}}, \citenamefont {{B{\"o}hm}},\ and\ \citenamefont
  {{Abstreiter}}}]{Finley_APL_1998}%
  \BibitemOpen
  \bibfield  {author} {\bibinfo {author} {\bibfnamefont {J.~J.}\ \bibnamefont
  {{Finley}}}, \bibinfo {author} {\bibfnamefont {M.}~\bibnamefont {{Skalitz}}},
  \bibinfo {author} {\bibfnamefont {M.}~\bibnamefont {{Arzberger}}}, \bibinfo
  {author} {\bibfnamefont {A.}~\bibnamefont {{Zrenner}}}, \bibinfo {author}
  {\bibfnamefont {G.}~\bibnamefont {{B{\"o}hm}}}, \ and\ \bibinfo {author}
  {\bibfnamefont {G.}~\bibnamefont {{Abstreiter}}},\ }\href {\doibase
  10.1063/1.122524} {\bibfield  {journal} {\bibinfo  {journal} {Appl. Phys.
  Lett.}\ }\textbf {\bibinfo {volume} {73}},\ \bibinfo {eid} {2618} (\bibinfo
  {year} {1998})}\BibitemShut {NoStop}%
\bibitem [{\citenamefont {{Yusa}}\ and\ \citenamefont
  {{Sakaki}}(1997)}]{Yusa_APL_1997}%
  \BibitemOpen
  \bibfield  {author} {\bibinfo {author} {\bibfnamefont {G.}~\bibnamefont
  {{Yusa}}}\ and\ \bibinfo {author} {\bibfnamefont {H.}~\bibnamefont
  {{Sakaki}}},\ }\href {\doibase 10.1063/1.119068} {\bibfield  {journal}
  {\bibinfo  {journal} {Appl. Phys. Lett.}\ }\textbf {\bibinfo {volume} {70}},\
  \bibinfo {pages} {345} (\bibinfo {year} {1997})}\BibitemShut {NoStop}%
\bibitem [{\citenamefont {{M{\"u}ller}}\ \emph {et~al.}(2008)\citenamefont
  {{M{\"u}ller}}, \citenamefont {{Worschech}}, \citenamefont {{Heinrich}},
  \citenamefont {{H{\"o}fling}},\ and\ \citenamefont
  {{Forchel}}}]{Muller_APL_2008}%
  \BibitemOpen
  \bibfield  {author} {\bibinfo {author} {\bibfnamefont {C.~R.}\ \bibnamefont
  {{M{\"u}ller}}}, \bibinfo {author} {\bibfnamefont {L.}~\bibnamefont
  {{Worschech}}}, \bibinfo {author} {\bibfnamefont {J.}~\bibnamefont
  {{Heinrich}}}, \bibinfo {author} {\bibfnamefont {S.}~\bibnamefont
  {{H{\"o}fling}}}, \ and\ \bibinfo {author} {\bibfnamefont {A.}~\bibnamefont
  {{Forchel}}},\ }\href {\doibase 10.1063/1.2967880} {\bibfield  {journal}
  {\bibinfo  {journal} {Appl. Phys. Lett.}\ }\textbf {\bibinfo {volume} {93}},\
  \bibinfo {eid} {063502} (\bibinfo {year} {2008})}\BibitemShut {NoStop}%
\bibitem [{\citenamefont {{Marquardt}}\ \emph {et~al.}(2009)\citenamefont
  {{Marquardt}}, \citenamefont {{Geller}}, \citenamefont {{Lorke}},
  \citenamefont {{Reuter}},\ and\ \citenamefont
  {{Wieck}}}]{Marquardt_APL_2009}%
  \BibitemOpen
  \bibfield  {author} {\bibinfo {author} {\bibfnamefont {B.}~\bibnamefont
  {{Marquardt}}}, \bibinfo {author} {\bibfnamefont {M.}~\bibnamefont
  {{Geller}}}, \bibinfo {author} {\bibfnamefont {A.}~\bibnamefont {{Lorke}}},
  \bibinfo {author} {\bibfnamefont {D.}~\bibnamefont {{Reuter}}}, \ and\
  \bibinfo {author} {\bibfnamefont {A.~D.}\ \bibnamefont {{Wieck}}},\ }\href
  {\doibase 10.1063/1.3175724} {\bibfield  {journal} {\bibinfo  {journal}
  {Appl. Phys. Lett.}\ }\textbf {\bibinfo {volume} {95}},\ \bibinfo {eid}
  {022113} (\bibinfo {year} {2009})}\BibitemShut {NoStop}%
\bibitem [{\citenamefont {Marquardt}\ \emph {et~al.}(2011)\citenamefont
  {Marquardt}, \citenamefont {Geller}, \citenamefont {Baxevanis}, \citenamefont
  {Pfannkuche}, \citenamefont {Wieck}, \citenamefont {Reuter},\ and\
  \citenamefont {Lorke}}]{Marquardt_NComm_2011}%
  \BibitemOpen
  \bibfield  {author} {\bibinfo {author} {\bibfnamefont {B.}~\bibnamefont
  {Marquardt}}, \bibinfo {author} {\bibfnamefont {M.}~\bibnamefont {Geller}},
  \bibinfo {author} {\bibfnamefont {B.}~\bibnamefont {Baxevanis}}, \bibinfo
  {author} {\bibfnamefont {D.}~\bibnamefont {Pfannkuche}}, \bibinfo {author}
  {\bibfnamefont {A.}~\bibnamefont {Wieck}}, \bibinfo {author} {\bibfnamefont
  {D.}~\bibnamefont {Reuter}}, \ and\ \bibinfo {author} {\bibfnamefont
  {A.}~\bibnamefont {Lorke}},\ }\href@noop {} {\bibfield  {journal} {\bibinfo
  {journal} {Nat. Commun.}\ }\textbf {\bibinfo {volume} {2}},\ \bibinfo {pages}
  {209} (\bibinfo {year} {2011})}\BibitemShut {NoStop}%
\bibitem [{\citenamefont {{Nowozin}}\ \emph {et~al.}(2014)\citenamefont
  {{Nowozin}}, \citenamefont {{Beckel}}, \citenamefont {{Bimberg}},
  \citenamefont {{Lorke}},\ and\ \citenamefont {{Geller}}}]{Nowozin_APL_2014}%
  \BibitemOpen
  \bibfield  {author} {\bibinfo {author} {\bibfnamefont {T.}~\bibnamefont
  {{Nowozin}}}, \bibinfo {author} {\bibfnamefont {A.}~\bibnamefont {{Beckel}}},
  \bibinfo {author} {\bibfnamefont {D.}~\bibnamefont {{Bimberg}}}, \bibinfo
  {author} {\bibfnamefont {A.}~\bibnamefont {{Lorke}}}, \ and\ \bibinfo
  {author} {\bibfnamefont {M.}~\bibnamefont {{Geller}}},\ }\href {\doibase
  10.1063/1.4864281} {\bibfield  {journal} {\bibinfo  {journal} {Appl. Phys.
  Lett.}\ }\textbf {\bibinfo {volume} {104}},\ \bibinfo {eid} {053111}
  (\bibinfo {year} {2014})}\BibitemShut {NoStop}%
\bibitem [{\citenamefont {Shields}\ \emph {et~al.}(2000)\citenamefont
  {Shields}, \citenamefont {O’Sullivan}, \citenamefont {Farrer},
  \citenamefont {Ritchie}, \citenamefont {Hogg}, \citenamefont {Leadbeater},
  \citenamefont {Norman},\ and\ \citenamefont
  {Pepper}}]{shields_detection_2000}%
  \BibitemOpen
  \bibfield  {author} {\bibinfo {author} {\bibfnamefont {A.~J.}\ \bibnamefont
  {Shields}}, \bibinfo {author} {\bibfnamefont {M.~P.}\ \bibnamefont
  {O’Sullivan}}, \bibinfo {author} {\bibfnamefont {I.}~\bibnamefont
  {Farrer}}, \bibinfo {author} {\bibfnamefont {D.~A.}\ \bibnamefont {Ritchie}},
  \bibinfo {author} {\bibfnamefont {R.~A.}\ \bibnamefont {Hogg}}, \bibinfo
  {author} {\bibfnamefont {M.~L.}\ \bibnamefont {Leadbeater}}, \bibinfo
  {author} {\bibfnamefont {C.~E.}\ \bibnamefont {Norman}}, \ and\ \bibinfo
  {author} {\bibfnamefont {M.}~\bibnamefont {Pepper}},\ }\href {\doibase
  10.1063/1.126745} {\bibfield  {journal} {\bibinfo  {journal} {Appl. Phys.
  Lett.}\ }\textbf {\bibinfo {volume} {76}},\ \bibinfo {pages} {3673} (\bibinfo
  {year} {2000})}\BibitemShut {NoStop}%
\bibitem [{\citenamefont {Rowe}\ \emph {et~al.}(2006)\citenamefont {Rowe},
  \citenamefont {Gansen}, \citenamefont {Greene}, \citenamefont {Hadfield},
  \citenamefont {Harvey}, \citenamefont {Su}, \citenamefont {Nam},
  \citenamefont {Mirin},\ and\ \citenamefont
  {Rosenberg}}]{rowe_single-photon_2006}%
  \BibitemOpen
  \bibfield  {author} {\bibinfo {author} {\bibfnamefont {M.~A.}\ \bibnamefont
  {Rowe}}, \bibinfo {author} {\bibfnamefont {E.~J.}\ \bibnamefont {Gansen}},
  \bibinfo {author} {\bibfnamefont {M.}~\bibnamefont {Greene}}, \bibinfo
  {author} {\bibfnamefont {R.~H.}\ \bibnamefont {Hadfield}}, \bibinfo {author}
  {\bibfnamefont {T.~E.}\ \bibnamefont {Harvey}}, \bibinfo {author}
  {\bibfnamefont {M.~Y.}\ \bibnamefont {Su}}, \bibinfo {author} {\bibfnamefont
  {S.~W.}\ \bibnamefont {Nam}}, \bibinfo {author} {\bibfnamefont {R.~P.}\
  \bibnamefont {Mirin}}, \ and\ \bibinfo {author} {\bibfnamefont
  {D.}~\bibnamefont {Rosenberg}},\ }\href {\doibase 10.1063/1.2403907}
  {\bibfield  {journal} {\bibinfo  {journal} {Appl. Phys. Lett.}\ }\textbf
  {\bibinfo {volume} {89}},\ \bibinfo {pages} {253505} (\bibinfo {year}
  {2006})}\BibitemShut {NoStop}%
\bibitem [{\citenamefont {Puri}\ \emph {et~al.}(2014)\citenamefont {Puri},
  \citenamefont {McMahon},\ and\ \citenamefont {Yamamoto}}]{Puri_PRB_2014}%
  \BibitemOpen
  \bibfield  {author} {\bibinfo {author} {\bibfnamefont {S.}~\bibnamefont
  {Puri}}, \bibinfo {author} {\bibfnamefont {P.~L.}\ \bibnamefont {McMahon}}, \
  and\ \bibinfo {author} {\bibfnamefont {Y.}~\bibnamefont {Yamamoto}},\ }\href
  {\doibase 10.1103/PhysRevB.90.155421} {\bibfield  {journal} {\bibinfo
  {journal} {Phys. Rev. B}\ }\textbf {\bibinfo {volume} {90}},\ \bibinfo
  {pages} {155421} (\bibinfo {year} {2014})}\BibitemShut {NoStop}%
\bibitem [{\citenamefont {Zhou}\ \emph {et~al.}(2000)\citenamefont {Zhou},
  \citenamefont {Leung},\ and\ \citenamefont {Chuang}}]{zhou00}%
  \BibitemOpen
  \bibfield  {author} {\bibinfo {author} {\bibfnamefont {X.}~\bibnamefont
  {Zhou}}, \bibinfo {author} {\bibfnamefont {D.}~\bibnamefont {Leung}}, \ and\
  \bibinfo {author} {\bibfnamefont {I.~L.}\ \bibnamefont {Chuang}},\
  }\href@noop {} {\bibfield  {journal} {\bibinfo  {journal} {Phys. Rev. A}\
  }\textbf {\bibinfo {volume} {62}},\ \bibinfo {pages} {052316} (\bibinfo
  {year} {2000})}\BibitemShut {NoStop}%
\bibitem [{\citenamefont {Delteil}\ \emph {et~al.}()\citenamefont {Delteil},
  \citenamefont {Sun}, \citenamefont {bo~Gao}, \citenamefont {Togan},
  \citenamefont {Faelt},\ and\ \citenamefont
  {Imamoglu}}]{imamoglu_entanglement}%
  \BibitemOpen
  \bibfield  {author} {\bibinfo {author} {\bibfnamefont {A.}~\bibnamefont
  {Delteil}}, \bibinfo {author} {\bibfnamefont {Z.}~\bibnamefont {Sun}},
  \bibinfo {author} {\bibfnamefont {W.}~\bibnamefont {bo~Gao}}, \bibinfo
  {author} {\bibfnamefont {E.}~\bibnamefont {Togan}}, \bibinfo {author}
  {\bibfnamefont {S.}~\bibnamefont {Faelt}}, \ and\ \bibinfo {author}
  {\bibfnamefont {A.}~\bibnamefont {Imamoglu}},\ }\href@noop {} {\ }\bibinfo
  {note} {ArXiv:1507.00465}\BibitemShut {NoStop}%
\bibitem [{\citenamefont {Sun}\ \emph {et~al.}()\citenamefont {Sun},
  \citenamefont {Kim}, \citenamefont {Solomon},\ and\ \citenamefont
  {Waks}}]{waks_strong}%
  \BibitemOpen
  \bibfield  {author} {\bibinfo {author} {\bibfnamefont {S.}~\bibnamefont
  {Sun}}, \bibinfo {author} {\bibfnamefont {H.}~\bibnamefont {Kim}}, \bibinfo
  {author} {\bibfnamefont {G.~S.}\ \bibnamefont {Solomon}}, \ and\ \bibinfo
  {author} {\bibfnamefont {E.}~\bibnamefont {Waks}},\ }\href@noop {} {\
  }\bibinfo {note} {ArXiv:1506.06036}\BibitemShut {NoStop}%
\bibitem [{\citenamefont {Xu}\ \emph {et~al.}(2009)\citenamefont {Xu},
  \citenamefont {Yao}, \citenamefont {Sun}, \citenamefont {Steel},
  \citenamefont {Bracker}, \citenamefont {Gammon},\ and\ \citenamefont
  {Sham}}]{xunat}%
  \BibitemOpen
  \bibfield  {author} {\bibinfo {author} {\bibfnamefont {X.}~\bibnamefont
  {Xu}}, \bibinfo {author} {\bibfnamefont {W.}~\bibnamefont {Yao}}, \bibinfo
  {author} {\bibfnamefont {B.}~\bibnamefont {Sun}}, \bibinfo {author}
  {\bibfnamefont {D.~G.}\ \bibnamefont {Steel}}, \bibinfo {author}
  {\bibfnamefont {A.~S.}\ \bibnamefont {Bracker}}, \bibinfo {author}
  {\bibfnamefont {D.}~\bibnamefont {Gammon}}, \ and\ \bibinfo {author}
  {\bibfnamefont {L.~J.}\ \bibnamefont {Sham}},\ }\href
  {http://dx.doi.org/10.1038/nature08120} {\bibfield  {journal} {\bibinfo
  {journal} {Nature}\ }\textbf {\bibinfo {volume} {459}},\ \bibinfo {pages}
  {1105} (\bibinfo {year} {2009})}\BibitemShut {NoStop}%
\bibitem [{\citenamefont {Latta}\ \emph {et~al.}(2009)\citenamefont {Latta},
  \citenamefont {Hogele}, \citenamefont {Zhao}, \citenamefont {Vamivakas},
  \citenamefont {Maletinsky}, \citenamefont {Kroner}, \citenamefont {Dreiser},
  \citenamefont {Carusotto}, \citenamefont {Badolato}, \citenamefont {Schuh},
  \citenamefont {Wegscheider}, \citenamefont {Atature},\ and\ \citenamefont
  {Imamoglu}}]{latta09}%
  \BibitemOpen
  \bibfield  {author} {\bibinfo {author} {\bibfnamefont {C.}~\bibnamefont
  {Latta}}, \bibinfo {author} {\bibfnamefont {A.}~\bibnamefont {Hogele}},
  \bibinfo {author} {\bibfnamefont {Y.}~\bibnamefont {Zhao}}, \bibinfo {author}
  {\bibfnamefont {A.~N.}\ \bibnamefont {Vamivakas}}, \bibinfo {author}
  {\bibfnamefont {P.}~\bibnamefont {Maletinsky}}, \bibinfo {author}
  {\bibfnamefont {M.}~\bibnamefont {Kroner}}, \bibinfo {author} {\bibfnamefont
  {J.}~\bibnamefont {Dreiser}}, \bibinfo {author} {\bibfnamefont
  {I.}~\bibnamefont {Carusotto}}, \bibinfo {author} {\bibfnamefont
  {A.}~\bibnamefont {Badolato}}, \bibinfo {author} {\bibfnamefont
  {D.}~\bibnamefont {Schuh}}, \bibinfo {author} {\bibfnamefont
  {W.}~\bibnamefont {Wegscheider}}, \bibinfo {author} {\bibfnamefont
  {M.}~\bibnamefont {Atature}}, \ and\ \bibinfo {author} {\bibfnamefont
  {A.}~\bibnamefont {Imamoglu}},\ }\href@noop {} {\bibfield  {journal}
  {\bibinfo  {journal} {Nat. Phys.}\ }\textbf {\bibinfo {volume} {5}},\
  \bibinfo {pages} {758} (\bibinfo {year} {2009})}\BibitemShut {NoStop}%
\bibitem [{\citenamefont {Ladd}\ \emph {et~al.}(2010)\citenamefont {Ladd},
  \citenamefont {Press}, \citenamefont {De~Greve}, \citenamefont {McMahon},
  \citenamefont {Friess}, \citenamefont {Schneider}, \citenamefont {Kamp},
  \citenamefont {H\"ofling}, \citenamefont {Forchel},\ and\ \citenamefont
  {Yamamoto}}]{laddprl}%
  \BibitemOpen
  \bibfield  {author} {\bibinfo {author} {\bibfnamefont {T.~D.}\ \bibnamefont
  {Ladd}}, \bibinfo {author} {\bibfnamefont {D.}~\bibnamefont {Press}},
  \bibinfo {author} {\bibfnamefont {K.}~\bibnamefont {De~Greve}}, \bibinfo
  {author} {\bibfnamefont {P.~L.}\ \bibnamefont {McMahon}}, \bibinfo {author}
  {\bibfnamefont {B.}~\bibnamefont {Friess}}, \bibinfo {author} {\bibfnamefont
  {C.}~\bibnamefont {Schneider}}, \bibinfo {author} {\bibfnamefont
  {M.}~\bibnamefont {Kamp}}, \bibinfo {author} {\bibfnamefont {S.}~\bibnamefont
  {H\"ofling}}, \bibinfo {author} {\bibfnamefont {A.}~\bibnamefont {Forchel}},
  \ and\ \bibinfo {author} {\bibfnamefont {Y.}~\bibnamefont {Yamamoto}},\
  }\href {\doibase 10.1103/PhysRevLett.105.107401} {\bibfield  {journal}
  {\bibinfo  {journal} {Phys. Rev. Lett.}\ }\textbf {\bibinfo {volume} {105}},\
  \bibinfo {pages} {107401} (\bibinfo {year} {2010})}\BibitemShut {NoStop}%
\bibitem [{\citenamefont {Stinaff}\ \emph {et~al.}(2006)\citenamefont
  {Stinaff}, \citenamefont {Scheibner}, \citenamefont {Bracker}, \citenamefont
  {Ponomarev}, \citenamefont {Korenev}, \citenamefont {Ware}, \citenamefont
  {Doty}, \citenamefont {Reinecke},\ and\ \citenamefont {Gammon}}]{Stinaff}%
  \BibitemOpen
  \bibfield  {author} {\bibinfo {author} {\bibfnamefont {E.~A.}\ \bibnamefont
  {Stinaff}}, \bibinfo {author} {\bibfnamefont {M.}~\bibnamefont {Scheibner}},
  \bibinfo {author} {\bibfnamefont {A.~S.}\ \bibnamefont {Bracker}}, \bibinfo
  {author} {\bibfnamefont {I.~V.}\ \bibnamefont {Ponomarev}}, \bibinfo {author}
  {\bibfnamefont {V.~L.}\ \bibnamefont {Korenev}}, \bibinfo {author}
  {\bibfnamefont {M.~E.}\ \bibnamefont {Ware}}, \bibinfo {author}
  {\bibfnamefont {M.~F.}\ \bibnamefont {Doty}}, \bibinfo {author}
  {\bibfnamefont {T.~L.}\ \bibnamefont {Reinecke}}, \ and\ \bibinfo {author}
  {\bibfnamefont {D.}~\bibnamefont {Gammon}},\ }\href {\doibase
  10.1126/science.1121189} {\bibfield  {journal} {\bibinfo  {journal}
  {Science}\ }\textbf {\bibinfo {volume} {311}},\ \bibinfo {pages} {636}
  (\bibinfo {year} {2006})}\BibitemShut {NoStop}%
\bibitem [{\citenamefont {Scheibner}\ \emph {et~al.}(2007)\citenamefont
  {Scheibner}, \citenamefont {Doty}, \citenamefont {Ponomarev}, \citenamefont
  {Bracker}, \citenamefont {Stinaff}, \citenamefont {Korenev}, \citenamefont
  {Reinecke},\ and\ \citenamefont {Gammon}}]{scheibner}%
  \BibitemOpen
  \bibfield  {author} {\bibinfo {author} {\bibfnamefont {M.}~\bibnamefont
  {Scheibner}}, \bibinfo {author} {\bibfnamefont {M.~F.}\ \bibnamefont {Doty}},
  \bibinfo {author} {\bibfnamefont {I.~V.}\ \bibnamefont {Ponomarev}}, \bibinfo
  {author} {\bibfnamefont {A.~S.}\ \bibnamefont {Bracker}}, \bibinfo {author}
  {\bibfnamefont {E.~A.}\ \bibnamefont {Stinaff}}, \bibinfo {author}
  {\bibfnamefont {V.~L.}\ \bibnamefont {Korenev}}, \bibinfo {author}
  {\bibfnamefont {T.~L.}\ \bibnamefont {Reinecke}}, \ and\ \bibinfo {author}
  {\bibfnamefont {D.}~\bibnamefont {Gammon}},\ }\href {\doibase
  10.1103/PhysRevB.75.245318} {\bibfield  {journal} {\bibinfo  {journal} {Phys.
  Rev. B}\ }\textbf {\bibinfo {volume} {75}},\ \bibinfo {pages} {245318}
  (\bibinfo {year} {2007})}\BibitemShut {NoStop}%
\bibitem [{\citenamefont {T\"ureci}\ \emph {et~al.}(2007)\citenamefont
  {T\"ureci}, \citenamefont {Taylor},\ and\ \citenamefont
  {Imamoglu}}]{tureciprb}%
  \BibitemOpen
  \bibfield  {author} {\bibinfo {author} {\bibfnamefont {H.~E.}\ \bibnamefont
  {T\"ureci}}, \bibinfo {author} {\bibfnamefont {J.~M.}\ \bibnamefont
  {Taylor}}, \ and\ \bibinfo {author} {\bibfnamefont {A.}~\bibnamefont
  {Imamoglu}},\ }\href {\doibase 10.1103/PhysRevB.75.235313} {\bibfield
  {journal} {\bibinfo  {journal} {Phys. Rev. B}\ }\textbf {\bibinfo {volume}
  {75}},\ \bibinfo {pages} {235313} (\bibinfo {year} {2007})}\BibitemShut
  {NoStop}%
\bibitem [{\citenamefont {Economou}\ \emph {et~al.}(2012)\citenamefont
  {Economou}, \citenamefont {Climente}, \citenamefont {Badolato}, \citenamefont
  {Bracker}, \citenamefont {Gammon},\ and\ \citenamefont {Doty}}]{economouprb}%
  \BibitemOpen
  \bibfield  {author} {\bibinfo {author} {\bibfnamefont {S.~E.}\ \bibnamefont
  {Economou}}, \bibinfo {author} {\bibfnamefont {J.~I.}\ \bibnamefont
  {Climente}}, \bibinfo {author} {\bibfnamefont {A.}~\bibnamefont {Badolato}},
  \bibinfo {author} {\bibfnamefont {A.~S.}\ \bibnamefont {Bracker}}, \bibinfo
  {author} {\bibfnamefont {D.}~\bibnamefont {Gammon}}, \ and\ \bibinfo {author}
  {\bibfnamefont {M.~F.}\ \bibnamefont {Doty}},\ }\href {\doibase
  10.1103/PhysRevB.86.085319} {\bibfield  {journal} {\bibinfo  {journal} {Phys.
  Rev. B}\ }\textbf {\bibinfo {volume} {86}},\ \bibinfo {pages} {085319}
  (\bibinfo {year} {2012})}\BibitemShut {NoStop}%
\bibitem [{\citenamefont {{Cohen}}(2015)}]{cohenarxiv}%
  \BibitemOpen
  \bibfield  {author} {\bibinfo {author} {\bibfnamefont {G.~Z.}\ \bibnamefont
  {{Cohen}}},\ }\href@noop {} {\bibfield  {journal} {\bibinfo  {journal}
  {arXiv:1501.01952}\ } (\bibinfo {year} {2015})}\BibitemShut {NoStop}%
\bibitem [{\citenamefont {Kim}\ \emph {et~al.}(2008)\citenamefont {Kim},
  \citenamefont {Economou}, \citenamefont {Badescu}, \citenamefont {Scheibner},
  \citenamefont {Bracker}, \citenamefont {Bashkansky}, \citenamefont
  {Reinecke},\ and\ \citenamefont {Gammon}}]{dkimprl}%
  \BibitemOpen
  \bibfield  {author} {\bibinfo {author} {\bibfnamefont {D.}~\bibnamefont
  {Kim}}, \bibinfo {author} {\bibfnamefont {S.~E.}\ \bibnamefont {Economou}},
  \bibinfo {author} {\bibfnamefont {S.}~\bibnamefont {Badescu}}, \bibinfo
  {author} {\bibfnamefont {M.}~\bibnamefont {Scheibner}}, \bibinfo {author}
  {\bibfnamefont {A.~S.}\ \bibnamefont {Bracker}}, \bibinfo {author}
  {\bibfnamefont {M.}~\bibnamefont {Bashkansky}}, \bibinfo {author}
  {\bibfnamefont {T.~L.}\ \bibnamefont {Reinecke}}, \ and\ \bibinfo {author}
  {\bibfnamefont {D.}~\bibnamefont {Gammon}},\ }\href {\doibase
  10.1103/PhysRevLett.101.236804} {\bibfield  {journal} {\bibinfo  {journal}
  {Phys. Rev. Lett.}\ }\textbf {\bibinfo {volume} {101}},\ \bibinfo {pages}
  {236804} (\bibinfo {year} {2008})}\BibitemShut {NoStop}%
\bibitem [{\citenamefont {Vamivakas}\ \emph {et~al.}(2010)\citenamefont
  {Vamivakas}, \citenamefont {Lu}, \citenamefont {Matthiesen}, \citenamefont
  {Zhao}, \citenamefont {Falt}, \citenamefont {Badolato},\ and\ \citenamefont
  {Atature}}]{vamivakasnat}%
  \BibitemOpen
  \bibfield  {author} {\bibinfo {author} {\bibfnamefont {A.~N.}\ \bibnamefont
  {Vamivakas}}, \bibinfo {author} {\bibfnamefont {C.~Y.}\ \bibnamefont {Lu}},
  \bibinfo {author} {\bibfnamefont {C.}~\bibnamefont {Matthiesen}}, \bibinfo
  {author} {\bibfnamefont {Y.}~\bibnamefont {Zhao}}, \bibinfo {author}
  {\bibfnamefont {S.}~\bibnamefont {Falt}}, \bibinfo {author} {\bibfnamefont
  {A.}~\bibnamefont {Badolato}}, \ and\ \bibinfo {author} {\bibfnamefont
  {M.}~\bibnamefont {Atature}},\ }\href {http://dx.doi.org/10.1038/nature09359}
  {\bibfield  {journal} {\bibinfo  {journal} {Nature}\ }\textbf {\bibinfo
  {volume} {467}},\ \bibinfo {pages} {297} (\bibinfo {year}
  {2010})}\BibitemShut {NoStop}%
\bibitem [{\citenamefont {Greilich}\ \emph {et~al.}(2011)\citenamefont
  {Greilich}, \citenamefont {Carter}, \citenamefont {Kim}, \citenamefont
  {Bracker},\ and\ \citenamefont {Gammon}}]{greilichnatphoton}%
  \BibitemOpen
  \bibfield  {author} {\bibinfo {author} {\bibfnamefont {A.}~\bibnamefont
  {Greilich}}, \bibinfo {author} {\bibfnamefont {S.~G.}\ \bibnamefont
  {Carter}}, \bibinfo {author} {\bibfnamefont {D.}~\bibnamefont {Kim}},
  \bibinfo {author} {\bibfnamefont {A.~S.}\ \bibnamefont {Bracker}}, \ and\
  \bibinfo {author} {\bibfnamefont {D.}~\bibnamefont {Gammon}},\ }\href
  {http://dx.doi.org/10.1038/nphoton.2011.237} {\bibfield  {journal} {\bibinfo
  {journal} {Nature Photon.}\ }\textbf {\bibinfo {volume} {5}},\ \bibinfo
  {pages} {702} (\bibinfo {year} {2011})}\BibitemShut {NoStop}%
\bibitem [{\citenamefont {Weiss}\ \emph {et~al.}(2012)\citenamefont {Weiss},
  \citenamefont {Elzerman}, \citenamefont {Delley}, \citenamefont
  {Miguel-Sanchez},\ and\ \citenamefont {Imamo\ifmmode~\breve{g}\else
  \u{g}\fi{}lu}}]{weissprl}%
  \BibitemOpen
  \bibfield  {author} {\bibinfo {author} {\bibfnamefont {K.~M.}\ \bibnamefont
  {Weiss}}, \bibinfo {author} {\bibfnamefont {J.~M.}\ \bibnamefont {Elzerman}},
  \bibinfo {author} {\bibfnamefont {Y.~L.}\ \bibnamefont {Delley}}, \bibinfo
  {author} {\bibfnamefont {J.}~\bibnamefont {Miguel-Sanchez}}, \ and\ \bibinfo
  {author} {\bibfnamefont {A.}~\bibnamefont {Imamo\ifmmode~\breve{g}\else
  \u{g}\fi{}lu}},\ }\href {\doibase 10.1103/PhysRevLett.109.107401} {\bibfield
  {journal} {\bibinfo  {journal} {Phys. Rev. Lett.}\ }\textbf {\bibinfo
  {volume} {109}},\ \bibinfo {pages} {107401} (\bibinfo {year}
  {2012})}\BibitemShut {NoStop}%
\bibitem [{\citenamefont {Elzerman}\ \emph {et~al.}(2011)\citenamefont
  {Elzerman}, \citenamefont {Weiss}, \citenamefont {Miguel-Sanchez},\ and\
  \citenamefont {Imamo\ifmmode~\check{g}\else \v{g}\fi{}lu}}]{elzermanprl}%
  \BibitemOpen
  \bibfield  {author} {\bibinfo {author} {\bibfnamefont {J.~M.}\ \bibnamefont
  {Elzerman}}, \bibinfo {author} {\bibfnamefont {K.~M.}\ \bibnamefont {Weiss}},
  \bibinfo {author} {\bibfnamefont {J.}~\bibnamefont {Miguel-Sanchez}}, \ and\
  \bibinfo {author} {\bibfnamefont {A.}~\bibnamefont
  {Imamo\ifmmode~\check{g}\else \v{g}\fi{}lu}},\ }\href {\doibase
  10.1103/PhysRevLett.107.017401} {\bibfield  {journal} {\bibinfo  {journal}
  {Phys. Rev. Lett.}\ }\textbf {\bibinfo {volume} {107}},\ \bibinfo {pages}
  {017401} (\bibinfo {year} {2011})}\BibitemShut {NoStop}%
\bibitem [{\citenamefont {Imamoglu}\ \emph {et~al.}(1999)\citenamefont
  {Imamoglu}, \citenamefont {Awschalom}, \citenamefont {Burkard}, \citenamefont
  {DiVincenzo}, \citenamefont {Loss}, \citenamefont {Sherwin},\ and\
  \citenamefont {Small}}]{imaprl99}%
  \BibitemOpen
  \bibfield  {author} {\bibinfo {author} {\bibfnamefont {A.}~\bibnamefont
  {Imamoglu}}, \bibinfo {author} {\bibfnamefont {D.~D.}\ \bibnamefont
  {Awschalom}}, \bibinfo {author} {\bibfnamefont {G.}~\bibnamefont {Burkard}},
  \bibinfo {author} {\bibfnamefont {D.~P.}\ \bibnamefont {DiVincenzo}},
  \bibinfo {author} {\bibfnamefont {D.}~\bibnamefont {Loss}}, \bibinfo {author}
  {\bibfnamefont {M.}~\bibnamefont {Sherwin}}, \ and\ \bibinfo {author}
  {\bibfnamefont {A.}~\bibnamefont {Small}},\ }\href {\doibase
  10.1103/PhysRevLett.83.4204} {\bibfield  {journal} {\bibinfo  {journal}
  {Phys. Rev. Lett.}\ }\textbf {\bibinfo {volume} {83}},\ \bibinfo {pages}
  {4204} (\bibinfo {year} {1999})}\BibitemShut {NoStop}%
\bibitem [{\citenamefont {Piermarocchi}\ \emph {et~al.}(2002)\citenamefont
  {Piermarocchi}, \citenamefont {Chen}, \citenamefont {Sham},\ and\
  \citenamefont {Steel}}]{shamprl02}%
  \BibitemOpen
  \bibfield  {author} {\bibinfo {author} {\bibfnamefont {C.}~\bibnamefont
  {Piermarocchi}}, \bibinfo {author} {\bibfnamefont {P.}~\bibnamefont {Chen}},
  \bibinfo {author} {\bibfnamefont {L.~J.}\ \bibnamefont {Sham}}, \ and\
  \bibinfo {author} {\bibfnamefont {D.~G.}\ \bibnamefont {Steel}},\ }\href
  {\doibase 10.1103/PhysRevLett.89.167402} {\bibfield  {journal} {\bibinfo
  {journal} {Phys. Rev. Lett.}\ }\textbf {\bibinfo {volume} {89}},\ \bibinfo
  {pages} {167402} (\bibinfo {year} {2002})}\BibitemShut {NoStop}%
\bibitem [{\citenamefont {Ladd}\ and\ \citenamefont
  {Yamamoto}(2011)}]{laddyamamotoprb}%
  \BibitemOpen
  \bibfield  {author} {\bibinfo {author} {\bibfnamefont {T.~D.}\ \bibnamefont
  {Ladd}}\ and\ \bibinfo {author} {\bibfnamefont {Y.}~\bibnamefont
  {Yamamoto}},\ }\href {\doibase 10.1103/PhysRevB.84.235307} {\bibfield
  {journal} {\bibinfo  {journal} {Phys. Rev. B}\ }\textbf {\bibinfo {volume}
  {84}},\ \bibinfo {pages} {235307} (\bibinfo {year} {2011})}\BibitemShut
  {NoStop}%
\bibitem [{\citenamefont {Van~Meter}\ \emph {et~al.}(2010)\citenamefont
  {Van~Meter}, \citenamefont {Ladd}, \citenamefont {Fowler},\ and\
  \citenamefont {Yamamoto}}]{van2010distributed}%
  \BibitemOpen
  \bibfield  {author} {\bibinfo {author} {\bibfnamefont {R.}~\bibnamefont
  {Van~Meter}}, \bibinfo {author} {\bibfnamefont {T.~D.}\ \bibnamefont {Ladd}},
  \bibinfo {author} {\bibfnamefont {A.~G.}\ \bibnamefont {Fowler}}, \ and\
  \bibinfo {author} {\bibfnamefont {Y.}~\bibnamefont {Yamamoto}},\ }\href@noop
  {} {\bibfield  {journal} {\bibinfo  {journal} {International Journal of
  Quantum Information}\ }\textbf {\bibinfo {volume} {8}},\ \bibinfo {pages}
  {295} (\bibinfo {year} {2010})}\BibitemShut {NoStop}%
\bibitem [{\citenamefont {Jones}\ \emph {et~al.}(2012)\citenamefont {Jones},
  \citenamefont {Van~Meter}, \citenamefont {Fowler}, \citenamefont {McMahon},
  \citenamefont {Kim}, \citenamefont {Ladd},\ and\ \citenamefont
  {Yamamoto}}]{jones2012layered}%
  \BibitemOpen
  \bibfield  {author} {\bibinfo {author} {\bibfnamefont {N.~C.}\ \bibnamefont
  {Jones}}, \bibinfo {author} {\bibfnamefont {R.}~\bibnamefont {Van~Meter}},
  \bibinfo {author} {\bibfnamefont {A.~G.}\ \bibnamefont {Fowler}}, \bibinfo
  {author} {\bibfnamefont {P.~L.}\ \bibnamefont {McMahon}}, \bibinfo {author}
  {\bibfnamefont {J.}~\bibnamefont {Kim}}, \bibinfo {author} {\bibfnamefont
  {T.~D.}\ \bibnamefont {Ladd}}, \ and\ \bibinfo {author} {\bibfnamefont
  {Y.}~\bibnamefont {Yamamoto}},\ }\href@noop {} {\bibfield  {journal}
  {\bibinfo  {journal} {Phys. Rev. X}\ }\textbf {\bibinfo {volume} {2}},\
  \bibinfo {pages} {031007} (\bibinfo {year} {2012})}\BibitemShut {NoStop}%
\bibitem [{\citenamefont {Petta}\ \emph {et~al.}(2005)\citenamefont {Petta},
  \citenamefont {Johnson}, \citenamefont {Taylor}, \citenamefont {Laird},
  \citenamefont {Yacoby}, \citenamefont {Lukin}, \citenamefont {Marcus},
  \citenamefont {Hanson},\ and\ \citenamefont {Gossard}}]{pettasci}%
  \BibitemOpen
  \bibfield  {author} {\bibinfo {author} {\bibfnamefont {J.~R.}\ \bibnamefont
  {Petta}}, \bibinfo {author} {\bibfnamefont {A.~C.}\ \bibnamefont {Johnson}},
  \bibinfo {author} {\bibfnamefont {J.~M.}\ \bibnamefont {Taylor}}, \bibinfo
  {author} {\bibfnamefont {E.~A.}\ \bibnamefont {Laird}}, \bibinfo {author}
  {\bibfnamefont {A.}~\bibnamefont {Yacoby}}, \bibinfo {author} {\bibfnamefont
  {M.~D.}\ \bibnamefont {Lukin}}, \bibinfo {author} {\bibfnamefont {C.~M.}\
  \bibnamefont {Marcus}}, \bibinfo {author} {\bibfnamefont {M.~P.}\
  \bibnamefont {Hanson}}, \ and\ \bibinfo {author} {\bibfnamefont {A.~C.}\
  \bibnamefont {Gossard}},\ }\href {\doibase 10.1126/science.1116955}
  {\bibfield  {journal} {\bibinfo  {journal} {Science}\ }\textbf {\bibinfo
  {volume} {309}},\ \bibinfo {pages} {2180} (\bibinfo {year}
  {2005})}\BibitemShut {NoStop}%
\bibitem [{\citenamefont {Folleti}\ \emph {et~al.}(2009)\citenamefont
  {Folleti}, \citenamefont {Bluhm}, \citenamefont {Mahalu}, \citenamefont
  {Umansky},\ and\ \citenamefont {Yacoby}}]{folleti}%
  \BibitemOpen
  \bibfield  {author} {\bibinfo {author} {\bibfnamefont {S.}~\bibnamefont
  {Folleti}}, \bibinfo {author} {\bibfnamefont {H.}~\bibnamefont {Bluhm}},
  \bibinfo {author} {\bibfnamefont {D.}~\bibnamefont {Mahalu}}, \bibinfo
  {author} {\bibfnamefont {V.}~\bibnamefont {Umansky}}, \ and\ \bibinfo
  {author} {\bibfnamefont {A.}~\bibnamefont {Yacoby}},\ }\href@noop {}
  {\bibfield  {journal} {\bibinfo  {journal} {Nature Phys.}\ }\textbf {\bibinfo
  {volume} {5}},\ \bibinfo {pages} {903} (\bibinfo {year} {2009})}\BibitemShut
  {NoStop}%
\bibitem [{\citenamefont {Bluhm}\ \emph {et~al.}(2011)\citenamefont {Bluhm},
  \citenamefont {Foletti}, \citenamefont {Neder}, \citenamefont {Rudner},
  \citenamefont {Mahalu}, \citenamefont {Umansky},\ and\ \citenamefont
  {Yacoby}}]{bluhm_dd}%
  \BibitemOpen
  \bibfield  {author} {\bibinfo {author} {\bibfnamefont {H.}~\bibnamefont
  {Bluhm}}, \bibinfo {author} {\bibfnamefont {S.}~\bibnamefont {Foletti}},
  \bibinfo {author} {\bibfnamefont {I.}~\bibnamefont {Neder}}, \bibinfo
  {author} {\bibfnamefont {M.}~\bibnamefont {Rudner}}, \bibinfo {author}
  {\bibfnamefont {D.}~\bibnamefont {Mahalu}}, \bibinfo {author} {\bibfnamefont
  {V.}~\bibnamefont {Umansky}}, \ and\ \bibinfo {author} {\bibfnamefont
  {A.}~\bibnamefont {Yacoby}},\ }\href {http://dx.doi.org/10.1038/nphys1856}
  {\bibfield  {journal} {\bibinfo  {journal} {Nature Phys.}\ }\textbf {\bibinfo
  {volume} {7}},\ \bibinfo {pages} {109} (\bibinfo {year} {2011})}\BibitemShut
  {NoStop}%
\bibitem [{\citenamefont {Maune}\ \emph {et~al.}(2012)\citenamefont {Maune},
  \citenamefont {Borselli}, \citenamefont {Huang}, \citenamefont {Ladd},
  \citenamefont {Deelman}, \citenamefont {Holabird}, \citenamefont {Kiselev},
  \citenamefont {Alvarado-Rodriguez}, \citenamefont {Ross}, \citenamefont
  {Schmitz}, \citenamefont {Sokolich}, \citenamefont {Watson}, \citenamefont
  {Gyure},\ and\ \citenamefont {Hunter}}]{maunenat}%
  \BibitemOpen
  \bibfield  {author} {\bibinfo {author} {\bibfnamefont {B.~M.}\ \bibnamefont
  {Maune}}, \bibinfo {author} {\bibfnamefont {M.~G.}\ \bibnamefont {Borselli}},
  \bibinfo {author} {\bibfnamefont {B.}~\bibnamefont {Huang}}, \bibinfo
  {author} {\bibfnamefont {T.~D.}\ \bibnamefont {Ladd}}, \bibinfo {author}
  {\bibfnamefont {P.~W.}\ \bibnamefont {Deelman}}, \bibinfo {author}
  {\bibfnamefont {K.~S.}\ \bibnamefont {Holabird}}, \bibinfo {author}
  {\bibfnamefont {A.~A.}\ \bibnamefont {Kiselev}}, \bibinfo {author}
  {\bibfnamefont {I.}~\bibnamefont {Alvarado-Rodriguez}}, \bibinfo {author}
  {\bibfnamefont {R.~S.}\ \bibnamefont {Ross}}, \bibinfo {author}
  {\bibfnamefont {A.~E.}\ \bibnamefont {Schmitz}}, \bibinfo {author}
  {\bibfnamefont {M.}~\bibnamefont {Sokolich}}, \bibinfo {author}
  {\bibfnamefont {C.~A.}\ \bibnamefont {Watson}}, \bibinfo {author}
  {\bibfnamefont {M.~F.}\ \bibnamefont {Gyure}}, \ and\ \bibinfo {author}
  {\bibfnamefont {A.~T.}\ \bibnamefont {Hunter}},\ }\href
  {http://dx.doi.org/10.1038/nature10707} {\bibfield  {journal} {\bibinfo
  {journal} {Nature}\ }\textbf {\bibinfo {volume} {481}},\ \bibinfo {pages}
  {344} (\bibinfo {year} {2012})}\BibitemShut {NoStop}%
\bibitem [{\citenamefont {Gaudreau}\ \emph {et~al.}(2012)\citenamefont
  {Gaudreau}, \citenamefont {Granger}, \citenamefont {Kam}, \citenamefont
  {Aers}, \citenamefont {Studenikin}, \citenamefont {Zawadzki}, \citenamefont
  {Pioro-Ladriere}, \citenamefont {Wasilewski},\ and\ \citenamefont
  {Sachrajda}}]{gaudreaunphys}%
  \BibitemOpen
  \bibfield  {author} {\bibinfo {author} {\bibfnamefont {L.}~\bibnamefont
  {Gaudreau}}, \bibinfo {author} {\bibfnamefont {G.}~\bibnamefont {Granger}},
  \bibinfo {author} {\bibfnamefont {A.}~\bibnamefont {Kam}}, \bibinfo {author}
  {\bibfnamefont {G.~C.}\ \bibnamefont {Aers}}, \bibinfo {author}
  {\bibfnamefont {S.~A.}\ \bibnamefont {Studenikin}}, \bibinfo {author}
  {\bibfnamefont {P.}~\bibnamefont {Zawadzki}}, \bibinfo {author}
  {\bibfnamefont {M.}~\bibnamefont {Pioro-Ladriere}}, \bibinfo {author}
  {\bibfnamefont {Z.~R.}\ \bibnamefont {Wasilewski}}, \ and\ \bibinfo {author}
  {\bibfnamefont {A.~S.}\ \bibnamefont {Sachrajda}},\ }\href
  {http://dx.doi.org/10.1038/nphys2149} {\bibfield  {journal} {\bibinfo
  {journal} {Nature Phys.}\ }\textbf {\bibinfo {volume} {8}},\ \bibinfo {pages}
  {54} (\bibinfo {year} {2012})}\BibitemShut {NoStop}%
\bibitem [{\citenamefont {Legero}\ \emph {et~al.}(2003)\citenamefont {Legero},
  \citenamefont {Wilk}, \citenamefont {Kuhn},\ and\ \citenamefont
  {Rempe}}]{legero03}%
  \BibitemOpen
  \bibfield  {author} {\bibinfo {author} {\bibfnamefont {T.}~\bibnamefont
  {Legero}}, \bibinfo {author} {\bibfnamefont {T.}~\bibnamefont {Wilk}},
  \bibinfo {author} {\bibfnamefont {A.}~\bibnamefont {Kuhn}}, \ and\ \bibinfo
  {author} {\bibfnamefont {G.}~\bibnamefont {Rempe}},\ }\href@noop {}
  {\bibfield  {journal} {\bibinfo  {journal} {Appl. Phys. B}\ }\textbf
  {\bibinfo {volume} {77}},\ \bibinfo {pages} {797} (\bibinfo {year}
  {2003})}\BibitemShut {NoStop}%
\bibitem [{\citenamefont {Ates}\ \emph {et~al.}(2012)\citenamefont {Ates},
  \citenamefont {Agha}, \citenamefont {Gulinatti}, \citenamefont {Rech},
  \citenamefont {Rakher}, \citenamefont {Badolato},\ and\ \citenamefont
  {Srinivasan}}]{PhysRevLett.109.147405}%
  \BibitemOpen
  \bibfield  {author} {\bibinfo {author} {\bibfnamefont {S.}~\bibnamefont
  {Ates}}, \bibinfo {author} {\bibfnamefont {I.}~\bibnamefont {Agha}}, \bibinfo
  {author} {\bibfnamefont {A.}~\bibnamefont {Gulinatti}}, \bibinfo {author}
  {\bibfnamefont {I.}~\bibnamefont {Rech}}, \bibinfo {author} {\bibfnamefont
  {M.~T.}\ \bibnamefont {Rakher}}, \bibinfo {author} {\bibfnamefont
  {A.}~\bibnamefont {Badolato}}, \ and\ \bibinfo {author} {\bibfnamefont
  {K.}~\bibnamefont {Srinivasan}},\ }\href {\doibase
  10.1103/PhysRevLett.109.147405} {\bibfield  {journal} {\bibinfo  {journal}
  {Phys. Rev. Lett.}\ }\textbf {\bibinfo {volume} {109}},\ \bibinfo {pages}
  {147405} (\bibinfo {year} {2012})}\BibitemShut {NoStop}%
\bibitem [{\citenamefont {{Borselli}}\ \emph {et~al.}(2014)\citenamefont
  {{Borselli}}, \citenamefont {{Eng}}, \citenamefont {{Ross}}, \citenamefont
  {{Hazard}}, \citenamefont {{Holabird}}, \citenamefont {{Huang}},
  \citenamefont {{Kiselev}}, \citenamefont {{Deelman}}, \citenamefont
  {{Warren}}, \citenamefont {{Milosavljevic}}, \citenamefont {{Schmitz}},
  \citenamefont {{Sokolich}}, \citenamefont {{Gyure}},\ and\ \citenamefont
  {{Hunter}}}]{Borselli_ArXiv_2014}%
  \BibitemOpen
  \bibfield  {author} {\bibinfo {author} {\bibfnamefont {M.~G.}\ \bibnamefont
  {{Borselli}}}, \bibinfo {author} {\bibfnamefont {K.}~\bibnamefont {{Eng}}},
  \bibinfo {author} {\bibfnamefont {R.~S.}\ \bibnamefont {{Ross}}}, \bibinfo
  {author} {\bibfnamefont {T.~M.}\ \bibnamefont {{Hazard}}}, \bibinfo {author}
  {\bibfnamefont {K.~S.}\ \bibnamefont {{Holabird}}}, \bibinfo {author}
  {\bibfnamefont {B.}~\bibnamefont {{Huang}}}, \bibinfo {author} {\bibfnamefont
  {A.~A.}\ \bibnamefont {{Kiselev}}}, \bibinfo {author} {\bibfnamefont {P.~W.}\
  \bibnamefont {{Deelman}}}, \bibinfo {author} {\bibfnamefont {L.~D.}\
  \bibnamefont {{Warren}}}, \bibinfo {author} {\bibfnamefont {I.}~\bibnamefont
  {{Milosavljevic}}}, \bibinfo {author} {\bibfnamefont {A.~E.}\ \bibnamefont
  {{Schmitz}}}, \bibinfo {author} {\bibfnamefont {M.}~\bibnamefont
  {{Sokolich}}}, \bibinfo {author} {\bibfnamefont {M.~F.}\ \bibnamefont
  {{Gyure}}}, \ and\ \bibinfo {author} {\bibfnamefont {A.~T.}\ \bibnamefont
  {{Hunter}}},\ }\href@noop {} {\bibfield  {journal} {\bibinfo  {journal}
  {arXiv:1408.0600}\ } (\bibinfo {year} {2014})}\BibitemShut {NoStop}%
\bibitem [{\citenamefont {Thon}\ \emph {et~al.}(2009)\citenamefont {Thon},
  \citenamefont {Rakher}, \citenamefont {Kim}, \citenamefont {Gudat},
  \citenamefont {Irvine}, \citenamefont {Petroff},\ and\ \citenamefont
  {Bouwmeester}}]{thon_apl}%
  \BibitemOpen
  \bibfield  {author} {\bibinfo {author} {\bibfnamefont {S.~M.}\ \bibnamefont
  {Thon}}, \bibinfo {author} {\bibfnamefont {M.~T.}\ \bibnamefont {Rakher}},
  \bibinfo {author} {\bibfnamefont {H.}~\bibnamefont {Kim}}, \bibinfo {author}
  {\bibfnamefont {J.}~\bibnamefont {Gudat}}, \bibinfo {author} {\bibfnamefont
  {W.~T.~M.}\ \bibnamefont {Irvine}}, \bibinfo {author} {\bibfnamefont {P.~M.}\
  \bibnamefont {Petroff}}, \ and\ \bibinfo {author} {\bibfnamefont
  {D.}~\bibnamefont {Bouwmeester}},\ }\href {\doibase
  http://dx.doi.org/10.1063/1.3103885} {\bibfield  {journal} {\bibinfo
  {journal} {Appl. Phys. Lett.}\ }\textbf {\bibinfo {volume} {94}},\ \bibinfo
  {eid} {111115} (\bibinfo {year} {2009})}\BibitemShut {NoStop}%
\bibitem [{\citenamefont {Schneider}\ \emph {et~al.}(2009)\citenamefont
  {Schneider}, \citenamefont {Huggenberger}, \citenamefont {Sünner},
  \citenamefont {Heindel}, \citenamefont {Strauss}, \citenamefont {Göpfert},
  \citenamefont {Weinmann}, \citenamefont {Reitzenstein}, \citenamefont
  {Worschech}, \citenamefont {Kamp}, \citenamefont {Höfling},\ and\
  \citenamefont {Forchel}}]{schnano}%
  \BibitemOpen
  \bibfield  {author} {\bibinfo {author} {\bibfnamefont {C.}~\bibnamefont
  {Schneider}}, \bibinfo {author} {\bibfnamefont {A.}~\bibnamefont
  {Huggenberger}}, \bibinfo {author} {\bibfnamefont {T.}~\bibnamefont
  {Sünner}}, \bibinfo {author} {\bibfnamefont {T.}~\bibnamefont {Heindel}},
  \bibinfo {author} {\bibfnamefont {M.}~\bibnamefont {Strauss}}, \bibinfo
  {author} {\bibfnamefont {S.}~\bibnamefont {Göpfert}}, \bibinfo {author}
  {\bibfnamefont {P.}~\bibnamefont {Weinmann}}, \bibinfo {author}
  {\bibfnamefont {S.}~\bibnamefont {Reitzenstein}}, \bibinfo {author}
  {\bibfnamefont {L.}~\bibnamefont {Worschech}}, \bibinfo {author}
  {\bibfnamefont {M.}~\bibnamefont {Kamp}}, \bibinfo {author} {\bibfnamefont
  {S.}~\bibnamefont {Höfling}}, \ and\ \bibinfo {author} {\bibfnamefont
  {A.}~\bibnamefont {Forchel}},\ }\href
  {http://stacks.iop.org/0957-4484/20/i=43/a=434012} {\bibfield  {journal}
  {\bibinfo  {journal} {Nanotechnology}\ }\textbf {\bibinfo {volume} {20}},\
  \bibinfo {pages} {434012} (\bibinfo {year} {2009})}\BibitemShut {NoStop}%
\bibitem [{\citenamefont {Vora}\ \emph {et~al.}(2015)\citenamefont {Vora},
  \citenamefont {Bracker}, \citenamefont {Carter}, \citenamefont {Sweeney},
  \citenamefont {Kim}, \citenamefont {Kim}, \citenamefont {Yang}, \citenamefont
  {Brereton}, \citenamefont {Economou},\ and\ \citenamefont
  {Gammon}}]{vorancomms}%
  \BibitemOpen
  \bibfield  {author} {\bibinfo {author} {\bibfnamefont {P.~M.}\ \bibnamefont
  {Vora}}, \bibinfo {author} {\bibfnamefont {A.}~\bibnamefont {Bracker}},
  \bibinfo {author} {\bibfnamefont {S.}~\bibnamefont {Carter}}, \bibinfo
  {author} {\bibfnamefont {T.}~\bibnamefont {Sweeney}}, \bibinfo {author}
  {\bibfnamefont {M.}~\bibnamefont {Kim}}, \bibinfo {author} {\bibfnamefont
  {C.~S.}\ \bibnamefont {Kim}}, \bibinfo {author} {\bibfnamefont
  {L.}~\bibnamefont {Yang}}, \bibinfo {author} {\bibfnamefont {P.}~\bibnamefont
  {Brereton}}, \bibinfo {author} {\bibfnamefont {S.}~\bibnamefont {Economou}},
  \ and\ \bibinfo {author} {\bibfnamefont {D.}~\bibnamefont {Gammon}},\
  }\href@noop {} {\bibfield  {journal} {\bibinfo  {journal} {Nat. Commun.}\
  }\textbf {\bibinfo {volume} {6:7665}} (\bibinfo {year} {2015})}\BibitemShut
  {NoStop}%
\bibitem [{\citenamefont {West}\ and\ \citenamefont
  {Fong}(2012)}]{westfongnjp}%
  \BibitemOpen
  \bibfield  {author} {\bibinfo {author} {\bibfnamefont {J.~R.}\ \bibnamefont
  {West}}\ and\ \bibinfo {author} {\bibfnamefont {B.~H.}\ \bibnamefont
  {Fong}},\ }\href@noop {} {\bibfield  {journal} {\bibinfo  {journal} {New J.
  Phys.}\ }\textbf {\bibinfo {volume} {14}},\ \bibinfo {pages} {083002}
  (\bibinfo {year} {2012})}\BibitemShut {NoStop}%
\bibitem [{\citenamefont {Benyoucef}\ and\ \citenamefont
  {Reithmaier}(2013)}]{benyoucef13}%
  \BibitemOpen
  \bibfield  {author} {\bibinfo {author} {\bibfnamefont {M.}~\bibnamefont
  {Benyoucef}}\ and\ \bibinfo {author} {\bibfnamefont {J.~P.}\ \bibnamefont
  {Reithmaier}},\ }\href@noop {} {\bibfield  {journal} {\bibinfo  {journal}
  {Semicond. Sci. Technol.}\ }\textbf {\bibinfo {volume} {28}},\ \bibinfo
  {pages} {094004} (\bibinfo {year} {2013})}\BibitemShut {NoStop}%
\bibitem [{\citenamefont {Wang}\ \emph {et~al.}(2014)\citenamefont {Wang},
  \citenamefont {Bishop}, \citenamefont {Barnes}, \citenamefont {Kestner},\
  and\ \citenamefont {Sarma}}]{wang_sarma_pra}%
  \BibitemOpen
  \bibfield  {author} {\bibinfo {author} {\bibfnamefont {X.}~\bibnamefont
  {Wang}}, \bibinfo {author} {\bibfnamefont {L.~S.}\ \bibnamefont {Bishop}},
  \bibinfo {author} {\bibfnamefont {E.}~\bibnamefont {Barnes}}, \bibinfo
  {author} {\bibfnamefont {J.~P.}\ \bibnamefont {Kestner}}, \ and\ \bibinfo
  {author} {\bibfnamefont {S.~D.}\ \bibnamefont {Sarma}},\ }\href {\doibase
  10.1103/PhysRevA.89.022310} {\bibfield  {journal} {\bibinfo  {journal} {Phys.
  Rev. A}\ }\textbf {\bibinfo {volume} {89}},\ \bibinfo {pages} {022310}
  (\bibinfo {year} {2014})}\BibitemShut {NoStop}%
\bibitem [{\citenamefont {Medford}\ \emph
  {et~al.}(2013{\natexlab{b}})\citenamefont {Medford}, \citenamefont {Beil},
  \citenamefont {Taylor}, \citenamefont {Rashba}, \citenamefont {Lu},
  \citenamefont {Gossard},\ and\ \citenamefont {Marcus}}]{medfordprl}%
  \BibitemOpen
  \bibfield  {author} {\bibinfo {author} {\bibfnamefont {J.}~\bibnamefont
  {Medford}}, \bibinfo {author} {\bibfnamefont {J.}~\bibnamefont {Beil}},
  \bibinfo {author} {\bibfnamefont {J.~M.}\ \bibnamefont {Taylor}}, \bibinfo
  {author} {\bibfnamefont {E.~I.}\ \bibnamefont {Rashba}}, \bibinfo {author}
  {\bibfnamefont {H.}~\bibnamefont {Lu}}, \bibinfo {author} {\bibfnamefont
  {A.~C.}\ \bibnamefont {Gossard}}, \ and\ \bibinfo {author} {\bibfnamefont
  {C.~M.}\ \bibnamefont {Marcus}},\ }\href {\doibase
  10.1103/PhysRevLett.111.050501} {\bibfield  {journal} {\bibinfo  {journal}
  {Phys. Rev. Lett.}\ }\textbf {\bibinfo {volume} {111}},\ \bibinfo {pages}
  {050501} (\bibinfo {year} {2013}{\natexlab{b}})}\BibitemShut {NoStop}%
\end{thebibliography}%
